\documentclass[twocolumn]{aastex701}

\usepackage[figuresright]{rotating} 
\usepackage{array}
\usepackage{lipsum}  

\newcolumntype{L}[1]{>{\raggedright\let\newline\\\arraybackslash\hspace{0pt}}m{#1}}
\newcolumntype{C}[1]{>{\centering\let\newline\\\arraybackslash\hspace{0pt}}m{#1}}
\newcolumntype{R}[1]{>{\raggedleft\let\newline\\\arraybackslash\hspace{0pt}}m{#1}}

\newcommand{\A}{\text{\normalfont\AA}}

\def \hbeta{H$\beta$}

\def \ebmvs{ E_{s}{\rm (B-V)} }
\def \ebmvn{ E_{n}{\rm (B-V)} }
\def \halpha{H$\alpha$}
\def \hgamma{H$\gamma$}
\def \Msol{{M}_{\odot}}

\def \logm{\log(M/\Msol)}
\def \lya{Ly$\alpha$}

\def \h2{{\rm H_{2}}}

\def \hbeta{H$\beta$}
\def \halpha{H$\alpha$}
\def \sii{[\ion{S}{2}]}

\def \cplus{[C\,{\sc ii}]}

\def \oii{[\ion{O}{2}]}
\def \oiii{[\ion{O}{3}]}
\def \nii{[\ion{N}{2}]}

\def \dn4000{D_{{\rm n}}(4000) }

\def \x{$\times$}

\def \Rtwothree{([\ion{O}{3}]$_{\rm 5007}$ + [\ion{O}{3}]$_{\rm 4959}$ + [\ion{O}{2}]$_{\rm 3727}$)/H$\beta$}
\def \Rthree{([\ion{O}{3}]$_{\rm 5007}$ + [\ion{O}{3}]$_{\rm 4959}$)/H$\beta$}
\def \Rtwo{[\ion{O}{2}]$_{\rm 3727}$/H$\beta$}
\def \Othreetwo{[\ion{O}{3}]$_{\rm 5007}$/[\ion{O}{2}]$_{\rm 3727}$}
\def \Ntwo{[\ion{N}{2}]$_{\rm 6585}$/H$\alpha$}
\def \Stwo{([\ion{S}{2}]$_{\rm 6732}$ + [\ion{S}{2}]$_{\rm 6718}$)/H$\alpha$}
\def \SR{[\ion{S}{2}]$_{\rm 6732}$/[\ion{S}{2}]$_{\rm 6718}$}

\graphicspath{{./}{figures/}}

\begin{document}

\title{The ALPINE-CRISTAL-JWST Survey:\\
JWST/IFU Optical Observations for 18 Main-Sequence Galaxies at \boldmath{$z=4-6$}}

\suppressAffiliations


\author[0000-0002-9382-9832]{A. L. Faisst}
\affiliation{IPAC, California Institute of Technology, 1200 E. California Blvd., Pasadena, CA 91125, USA}
\email{afaisst@caltech.edu}
\correspondingauthor{Andreas L. Faisst}

\author[0000-0001-7201-5066]{S. Fujimoto}
\affiliation{Department of Astronomy, The University of Texas at Austin, Austin, TX 78712, USA}
\affiliation{David A. Dunlap Department of Astronomy and Astrophysics, University of Toronto, 50 St. George Street, Toronto, Ontario, M5S 3H4, Canada}
\email{}

\author[0000-0002-0498-5041]{A. Tsujita}
\affiliation{Department of Astronomy, School of Science, SOKENDAI (The Graduate University for Advanced Studies), 2-21-1 Osawa, Mitaka, Tokyo 181-8588, Japan}
\email{}

\author[0000-0002-7964-6749]{W. Wang}
\affiliation{Caltech/IPAC, 1200 E. California Blvd. Pasadena, CA 91125, USA}
\email{}


\author[0000-0002-7755-8649]{N. Nezhad}
\affiliation{Department of Physics and Astronomy, University of California, Riverside, 900 University Avenue, Riverside, CA 92521, USA}
\email{}

\author[0000-0002-8858-6784]{F. Loiacono}
\affiliation{INAF – Osservatorio di Astrofisica e Scienza dello Spazio di Bologna, Via Gobetti 93/3, 40129 Bologna, Italy}
\email{}

\author[0000-0003-4891-0794]{H. \"Ubler}
\affiliation{Max-Planck-Institute f\"ur extratarrestrische Physik, Giessenbachstrasse 1, 85748 Garching, Germany}
\email{}

\author[0000-0002-3915-2015]{M. B\'ethermin}
\affiliation{Universit\'e de Strasbourg, CNRS, Observatoire astronomique de Strasbourg, UMR 7550, 67000 Strasbourg, France}
\email{}

\author[0000-0002-6716-4400]{P. Cassata}
\affiliation{Dipartimento di Fisica e Astronomia Galileo Galilei Universit{\`a} degli Studi di Padova, Vicolo dell’Osservatorio 3, 35122 Padova Italy}
\affiliation{INAF - Osservatorio Astronomico di Padova, Vicolo dell’Osservatorio 5, I-35122, Padova, Italy}
\email{}

\author[0000-0003-0348-2917]{M. Dessauges-Zavadsky}
\affiliation{Observatoire de Gen\`eve, Universit\'e de Gen\`eve, Chemin Pegasi 51, 1290 Versoix, Switzerland}
\email{}

\author[0000-0002-2775-0595]{R. Herrera-Camus}
\affiliation{Departamento de Astronom{\'i}a, Universidad de Concepci{\'o}n, Barrio Universitario, Concepci{\'o}n, Chile}
\affiliation{Millenium Nucleus for Galaxies (MINGAL), Av. Ej\'ercito 441, Santiago 8370191, Chile}
\email{}

\author[0000-0001-7144-7182]{D. Schaerer}
\affiliation{Observatoire de Gen\`eve, Universit\'e de Gen\`eve, Chemin Pegasi 51, 1290 Versoix, Switzerland}
\email{}

\author[0000-0002-0000-6977]{J. Silverman}
\affiliation{Kavli Institute for the Physics and Mathematics of the Universe (Kavli IPMU, WPI), UTIAS, Tokyo Institutes for Advanced Study, University of Tokyo, Chiba, 277-8583, Japan}
\affiliation{Department of Astronomy, School of Science, The University of Tokyo, 7-3-1 Hongo, Bunkyo, Tokyo 113-0033, Japan}
\affiliation{Center for Data-Driven Discovery, Kavli IPMU (WPI), UTIAS, The University of Tokyo, Kashiwa, Chiba 277-8583, Japan}
\affiliation{Center for Astrophysical Sciences, Department of Physics \& Astronomy, Johns Hopkins University, Baltimore, MD 21218, USA}
\email{}

\author[0000-0003-1710-9339]{L. Yan}
\affil{Caltech Optical Observatories, California Institute of Technology, Pasadena, CA 91125, USA}
\email{}

\author[0000-0002-6290-3198]{M. Aravena}
\affiliation{Instituto de Estudios Astrof\'isicos, Facultad de Ingenier\'ia y Ciencias, Universidad Diego Portales, Av. Ej\'ercito Libertador 441, Santiago 8370191, Chile}
\affiliation{Millenium Nucleus for Galaxies (MINGAL), Av. Ej\'ercito 441, Santiago 8370191, Chile}
\email{}

\author[0000-0001-9419-6355]{I. De Looze}
\affiliation{Sterrenkundig Observatorium, Ghent University, Krijgslaan 281 - S9, B-9000 Gent, Belgium}
\email{}

\author[0000-0003-4264-3381]{N. M. F\"orster Schreiber}
\affiliation{Max-Planck-Institute f\"ur extratarrestrische Physik, Giessenbachstrasse 1, 85748 Garching, Germany}
\email{}

\author{J. Gonz\'alez-L\'opez}
\affiliation{Instituto de Astrof\'isica, Facultad de Física, Pontificia Universidad Cat\'olica de Chile, Santiago 7820436, Chile 3 Las Campanas}
\affiliation{Observatory, Carnegie Institution of Washington, Ra\'u Bitr\'an 1200, La Serena, Chile}
\email{}

\author[0000-0003-3256-5615]{J. Spilker}
\affiliation{Department of Physics and Astronomy and George P. and Cynthia Woods Mitchell Institute for Fundamental Physics and Astronomy, Texas A\&M University, 4242}
\email{}

\author[0000-0001-9728-8909]{K. Tadaki}
\affiliation{Faculty of Engineering, Hokkai-Gakuen University, Toyohira-ku, Sapporo 062-8605, Japan}
\email{}

\author[0000-0002-0930-6466]{C. M. Casey}
\affiliation{Department of Physics, University of California, Santa Barbara, Santa Barbara, CA 93109, USA}
\affiliation{Department of Astronomy, The University of Texas at Austin, Austin, TX 78712, USA}
\affiliation{Cosmic Dawn Center (DAWN), Copenhagen, Denmark}
\email{}

\author[0000-0002-3560-8599]{M. Franco}
\affiliation{Universit\'e Paris-Saclay, Universit\'e Paris Cit\'e, CEA, CNRS, AIM, 91191 Gif-sur-Yvette, France}
\email{}

\author[0000-0003-0129-2079]{S. Harish}
\affiliation{Laboratory for Multiwavelength Astrophysics, School of Physics and Astronomy, Rochester Institute of Technology, 84 Lomb Memorial Drive, Rochester, NY 14623, USA}
\email{}

\author[0000-0002-9489-7765]{H. J. McCracken}
\affiliation{Institut d’Astrophysique de Paris, UMR 7095, CNRS, and Sorbonne Universit\'e, 98 bis boulevard Arago, F-75014 Paris, France}
\email{}

\author[0000-0001-9187-3605]{J. S. Kartaltepe}
\affiliation{Laboratory for Multiwavelength Astrophysics, School of Physics and Astronomy, Rochester Institute of Technology, 84 Lomb Memorial Drive, Rochester, NY 14623, USA}
\email{}

\author[0000-0002-6610-2048]{A. M. Koekemoer}
\affiliation{Space Telescope Science Institute, 3700 San Martin Dr., Baltimore, MD 21218, USA} 
\email{}

\author[0000-0002-0101-336X]{A. A. Khostovan}
\affiliation{Department of Physics and Astronomy, University of Kentucky, 505 Rose Street, Lexington, KY 40506, USA}
\affiliation{Laboratory for Multiwavelength Astrophysics, School of Physics and Astronomy, Rochester Institute of Technology, 84 Lomb Memorial Drive, Rochester, NY 14623, USA}
\email{}

\author[0000-0001-9773-7479]{D. Liu}
\affiliation{Purple Mountain Observatory, Chinese Academy of Sciences, 10 Yuanhua Road, Nanjing 210023, China}
\email{}

\author[0000-0002-4485-8549]{J. Rhodes}
\affiliation{Jet Propulsion Laboratory, California Institute of Technology, 4800 Oak Grove Drive, Pasadena, CA 91001, USA}
\email{}

\author[0000-0002-4271-0364]{B. E. Robertson}
\affiliation{Department of Astronomy and Astrophysics, University of California, Santa Cruz, 1156 High Street, Santa Cruz, CA 95064, USA}
\email{}

\author[0000-0001-5758-1000]{R. Amorin}
\affiliation{Instituto de Investigaci\`on Multidisciplinar en Ciencia y Tecnología, Universidad de La Serena, Ra\'ul Bitr\'an 1305, La Serena, Chile}
\affiliation{Departamento de Astronom\'ia, Universidad de La Serena, Av. Juan Cisternas 1200 Norte, La Serena, Chile}
\email{}

\author[0000-0002-9508-3667]{R. J. Assef}
\affiliation{Instituto de Estudios Astrof\'isicos, Facultad de Ingenier\'ia y Ciencias, Universidad Diego Portales, Av. Ej\'ercito Libertador 441, Santiago 8370191, Chile}
\email{}

\author[0000-0003-4569-2285]{A. J. Battisti}
\affiliation{International Centre for Radio Astronomy Research (ICRAR), The University of Western Australia, M468, 35 Stirling Highway, Crawley, WA 6009, Australia}
\affiliation{Research School of Astronomy and Astrophysics, Australian National University, Cotter Road, Weston Creek, ACT 2611, Australia}
\affiliation{ARC Center of Excellence for All Sky Astrophysics in 3 Dimensions (ASTRO 3D), Australia}
\email{}

\author[0000-0002-3272-7568]{J. E. Birkin}
\affiliation{Department of Physics and Astronomy and George P. and Cynthia Woods Mitchell Institute for Fundamental Physics and Astronomy, Texas A\&M University, 4242}
\email{}

\author[0000-0003-0946-6176]{M. Boquien}
\affil{Université Côte d'Azur, Observatoire de la Côte d'Azur, CNRS, Laboratoire Lagrange, 06000, Nice, France}
\email{}

\author[0000-0001-9759-4797]{E. Da Cunha}
\affiliation{International Centre for Radio Astronomy Research (ICRAR), The University of Western Australia, M468, 35 Stirling Highway, Crawley, WA 6009, Australia}
\affiliation{ARC Center of Excellence for All Sky Astrophysics in 3 Dimensions (ASTRO 3D), Australia}
\email{}

\author[0009-0007-7842-9930]{P. Dam}
\affiliation{Dipartimento di Fisica e Astronomia Galileo Galilei Universit{\`a} degli Studi di Padova, Vicolo dell’Osservatorio 3, 35122 Padova Italy}
\email{}

\author[0000-0002-3324-4824]{R. L. Davies}
\affiliation{Centre for Astrophysics and Supercomputing, Swinburne University of Technology, Hawthorn, Victoria 3122, Australia}
\email{}

\author[0009-0008-1835-7557]{D. A. G\'{o}mez-Espinoza}
\affiliation{Instituto de F\'{i}sica y Astronom\'{i}a, Universidad de Valpara\'{i}so, Avda. Gran Breta\~{n}a 1111, Valpara\'{i}so, Chile}
\email{}

\author[0000-0002-9400-7312]{A. Ferrara}
\affiliation{Scuola Normale Superiore, Piazza dei Cavalieri 7, I-56126 Pisa, Italy}
\email{}

\author[0000-0001-7440-8832]{Y. Fudamoto} 
\affiliation{Center for Frontier Science, Chiba University, 1-33 Yayoi-cho, Inage-ku, Chiba 263-8522, Japan}
\email{}

\author[0000-0001-9885-4589]{S. Gillman}
\affiliation{Cosmic Dawn Center (DAWN), Copenhagen, Denmark}
\affiliation{DTU-Space, Technical University of Denmark, Elektrovej 327, DK-2800 Kgs. Lyngby, Denmark}
\email{}

\author[0000-0002-9122-1700]{M. Ginolfi}
\affiliation{Universit\`a di Firenze, Dipartimento di Fisica e Astronomia, via G. Sansone 1, 50019 Sesto Fiorentino, Florence, Italy}
\affiliation{INAF --- Arcetri Astrophysical Observatory, Largo E. Fermi 5, I-50125, Florence, Italy}
\email{}

\author[0000-0002-0236-919X]{G. Gozaliasl}
\affiliation{Department of Computer Science, Aalto University, P.O. Box 15400, FI-00076 Espoo, Finland}
\affiliation{Department of Physics, University of, P.O. Box 64, FI-00014 Helsinki, Finland}
\email{}

\author[0000-0002-5836-4056]{C. Gruppioni}
\affiliation{INAF – Osservatorio di Astrofisica e Scienza dello Spazio di Bologna, Via Gobetti 93/3, 40129 Bologna, Italy}
\email{}

\author[0009-0003-3097-6733]{A. Hadi}
\affiliation{Department of Physics and Astronomy, University of California, Riverside, 900 University Avenue, Riverside, CA 92521, USA}
\email{}

\author[0000-0001-6145-5090]{N. Hathi}
\affiliation{Space Telescope Science Institute, 3700 San Martin Dr., Baltimore, MD 21218, USA} 
\email{}

\author[0009-0008-9801-2224]{E. Ibar}
\affiliation{Instituto de F\'{i}sica y Astronom\'{i}a, Universidad de Valpara\'{i}so, Avda. Gran Breta\~{n}a 1111, Valpara\'{i}so, Chile}
\affiliation{Millenium Nucleus for Galaxies (MINGAL), Av. Ej\'ercito 441, Santiago 8370191, Chile}
\email{}

\author[0000-0002-2634-9169]{R. Ikeda}
\affiliation{Department of Astronomy, School of Science, SOKENDAI (The Graduate University for Advanced Studies), 2-21-1 Osawa, Mitaka, Tokyo 181-8588, Japan}
\affiliation{National Astronomical Observatory of Japan, 2-21-1 Osawa, Mitaka, Tokyo 181-8588, Japan}
\email{}

\author[0000-0003-4268-0393]{H. Inami}
\affiliation{Hiroshima Astrophysical Science Center, Hiroshima University, 1-3-1 Kagamiyama, Higashi-Hiroshima, Hiroshima 739-8526, Japan}
\email{}

\author[0000-0002-0267-9024]{G. C. Jones}
\affiliation{Kavli Institute for Cosmology, University of Cambridge, Madingley Road, Cambridge CB3 0HA, UK}
\affiliation{Cavendish Laboratory, University of Cambridge, 19 JJ Thomson Avenue, Cambridge CB3 0HE, UK}
\email{}

\author[0000-0003-1041-7865]{M. Kohandel}
\affiliation{Scuola Normale Superiore, Piazza dei Cavalieri 7, I-56126 Pisa, Italy}
\email{}

\author{Y. Li}
\affiliation{Department of Physics and Astronomy and George P. and Cynthia Woods Mitchell Institute for Fundamental Physics and Astronomy, Texas A\&M University, 4242}
\email{}

\author[0000-0001-8792-3091]{Y-H. Lin}
\affiliation{Caltech/IPAC, 1200 E. California Blvd. Pasadena, CA 91125, USA}
\email{}

\author[0000-0002-9252-114X]{Z. Liu}
\affiliation{Kavli Institute for the Physics and Mathematics of the Universe (Kavli IPMU, WPI), UTIAS, Tokyo Institutes for Advanced Study, University of Tokyo, Chiba, 277-8583, Japan}
\affiliation{Department of Astronomy, School of Science, The University of Tokyo, 7-3-1 Hongo, Bunkyo, Tokyo 113-0033, Japan}
\affiliation{Center for Data-Driven Discovery, Kavli IPMU (WPI), UTIAS, The University of Tokyo, Kashiwa, Chiba 277-8583, Japan}
\affiliation{Universit\'e Paris-Saclay, Universit\'e Paris Cit\'e, CEA, CNRS, AIM, 91191 Gif-sur-Yvette, France}
\email{}

\author[0009-0004-1270-2373]{L-J. Liu}
\affiliation{Physics Department, California Institute of Technology, 1200 E. California Blvd., Pasadena, CA 91125, USA}
\email{}

\author[0000-0002-7530-8857]{A. S. Long}
\affiliation{Department of Astronomy, The University of Washington, Seattle, WA 98195, USA}
\email{}

\author[0000-0002-4872-2294]{G. Magdis}
\affiliation{Cosmic Dawn Center (DAWN), Copenhagen, Denmark} 
\affiliation{DTU-Space, Technical University of Denmark, Elektrovej 327, DK-2800 Kgs. Lyngby, Denmark}
\affiliation{Niels Bohr Institute, University of Copenhagen, Jagtvej 128, DK-2200, Copenhagen, Denmark}
\email{}

\author[0000-0001-7711-3677]{C. Maraston}
\affiliation{Institute of Cosmology and Gravitation, University of Portsmouth, Dennis Sciama Building, Burnaby Road, Portsmouth, PO13FX, United Kingdom}
\email{}

\author[0000-0001-9189-7818]{C. L. Martin}
\affil{Department of Physics, University of California, Santa Barbara, Santa Barbara, CA 93109, USA}
\email{}

\author[0000-0001-7300-9450]{I. Mitsuhashi}
\affiliation{Department for Astrophysical \& Planetary Science, University of Colorado, Boulder, CO 80309, USA}
\email{}

\author[0000-0001-5846-4404]{B. Mobasher}
\affiliation{Department of Physics and Astronomy, University of California, Riverside, 900 University Avenue, Riverside, CA 92521, USA}
\email{}

\author[0000-0002-8136-8127]{J. Molina}
\affiliation{Instituto de F\'{i}sica y Astronom\'{i}a, Universidad de Valpara\'{i}so, Avda. Gran Breta\~{n}a 1111, Valpara\'{i}so, Chile}
\affiliation{Millenium Nucleus for Galaxies (MINGAL), Av. Ej\'ercito 441, Santiago 8370191, Chile}
\email{}

\author[0000-0001-2345-6789]{A. Nanni}
\affiliation{National Centre for Nuclear Research, ul. Pasteura 7, 02-093 Warsaw, Poland}
\affiliation{INAF - Osservatorio astronomico d'Abruzzo, Via Maggini SNC, 64100, Teramo, Italy}
\email{}

\author[0000-0002-3574-9578]{M. Palla}
\affiliation{Dipartimento di Fisica e Astronomia ``Augusto Righi'', Alma Mater Studiorum, Universi\'a di Bologna, Via Gobetti 93/2, 40129 Bologna, Italy}
\affiliation{INAF – Osservatorio di Astrofisica e Scienza dello Spazio di Bologna, Via Gobetti 93/3, 40129 Bologna, Italy}
\email{}

\author[0000-0002-7129-5761]{A. Pallottini}
\affiliation{Dipartimento di Fisica ``Enrico Fermi'', Universit\'{a} di Pisa, Largo Bruno Pontecorvo 3, Pisa I-56127, Italy}
\email{}

\author[0000-0002-7412-647X]{F. Pozzi}
\affiliation{Dipartimento di Fisica e Astronomia ``Augusto Righi'', Alma Mater Studiorum, Universi\'a di Bologna, Via Gobetti 93/2, 40129 Bologna, Italy}
\affiliation{INAF – Osservatorio di Astrofisica e Scienza dello Spazio di Bologna, Via Gobetti 93/3, 40129 Bologna, Italy}
\email{}

\author[0000-0003-1682-1148]{M. Relano}
\affiliation{Dept. Física Te\'{o}rica y del Cosmos, Campus de Fuentenueva, Edificio Mecenas, Universidad de Granada, E-18071, Granada, Spain}
\affiliation{Instituto Universitario Carlos I de Física Te\'{o}rica y Computacional, Universidad de Granada, 18071, Granada, Spain}
\email{}

\author[0000-0002-3742-6609]{W. Ren}
\affiliation{School of Astronomy and Space Science, University of Science and Technology of China, Hefei 230026, China}
\affiliation{Kavli Institute for the Physics and Mathematics of the Universe (Kavli IPMU, WPI), UTIAS, Tokyo Institutes for Advanced Study, University of Tokyo, Chiba, 277-8583, Japan}
\email{}

\author[0000-0001-9585-1462]{D. A. Riechers}
\affiliation{Institut f\"ur Astrophysik, Universit\"at zu K\"oln, Z\"ulpicher Stra{\ss}e 77, D-50937 K\"oln, Germany}
\email{}

\author[0000-0002-9948-3916]{M. Romano}
\affiliation{Max-Planck-Institut für Radioastronomie, Auf dem H\"ugel 69, 53121 Bonn, Germany}
\affiliation{INAF - Osservatorio Astronomico di Padova, Vicolo dell’Osservatorio 5, I-35122, Padova, Italy}
\email{}

\author[0000-0002-1233-9998]{D. B. Sanders}
\affiliation{Institute for Astronomy, University of Hawaii, 2680 Woodlawn Drive, Honolulu, HI 96822, USA}
\email{}

\author[0000-0002-0498-8074]{P. Sawant}
\affiliation{National Centre for Nuclear Research, ul. Pasteura 7, 02-093 Warsaw, Poland}
\email{}

\author[0000-0002-7087-0701]{M. Shuntov}
\affiliation{Cosmic Dawn Center (DAWN), Copenhagen, Denmark}
\affiliation{Niels Bohr Institute, University of Copenhagen, Jagtvej 128, DK-2200, Copenhagen, Denmark}
\affiliation{University of Geneva, 24 rue du G\'en\'eral-Dufour, 1211 Gen\`eve 4, Switzerland}
\email{}

\author[0000-0001-8034-7802]{R. Smit}
\affiliation{Astrophysics Research Institute, Liverpool John Moores University, 146 Brownlow Hill, Liverpool L3 5RF, UK}
\email{}

\author[0000-0002-2906-2200]{L. Sommovigo}
\affiliation{Center for Computational Astrophysics, Flatiron Institute, 162 5th Avenue, New York, NY 10010, USA}
\email{}

\author[0000-0003-4352-2063]{M. Talia}
\affiliation{Dipartimento di Fisica e Astronomia ``Augusto Righi'', Alma Mater Studiorum, Universi\'a di Bologna, Via Gobetti 93/2, 40129 Bologna, Italy}
\affiliation{INAF – Osservatorio di Astrofisica e Scienza dello Spazio di Bologna, Via Gobetti 93/3, 40129 Bologna, Italy}
\email{}

\author[0000-0002-3258-3672]{L. Vallini}
\affiliation{INAF – Osservatorio di Astrofisica e Scienza dello Spazio di Bologna, Via Gobetti 93/3, 40129 Bologna, Italy}
\email{}

\author[0009-0007-1304-7771]{E. Veraldi}
\affiliation{Scuola Internazionale Superiore Studi Avanzati (SISSA), Physics Area, Via Bonomea 265, 34136 Trieste, Italy}
\email{}

\author[0000-0003-0898-2216]{D. Vergani}
\affiliation{INAF – Osservatorio di Astrofisica e Scienza dello Spazio di Bologna, Via Gobetti 93/3, 40129 Bologna, Italy}
\email{}

\author[0000-0002-1905-4194]{A. P. Vijayan}
\affiliation{Astronomy Centre, University of Sussex, Falmer, Brighton BN1 9QH, UK} 
\email{}

\author[0000-0002-5877-379X]{V. Villanueva}
\affiliation{Departamento de Astronom{\'i}a, Universidad de Concepci{\'o}n, Barrio Universitario, Concepci{\'o}n, Chile}
\email{}

\author[0000-0002-2318-301X]{G. Zamorani}
\affiliation{INAF – Osservatorio di Astrofisica e Scienza dello Spazio di Bologna, Via Gobetti 93/3, 40129 Bologna, Italy}
\email{}

\collaboration{all}{(Affiliations can be found after the references)}






\begin{abstract}
To fully characterize the formation and evolution of galaxies, we need to observe their stars, gas, and dust on resolved spatial scales. We present the ALPINE-CRISTAL-JWST survey, which combines kpc-resolved imaging and spectroscopy from HST, JWST, and ALMA for $18$ representative main-sequence galaxies at $z=4-6$ and $\rm \log(M_*/{\rm M_\odot}) > 9.5$ to study their star formation, chemical properties, and extended gas reservoirs. The co-spatial measurements resolving the ionized gas, molecular gas, stars, and dust on $1-2\,{\rm kpc}$ scales make this a unique benchmark sample for the study of galaxy formation and evolution at $z\sim5$, connecting the Epoch of Reionization with the cosmic noon. 
In this paper, we outline the survey goals and sample selection, and present a summary of the available data for the $18$ galaxies. In addition, we measure spatially integrated quantities (such as global gas metallicity), test different star formation rate indicators, and quantify the presence of \halpha~halos. Our targeted galaxies are relatively metal rich ($10-70\%$ solar), complementary to JWST samples at lower stellar mass, and there is broad agreement between different star formation indicators. One galaxy has the signature of an active galactic nuclei (AGN) based on its emission line ratios. Six show broad \halpha~emission suggesting type 1 AGN candidates. We conclude with an outlook on the exciting science that will be pursued with this unique sample in forthcoming papers. 
\end{abstract}




\section{Introduction} \label{sec:intro}

A top priority in modern astrophysics is to unveil the details of the formation of galaxies in the early Universe and their evolution to the present time, leading to the local cosmic environment we live in and ultimately our own Milky Way.
This includes understanding what sets the star formation efficiency and mass assembly rates, the formation of the most massive and dusty galaxies, the emergence of quiescent galaxies, the contribution of mergers to galaxy growth, the co-evolution of black holes and their hosts, and importantly the properties of the first stars \citep[for useful reviews on these topics see][]{kormendy13,delucia14a,dayal18,tacconi20,robertson22,klessen23}.

\begin{figure*}[t!]
\centering
\includegraphics[angle=0,width=\textwidth]{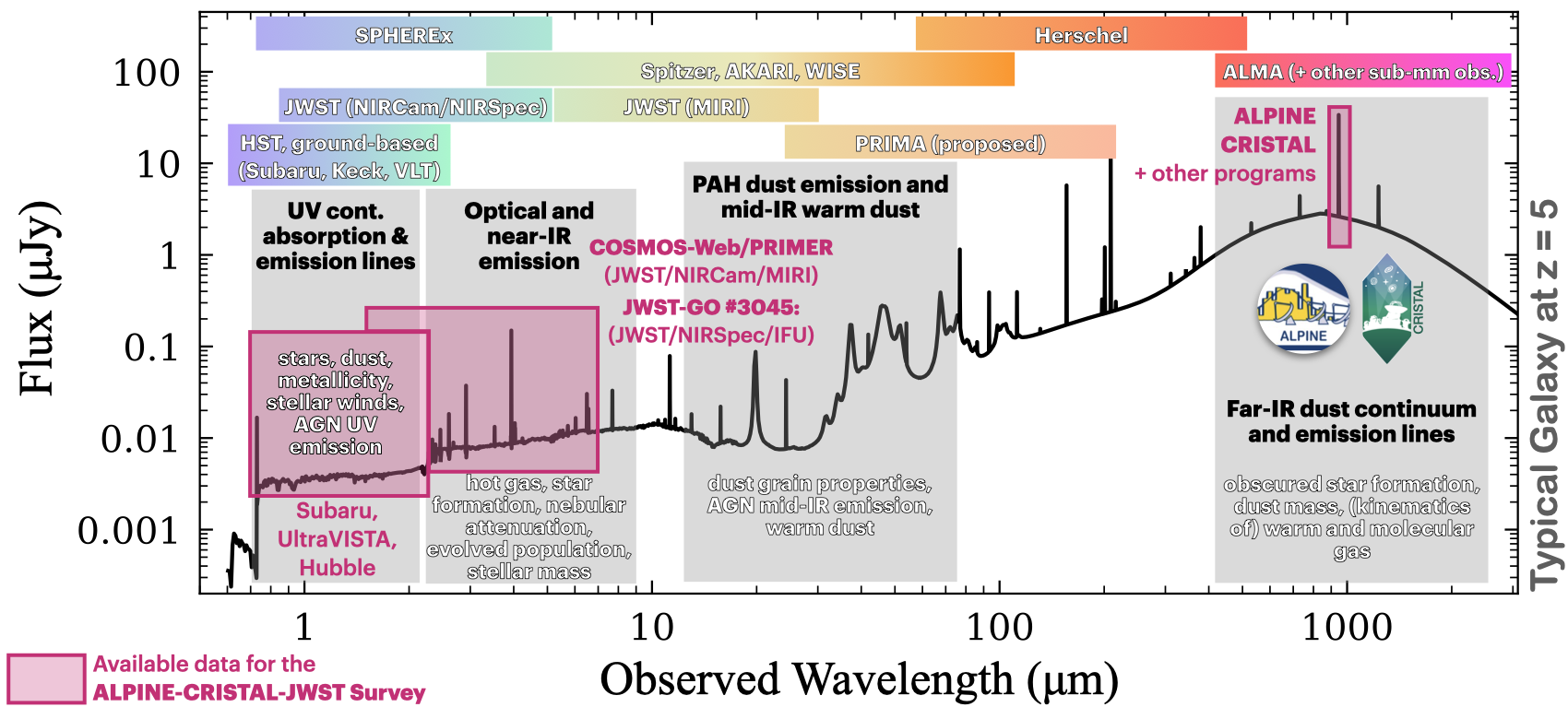}\vspace{-3mm}
\caption{The spectrum of a typical $z=5$ galaxy overlaid with wavelength regimes of physical interest. A selection of ground and space-based facilities are highlighted (including past missions like Spitzer and Herschel). The wavelength regions that are observed by the available data of the ALPINE-CRISTAL-JWST survey sample are shown as magenta boxes. We note the lack of observations at mid-IR and far-IR wavelength, covering warm dust emission and Polycyclic Aromatic Hydrocarbons (PAHs), important tracers of the dust grain distribution, at high redshifts. This wavelength range may be filled in by future facilities such as PRIMA \citep{glenn25}.
\label{fig:sed}}
\end{figure*}

Star-forming galaxies spend most of their life on the galaxy star-forming main-sequence, a tight relation between stellar mass and star formation rate (SFR) \citep{brinchmann04,noeske07,daddi07}, which has now been robustly identified through ultraviolet (UV) and far-infrared (far-IR) observations out to $z\sim 5$ in large samples of galaxies \citep[e.g.,][]{steinhardt14, caputi17, khusanova21,cole23,clarke24,cole25}.
During that time, the evolution of galaxies is determined by a balance between the consumption or removal of gas through star formation or outflows and the replenishment of gas through accretion or minor mergers onto the galaxy's disk \citep[e.g.,][]{dave12,lilly13,feldmann15,tacconi20}.
Spatially integrated galaxy properties have been measured with large statistical samples across billions of years of galaxy history to understand the evolution of galaxies overall. But we still lack the understanding of the underlying processes that govern the sub-galactic spatial scales, especially in galaxies at high redshifts.
For example, we do not yet have a consistent picture of the details of metal enrichment and star formation efficiency \citep[e.g.,][]{dayal13,popping17,vijayan19,katz22,choban24,pallottini24}, dust distribution \citep[e.g.,][]{ferrara21,sommovigo22,ziparo23,triani21}, origin and impact of outflows \citep[e.g.,][]{pizzati20,gallerani18}, or the growth of supermassive black holes and the structure formation in the early Universe \citep[e.g.,][]{dayal18,volonteri21}.
While much progress has been made recently in characterizing galaxies in the Epoch of Reionization (EoR, $z\gtrsim6$), it is of fundamental importance to study the connection between this primordial epoch and mature galaxies at cosmic noon ($z=2-3$). 
Understanding the intertwined connections between these epochs and processes requires the study of baryons (stars, gas, dust) on galactic-scales in the interstellar and circumgalactic medium (ISM and CGM, respectively). With their energy output spanning more than three orders of magnitude in wavelength (Figure~\ref{fig:sed}), these processes can only be characterized through a multi-wavelength photometric and spectroscopic study using a combination of ground- and space-based observatories.

Prior to the James Webb Space Telescope (JWST), such multi-wavelength spectroscopic studies at resolved spatial scales have only been possible for (unlensed) galaxies at cosmic noon and later epochs.
One notable example is the {\it Spectroscopic Imaging survey in the Near-infrared with SINFONI} \citep[SINS;][]{forsterschreiber09} and later, in combination with the zCOSMOS survey \citep{lilly07}, the SINS/zC-SINF survey \citep{forsterschreiber18}.
These surveys combine ground-based observations from the Very Large Telescopes' (VLT) SINFONI AO-assisted spectrograph in the rest-frame optical, and {\it Hubble} Space Telescope (HST) rest-frame UV/optical observations of a large sample of spectroscopically confirmed $z\sim2$ galaxies.
The combination of these instruments enabled breakthroughs in our understanding of the buildup of such galaxies, including the properties of their star-forming clumps, kinematics, and the origin of outflows \citep[e.g.,][]{genzel11,newman12,davies19} --- ultimately emphasizing the need of such studies at higher redshifts.
Similarly, the SHiZELS survey provided resolved emission line maps of narrow-band selected galaxies at $z=0.8-2.2$ with SINFONI-AO \citep{swinbank12}.
Significantly larger samples with similar measurements at $0.7 < z < 2.7$ but more statistical power have been realized later by the KMOS$^{\rm 3D}$ \citep{forsterschreiber19,wisnioski19} and the KROSS survey \citep{stott16}.
In parallel to that, large spectroscopic campaigns such as the {\it MOSFIRE Deep Evolution Field} survey \citep[MOSDEF;][]{kriek15} or the  {\it Keck Baryonic Structure Survey} \citep[KBSS;][]{rudie12,steidel14} delivered integrated spectroscopic data revealing the evolution of chemical properties at $z<3$ \citep[e.g.,][]{strom17,sanders21}.

At higher redshifts, such comparable studies have not been possible due to two reasons; the lack of multi-wavelength coverage at sufficiently long wavelengths into the IR to probe the rest-frame optical emission lines at $z>3$ and the lack of sufficient spatial resolving power and/or sensitivity. With JWST up and running, such detailed analyses are now possible for the first time, covering the galaxies' history deep into the EoR.
Large photometric and spectroscopic surveys in prominent JWST fields such as JADES \citep{eisenstein23} and CEERS \citep{finkelstein25} now provide integrated spectroscopy of galaxies at high redshifts to study their chemical compositions and the abundances of active galactic nuclei (AGN). Designated surveys such as the {\it Chemical Evolution Constrained using Ionized Lines in Interstellar Aurorae} survey \citep[CECILIA;][]{strom23} or the {\it Assembly of Ultradeep Rest-optical Observations Revealing Astrophysics} (AURORA) survey \citep{shapley25} use the power of deep spectroscopic observations at rest-frame UV and optical wavelengths for detailed studies of chemical and ionization properties of such galaxies.

However, all of these large surveys provide spatially integrated measurements, making it challenging to disentangle the different physical processes defined at smaller spatial scales and to compare the observations at high redshifts to local studies at similar physical scales. Furthermore, complementary observations at rest-frame far-IR wavelengths are necessary to compare the ionized phases with those of cold gas and dust.
It is therefore of profound importance to establish a proper, well-studied multi-wavelength benchmark sample at $z=4-6$ between the EoR and cosmic noon to
{\it (i)} measure the chemical and structural properties of this galaxy population,
{\it (ii)} define a comparison sample for similar studies at lower and even higher redshifts with future facilities, and
{\it (iii)} anchor state-of-the-art cosmological simulations.

\begin{figure*}[t!]
\centering
\includegraphics[angle=0,width=\textwidth]{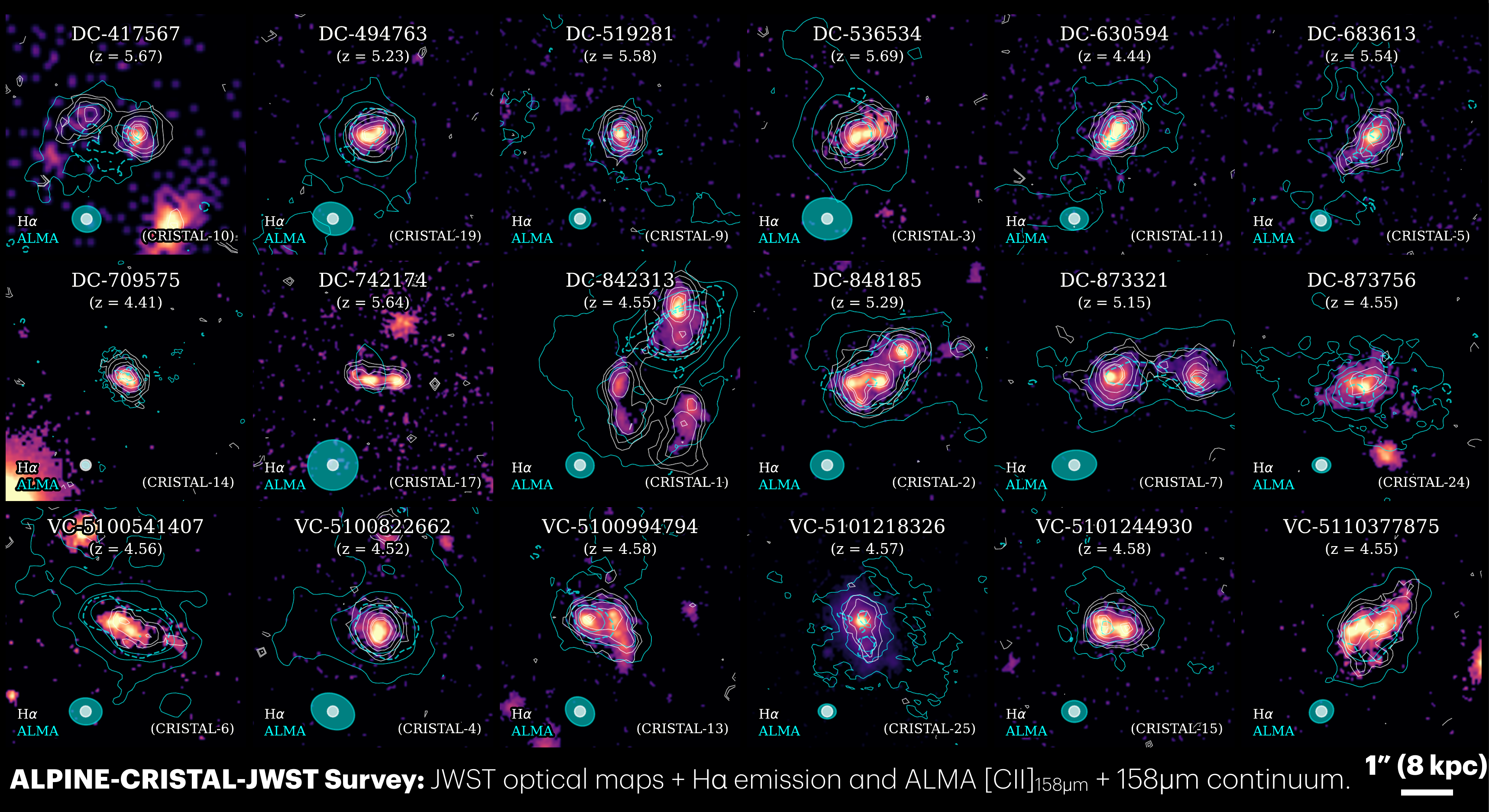}\vspace{-0mm}
\caption{Collage of the $18$ targets part of the ALPINE-CRISTAL-JWST sample. For each panel, the background shows the COSMOS-Web JWST/NIRCam F277W image (except for {\it DC-417567} for which we show HST WFC3/F160W). The \halpha~map from JWST/NIRSpec IFU observations is indicated with the white contours. The CRISTAL ALMA data are shown as cyan contours (\cplus: solid; $158\,{\rm \mu m}$ continuum: dashed). The contours show 3, 5, 10, 20, 40, and 50$\sigma$ levels for JWST and 3, 10, 20$\sigma$ levels for ALMA data. The ALMA (cyan) and NIRSpec/NIRCam (white) beam/PSF are indicated in all panels.
\label{fig:collage}}
\end{figure*}

The {\it Atacama Large (Sub) Millimeter Array} (ALMA) changed the landscape for directly detecting rest-frame far-IR light in high-redshift galaxies. It enabled a first look at rest-frame far-IR wavelengths to constrain the dust content via the far-IR continuum and cold gas properties through the measurement of singly ionized carbon (\cplus) emission at $158\,{\rm \mu m}$ for typical main-sequence galaxies at high redshifts \citep{capak15,riechers14,carniani20,hashimoto19,harikane20,wong22}.
Prior to ALMA, \cplus~had only been observed at high redshifts for single starburst galaxies and not spatially resolved \citep[e.g.,][]{iono06,maiolino09}.
The ALPINE ALMA large program \citep[\#2017.1.00428.L, PI: Le F\`evre;][]{lefevre20,bethermin20,faisst20b} was the first step towards achieving a multi-wavelength post-EoR galaxy sample.
ALPINE combined rest-frame UV to optical photometry with ALMA observations of the far-IR dust continuum and \cplus~at rest-frame $158\,{\rm \mu m}$ for $118$ typical post-EoR main-sequence galaxies at $z=4-6$. ALPINE has led to several breakthroughs in the study of dust, gas, kinematics, and outflows of post-EoR galaxies \citep[see comprehensive ALPINE review paper by][]{faisst22}.
Specifically, it shows that the relation between \cplus~and SFR is in place out to $z\sim5$ \citep{schaerer20} and showcases the use of \cplus~for inferring the dynamics, molecular gas mass, and depletion times \citep{dessauges20}. ALPINE data also suggest that dust and UV emission may be spatially offset, which could lead to significant underestimation of the dust components based on rest-frame UV light only \citep{faisst17b,sommovigo22,killi24}. Furthermore, outflows (which may be related to the creation of \cplus~halos) are common in these star-forming high-redshift galaxies \citep{ginolfi20,pizzati23,birkin25,fujimoto19}.
But its success also came with new questions about
{\it (i)} the dust production/destruction in early galaxies,
{\it (ii)} the details of cold and hot gas kinematics,
{\it (iii)} the frequency of super massive black holes (SMBHs) residing in typical high-redshift galaxies, and 
{\it (iv)} the origin of the reported \cplus~halos extending out to $10\,{\rm kpc}$ into the CGM.
These are key questions to be answered to understand early galaxy evolution.

{\it To answer these questions, we must go one step further and obtain far-IR observations at higher spatial resolutions and fill in the gap of missing spatial resolved spectroscopy at rest-frame optical wavelengths.}

The ALMA-CRISTAL survey \citep{herreracamus25} achieved the former by providing high-resolution far-IR follow-up observations of $19$ ALPINE galaxies, thus enabling the first study of the gas and dust properties of these galaxies on kpc scales.

The aim of this new {\it ALPINE-CRISTAL-JWST} survey, which science goals, data, sample selection, and basic measurements are presented in this paper, is to leverage these excellent data with rest-frame optical spectroscopy. This program uniquely combines the capabilities of JWST \mbox{NIRSpec/IFU} spatially resolved rest-frame optical spectroscopy with JWST \mbox{NIRCam} imaging, HST rest-frame UV data, and high-resolution ($0.3\arcsec$) ALMA \cplus~and dust continuum observations for a sample of 18 massive ($\rm \log(M_*/{M_\odot}) > 9.5$) main-sequence galaxies at $z=4-6$ (Figure~\ref{fig:sed}). This post-EoR multi-wavelength photometric and spectroscopic survey establishes a benchmark sample to study in detail early galaxy formation and evolution and to connect the EoR with mature galaxy evolution at cosmic noon.

In this paper we give an overview of the ALPINE-CRISTAL-JWST survey and present a few key results based on spatially integrated measurements.
We note that the presentation of the detailed data reduction and resolved measurements is beyond the scope of this paper and would distract from its main purpose to present the general purpose of this survey. Specifically, the details of the data reduction of the JWST/NIRSpec IFU data will be presented in \citet{fujimoto25}.
In Section~\ref{sec:design}, we introduce the science goals of the survey and describe the sample selection. In Section~\ref{sec:data}, we detail the available multi-wavelength data.
Section~\ref{sec:measurements} includes basic measurements for our sample, which are compared to other observations from the literature at lower and higher redshifts.
We conclude in Section~\ref{sec:conclusions} with an outlook of upcoming science results from this survey.

Throughout this work, we assume a $\Lambda$CDM cosmology with $H_0 = 70\,{\rm km\,s^{-1}\,Mpc^{-1}}$, $\Omega_\Lambda = 0.7$, and $\Omega_{\rm m} = 0.3$ and magnitudes are given in the AB system \citep{oke74}. We use a Chabrier initial mass function \citet[IMF;][]{chabrier03} for stellar masses and SFRs.

All measurements derived work are tabulated in the Appendix~\ref{sec:tables} and catalogs containing these data will be accessible to the public upon request.
For details on the data reduction of the JWST/NIRSpec IFU observations we refer to the companion paper by \citet{fujimoto25}.

\section{The ALPINE-CRISTAL-JWST Survey} \label{sec:design}

\subsection{Motivation and Science Goals}

A careful study of galaxy evolution and to answer the above questions requires the measurement of all baryon components constituting a galaxy (stars, dust, and gas), which is only possible with the combined observations of different telescopes in both imaging and spectroscopic modes across the spectrum (Figure~\ref{fig:sed}).

Although the ALPINE survey provided a valuable multi-wavelength picture of post-EoR galaxies, several key observations were missing, which led to several fundamental unknowns including ionized gas properties and kinematics, chemical properties such as metal abundance, star formation, and black hole growth.
For example, no deep rest-frame optical imaging was available \citep[see][]{faisst20b}, ALMA observations were too coarse to identify the finer structure of dust and \cplus~emission, and spectroscopy was limited to the rest-frame UV wavelength and lacked spatial resolution \citep{lefevre13,bouwens15,hasinger18}.
Observations using the Spitzer Space Telescope made headway to studying the rest-frame optical emission of galaxies \citep{steinhardt14,caputi17} and yield the first measurements of \halpha~emission at $z\sim 5$ including the ALPINE sample \citep[e.g.,][]{shim11,stark13,faisst19}, but at a resolution and depth not enough to perform resolved studies or spectroscopy.
Similarly, Herschel's sensitivity was only sufficient to detect the mid-/far-IR continuum in stacks of typical main-sequence galaxies at $z>4$ such as the ALPINE galaxies \citep[e.g.,][]{bethermin20}.

With HST, JWST, and ALMA operating at the same time, it is now finally possible to overcome these limitations and provide a multi-wavelength photometric and spectroscopic benchmark sample to study the connection between primordial galaxy formation in the EoR and the galaxy population at cosmic noon.
The {\it ALPINE-CRISTAL-JWST survey}, presented in this work, combines observations with HST and ALMA with the unprecedented sensitivity and spatial resolution of JWST imaging and integral field spectroscopy for a unique sample of $18$ main-sequence galaxies at $z=4-6$ (Section~\ref{sec:sampleselection}).
Figure~\ref{fig:collage} showcases the available data for each of the $18$ targets from JWST imaging (background image in NIRCam/F277W), JWST spectroscopy (white contours showing the \halpha~emission measured by the NIRSpec/IFU), and ALMA (cyan solid and dashed contours for \cplus~and dust continuum emission).
Current observatories are uniquely suited to provide a comprehensive multi-wavelength view of galaxies in the redshift range of $4 < z < 6$.
Note that JWST would not cover \halpha~in galaxies at $z\gtrsim 7$ unless using MIRI \citep{zavala25}, which however is significantly less sensitive than NIRSpec. 
The redshift range $4 < z < 6$ is crucial to understand the connection between primordial galaxy formation in the EoR and late galaxy evolution at cosmic noon. It is the epoch in time where galaxies significantly grow their stellar mass, enrich their ISM and CGM with metals, and establish their structures before they reach the peak of their star formation and eventually become quiescent. Importantly, the results from this program can be compared directly to similar efforts at $z>6.5$ \citep[REBELS-IFU;][]{bouwens22,rowland25} and lower redshifts (e.g., SINS at $z\sim2$ and others mentioned earlier) to render a complete picture of galaxy evolution.
Therefore, the main goal of the ALPINE-CRISTAL-JWST survey is to establish a multi-wavelength kpc-resolution post-EoR benchmark sample at the highest redshifts to date to connect early and late galaxies. This survey enables the study of how the first galaxies mature and grow in structure and chemical composition to become the systems we observe at later cosmic times.

The ALPINE-CRISTAL-JWST survey is designed to address the following questions:
\begin{itemize}
    \item {\bf How and where are stars being formed?}~Constraining the modes of star formation in conjunction with gas and dust abundances and a detailed study of star formation histories using combined imaging and spectroscopic data.\vspace{-2mm}
    
    \item {\bf How is the ISM enriched?}~Studying the interplay between star formation, gas, metals, and dust by detailed measurements of (spatially resolved) metal abundances and photoionization properties.\vspace{-2mm}
    
    \item {\bf How is dust produced?}~Deciphering the dominant modes of dust production and variations in dust content by the combination of chemical and dust evolution models and the spatial distribution of rest-frame UV, optical, and far-IR light.\vspace{-2mm}
    
    \item {\bf What is the contribution of AGN to the typical galaxy population?}~Understanding the occurrence of AGN in typical main-sequence galaxies via broad optical emission lines and line ratio diagnostics to investigate the co-evolution of SMBHs and host galaxies in relation to AGN found by recent JWST surveys.\vspace{-2mm}
    
    \item {\bf What is the relation between the ISM and the CGM?}~Constraining the baryon cycle and enrichment of the CGM and solving the origin of the \cplus~halos by measuring the ionized properties and outflows of the surrounding gas.

    \item {\bf How does the kinematics of hot and cold gas compare?}~Study the kinematics of the stellar$+$ionized and gas disk via the velocity fields of \halpha~and \cplus~at redshifts in the midst of cosmic disk formation.
\end{itemize}

The ALPINE-CRISTAL-JWST survey will answer these questions within a representative sample of main-sequence galaxies at $z=4-6$ by combining observations from HST, JWST, and ALMA.

\begin{figure}[t!]
\centering
\includegraphics[angle=0,width=\columnwidth]{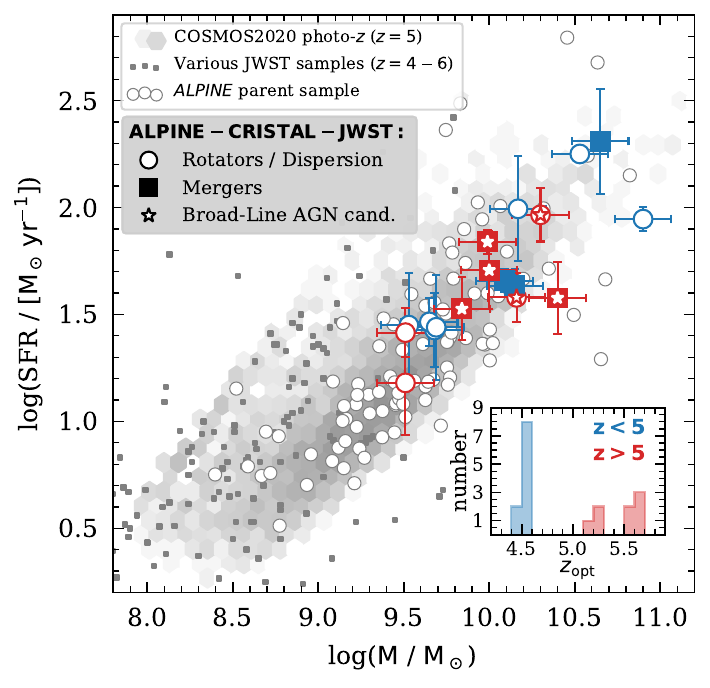}\vspace{-3mm}
\caption{The $18$ targets of the ALPINE-CRISTAL-JWST survey on the stellar mass vs.~SFR plane. The targets are shown with large symbols, split in $z<5$ (blue) and $z>5$ (red) sub-samples. The redshift distribution is shown in the inset.
Furthermore, we indicate rotators/dispersion-dominated (open circles) and mergers (solid squares) identified by low-resolution $\sim1\arcsec$ \cplus~kinematics data \citep{romano21}. Candidate type 1 (broad-line) AGN are indicated with a star \citep{ren25}. The ALPINE parent sample is shown as small white circles \citep{faisst20b}.
In addition, we show galaxies at $z\sim5$ from the COSMOS2020 catalog \citep{weaver22} as well as various JWST-detected galaxies at $z=4-6$ from the literature (small gray squares; see text for references).
\label{fig:sfrmass}}
\end{figure}

\begin{table*}
\footnotesize
\centering
\setlength{\tabcolsep}{3pt}\vspace{-3mm}
\caption{Identifications and available HST and JWST data for the targets. Note that all targets have rest-frame optical IFU spectroscopy from JWST/NIRSpec and $0.3\arcsec$ resolution ALMA $158\,{\rm \mu m}$ continuum and \cplus~spectroscopic data.}
\label{tab:obsdata}
\begin{tabular}{cccc c c c c c c c c c c c c  c  c c c c c c c c c c c  } 
\hline \hline
\multicolumn{3}{c}{Name} && z$_{\rm opt}$ & R.A. & Decl. && \multicolumn{7}{c}{HST} && \multicolumn{11}{c}{JWST}\\ \cline{1-3} \cline{9-16} \cline{18-28}
ALPINE & CRISTAL & Others && ~ & (J2000) & (J2000) && ACS &&&  \multicolumn{5}{c}{ WFC3 } && \multicolumn{8}{c}{ NIRCam } && \multicolumn{2}{c}{ MIRI }    \\  \cline{9-10}  \cline{12-16} \cline{18-25} \cline{27-28}
~ & ~ & ~ && ~ & ~ & ~ && \rotatebox[origin=c]{90}{~F814W~} &&& \rotatebox[origin=c]{90}{~F105W~} & \rotatebox[origin=c]{90}{~F110W~} & \rotatebox[origin=c]{90}{~F125W~} & \rotatebox[origin=c]{90}{~F140W~} & \rotatebox[origin=c]{90}{~F160W~} &&  \rotatebox[origin=c]{90}{~F090W~} & \rotatebox[origin=c]{90}{~F115W~} & \rotatebox[origin=c]{90}{~F150W~} & \rotatebox[origin=c]{90}{~F200W~} & \rotatebox[origin=c]{90}{~F277W~} & \rotatebox[origin=c]{90}{~F356W~} & \rotatebox[origin=c]{90}{~F410M~} & \rotatebox[origin=c]{90}{~F444W~} && \rotatebox[origin=c]{90}{~F770W~} & \rotatebox[origin=c]{90}{~F1800W~}  \\ \hline
{\bf DC-417567}$^{\rm a,b}$ & C-10 & HZ2$^{\rm c}$ && 5.6700 & 150.51701 &	1.92889 && \x &&& \x & & \x & & \x && & & & & & & & && &  \\
{\bf DC-494763}$^{\rm b}$ & C-19 & -- && 5.2337 & 150.02127 &	2.05338 && \x &&& \x & & \x & \x & \x && & \x & \x & & \x & & & \x && \x &  \\
{\bf DC-519281}$^{\rm a,b}$ & C-09 & -- && 5.5759 & 149.75372 &	2.09100 && \x &&& & \x & & & \x && & \x & \x & & \x & & & \x && &  \\
{\bf DC-536534}$^{\rm a,b}$ & C-03 & HZ1$^{\rm c}$ && 5.6886 & 149.97183 &	2.11818 && \x &&& \x & \x & \x & & \x && & \x & \x & & \x & & & \x && \x &  \\
{\bf DC-630594}$^{\rm b}$ & C-11 & -- && 4.4403 & 150.13583 &	2.25788 && \x &&& & & \x & \x & \x && \x & \x & \x & \x & \x & \x & \x & \x && \x &  \\
{\bf DC-683613}$^{\rm a,b}$ & C-05 & HZ3$^{\rm c}$ && 5.5420 & 150.03926 &	2.33718 && \x &&& \x & & \x & & \x && & \x & \x & & \x & & & \x && \x & \x  \\
DC-709575$^{\rm b}$ & C-14 & -- && 4.4121 & 149.94610 &	2.37579 && \x &&& & & & & \x && & \x & \x & & \x & & & \x && &  \\
DC-742174$^{\rm b}$ & C-17 & -- && 5.6360 & 150.16302 &	2.42560 && \x &&& & & \x & \x & \x && \x & \x & \x & \x & \x & \x & \x & \x && \x &  \\
{\bf DC-842313}$^{\rm b}$ & C-01 & -- && 4.5537 & 150.22721 &	2.57638 && \x &&& \x & & \x & \x & \x && & \x & \x & & \x & & & \x && &  \\
{\bf DC-848185}$^{\rm a,b,e}$ & C-02 & HZ6$^{\rm c}$/LBG-1$^{\rm d}$ && 5.2931 & 150.08963 &	2.58635 && \x &&& \x & & \x & & \x && & \x & \x & & \x & & & \x && &  \\
{\bf DC-873321}$^{\rm a,b}$ & C-07 & HZ8$^{\rm c}$ && 5.1542 & 150.01690 &	2.62661 && \x &&& \x & & \x & & \x && & \x & \x & & \x & & & \x && &  \\
{\bf DC-873756}$^{\rm e,f,g}$ & C-24 & -- && 4.5457 & 150.01132 &	2.62781 && \x &&& \x & \x & \x & \x & \x && & \x & \x & & \x & & & \x && &  \\
{\bf VC-5100541407}$^{\rm b}$ & C-06 & -- && 4.5630 & 150.25382 &	1.80935 && \x &&& & \x & & \x & \x && & \x & \x & & \x & & & \x && &  \\
{\bf VC-5100822662}$^{\rm b}$ & C-04 & -- && 4.5205 & 149.74130 & 2.08094 && \x &&& & \x & & & \x && & \x & \x & & \x & & & \x && &  \\
{\bf VC-5100994794}$^{\rm b}$ & C-13 & -- && 4.5802 & 150.17143 &	2.28726 && \x &&& & & \x & \x & \x && \x & \x & \x & \x & \x & \x & \x & \x && \x &  \\
{\bf VC-5101218326}$^{\rm e,f}$ & C-25 & -- && 4.5739 & 150.30207 &	2.31461 && \x &&& \x & & & & \x && & \x & \x & & \x & & & \x && &  \\
VC-5101244930$^{\rm b}$ & C-15 & -- && 4.5803 & 150.19856 &	2.30058 && \x &&& & & \x & \x & \x && \x & \x & \x & \x & \x & \x & \x & \x && \x &  \\
VC-5110377875$^{\rm h}$ & -- & -- && 4.5505 & 150.38475 & 2.40842 && \x &&& & & & & \x && & \x & \x & & \x & & & \x && &  \\ \hline
\end{tabular}
\tablecomments{\raggedright All galaxies are observed by the ALPINE ALMA program \citep[2017.1.00428.L, PI: Le F\`evre;][]{lefevre20} and detected in \cplus. Names in {\bf bold} indicate galaxy systems with additional $150\,{\rm \mu m}$ dust continuum detection \citep{mitsuhashi24,bethermin20}.\\
$^{\rm a}$ Type-1 AGN candidates from \cite{ren25}.\\
$^{\rm b}$ Observed by the ALMA--CRISTAL survey \citep[2021.1.00280.L, PI: Herrera-Camus;][]{herreracamus25}.\\
$^{\rm c}$ See \citet{capak15}. \\
$^{\rm d}$ See \citet{riechers14}. \\
$^{\rm e}$ Band 4 observations available (2024.0.01401.S, PI: Herrera-Camus).\\
$^{\rm f}$ Observed by ALMA program 2019.1.00226.S (PI: Ibar).\\
$^{\rm g}$ Band 9 observations available (2024.0.01401.S, PI: Herrera-Camus).\\
$^{\rm h}$ Observed by ALMA program 2022.1.01118.S (PI: B\'ethermin).\\
}
\end{table*}


\subsection{Sample Selection}\label{sec:sampleselection}

The ALPINE-CRISTAL-JWST survey consists of 18 galaxies selected from the combined survey data of the larger ALPINE ALMA program \citep{lefevre20,bethermin20,faisst20b} and the CRISTAL survey \citep{herreracamus25}. ALPINE comprises a set of 118 galaxies selected from the COSMOS \citep{scoville07} and GOODS-S \citep{giavalisco04} fields with ALMA observations in band 7, targeting \cplus~and the rest-frame $\sim158\,{\rm \mu m}$ dust continuum emission.
The ALPINE sample was selected by requiring secure spectroscopic redshifts at $z=4-6$ from Keck/DEIMOS and VLT/VIMOS observations \citep{lefevre13,hasinger18,khostovan25} with a cut applied in \cplus~luminosity based on the $\rm M_{UV} - L_{[C\,II]}$ relation derived from the pilot sample by \citet{capak15}. The spectroscopic redshifts are derived from \lya~emission as well as UV absorption lines to minimize a sample bias towards strong emission line galaxies. X-ray or UV line detected AGN are excluded from this sample to focus solely on typical main-sequence galaxies.

The ALPINE-CRISTAL-JWST sample is then constructed from the ALPINE sample with the following requirements:

\begin{enumerate}
    \item a robust ($>5\sigma$ integrated) \cplus~detection, \vspace{-2mm}
    
    \item follow-up observations by ALMA of \cplus~emission at a spatial resolution of $\sim 0.3\arcsec$ close to the pixel size of the JWST/NIRSpec IFU optical spectroscopy, \vspace{-2mm}
    
    \item imaging observations with JWST/NIRCam at $1-5\,{\rm \mu m}$ in at least four bands, \vspace{-2mm}
    
    \item coverage with HST in ACS/F814W and at least one redder HST band, and \vspace{-2mm}
    
    \item detectability ($>\,$5$\sigma$) of all strong optical emission lines (such as \hbeta, \oii$_{3727}$ (unresolved), \oiii$_{5007}$, \halpha, \nii$_{6585}$, and \sii$_{6718}$) with the JWST/NIRSpec IFU in less than two hours on-source time. \vspace{-2mm}
\end{enumerate}

To satisfy the requirement of high resolution \cplus~observations, we chose galaxies observed by the ALMA-CRISTAL survey as well as ALMA programs \#2022.1.01118.S (PI: B\'ethermin) and \#2019.1.00226.S (PI: Ibar) providing beam sizes on the order of $0.3\arcsec$.
The optical emission line signal-to-noise (SNR) selection was driven by faint lines such as \nii, \sii~and \oii~whereas the other lines, which are intrinsically brighter, were observed at higher SNR. The disperser/filter combinations G235M/F170LP and G395M/F290LP were used to cover these lines for the redshift range of $z=4.4 - 5.9$.
The \halpha~emission line fluxes to estimate the JWST/NIRSpec exposure times were predicted from the SFRs of the galaxies, which themselves were derived from the combined UV and far-IR emission \citep{schaerer20} or SED fitting \citep{faisst20b}. The remaining line fluxes were predicted based on their ratios with respect to \halpha~for conservative assumptions of gas phase metallicities \citep[derived from the expected mass-metallicity relation;][]{maiolino08} and dust obscuration \citep[derived from SED fitting to the COSMOS photometry;][]{faisst20b}.
The SNRs were estimated using the JWST exposure time calculator \texttt{Pandeia} \citep{pontoppidan16} in the given disperser/filter configurations.
Overall, this technique worked well and we achieved the expected SNRs for all galaxies except three. For the latter, we likely underpredicted the dust attenuation, leading to a lower-than-expected SNR in the blue grating.

The final sample consists of $18$ galaxies with HST, JWST (imaging and IFU spectroscopy), and $0.3\arcsec$ beam-size ALMA observations. We note that one galaxy ({\it DC-417567}/{CRISTAL-10}) is not covered by JWST photometric observations but has HST ACS/F814W as well as WCF3/F105W, F125W and F160W data. This galaxy was included in Phase 2 of the JWST proposal process to replace a galaxy with weak \cplus~emission suggested by the new CRISTAL observations \citep[see][]{herreracamus25}.
We emphasize that no selection is imposed on the far-IR continuum emission measured by ALPINE or CRISTAL. In the final sample, however, $15$ out of the $18$ galaxy systems are dust continuum detected at $>3\sigma$ \citep{bethermin20,herreracamus25}.
In this paper, we follow the naming convention of ALPINE, but we list the corresponding CRISTAL names of the targets in Table~\ref{tab:obsdata} and Figure~\ref{fig:collage}.

Figure~\ref{fig:sfrmass} shows the final sample of $18$ galaxies on the stellar mass vs. SFR plane. In this case, the SFRs are derived from the total UV$+$far-IR luminosity, or, if no ALMA far-IR continuum is available, from SED fitting (see Section~\ref{sec:physical}). The sample is compared to the parent ALPINE sample (white circles) and galaxies selected from the COSMOS2020 \citep{weaver22} catalog at $z\sim5$, as well as other JWST-detected sources at $z=4-6$ \citep{morishita24,nakajima23,curti23,sarkar25}.
Galaxies indicated by filled squares are mergers identified by \citet{romano21} based on their low-resolution ($1\arcsec$) \cplus~kinematic analysis \citep[see also similar analysis by][]{ikeda25}. This includes the targets {\it DC-417567}, {\it DC-519281}, {\it DC-536534}, {\it DC-842313}, {\it DC-873321}, {\it VC-5100541407}, and {\it VC-5100822662}.
However, note that the higher resolution ALMA data from CRISTAL suggest that also {\it DC-683613} \citep{posses24} and {\it DC-848185} \citep{davies25} are likely interacting/merging.

We also indicate type 1 broad-line AGN candidates selected by a rigorous logic involving a comparison between the Bayesian Information Criteria (BIC) for a fit with and without broad lines and a high SNR broad-line detection \citep[for details we refer to][]{ren25}. The \halpha~broad line widths for the final candidates range from $600\,{\rm km\,s^{-1}}$ (least robust candidate) to $2800\,{\rm km\,s^{-1}}$ (most robust candidate). 

This type 1 AGN candidate sample include the targets {\it DC-417567}, {\it DC-519281}, {\it DC-536534}, {\it DC-683613}, {\it DC-848185}, and {\it DC-873321}.
Note that none of the targets, by construction, are specifically identified as AGN based on their X-ray or radio emission.
Furthermore, we find that none of our targets (except for {\it DC-873756}) lie in the AGN region of the Baldwin, Phillips \& Terlevich (BPT) diagram \citep{baldwin81} derived from the new JWST data presented here (see Section~\ref{sec:measurements}).

The redshift distribution of the sample spans the range $z=4.4-5.7$ with a gap between $z=4.6-5.1$. This gap originates from the original ALPINE selection to avoid a low atmospheric transmission at the observed frequency of \cplus. The sample concentrates at $\logm > 9.5$, hence populates the more massive end of the $z\sim5$ galaxy population but two orders of magnitude less massive than the knee of the mass function at $z=5$ \citep{davidzon17}.
It therefore complements optimally other JWST-detected galaxy samples at lower stellar masses.

\section{Multi-Wavelength Observational Data}\label{sec:data}

In the following sections we provide a summary of the acquisition and reduction of the data for the $18$ targets observed by different observatories (Table~\ref{tab:obsdata}). Details about the observing programs are presented in Table~\ref{tab:obsdataext}. Figure~\ref{fig:collage} shows a portrait the available data for all 18 targets in the sample.

\begin{figure*}[t!]
\centering
\includegraphics[angle=0,width=\textwidth]{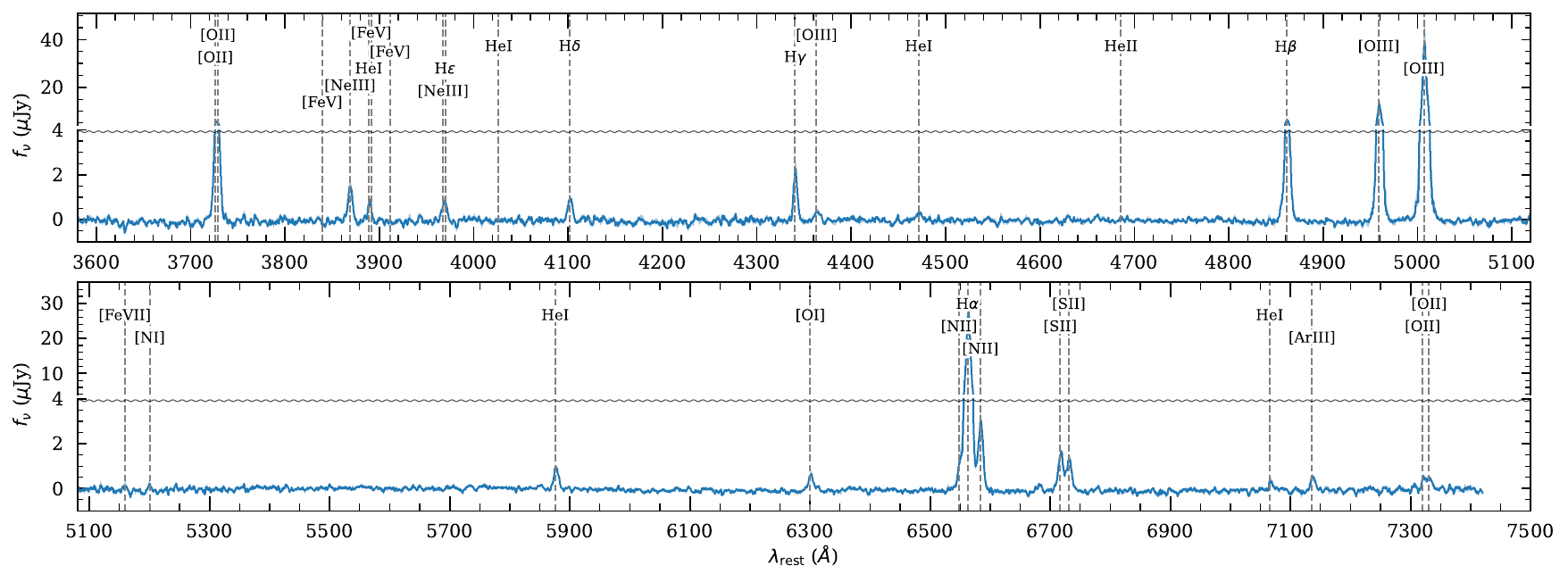}\vspace{-3mm}
\caption{Stacked spectrum of all $18$ ALPINE-CRISTAL-JWST survey targets (note that the $y$-axis is broken at $4\,{\rm \mu Jy}$ to make the higher fluxes better visible). Some prominent and common emission lines are indicated by the vertical dashed lines. The spectra have been continuum subtracted before stacking. More details on the derivation of the spectra can be found in Section~\ref{sec:measurements}.
\label{fig:specstack}}
\end{figure*}

\subsection{JWST NIRSpec/IFU Observations}\label{sec:dataifu}

The JWST/NIRSpec IFU cycle 2 GO program \#3045 (PI: Faisst) builds the backbone of the ALPINE-CRISTAL-JWST survey data. The technical details, especially the details of the data reduction process, of this program are presented in \citet{fujimoto25}. Here, we provide a brief overview over the technical details of the program and the reduction of the NIRSpec/IFU data.

The observations were carried out during cycle 2 between April and May
2024. The targets were observed with the NIRSpec IFU in configurations G235M/F170LP and G395M/F290LP at medium spectral resolution ($R\sim1000$) to cover all strong optical emission lines. However, target {\it DC-842313} was observed in G395H/F290LP by the cycle 2 program \#4265 \citep[PIs Gonz\'alez-L\'opez \& Aravena, see also][]{solimano25}. These data are also included in this work. 
For program \#3045, the integration times vary from $500\,{\rm s}$ to $7200\,{\rm s}$ per target per configuration, with a 2-point sparse cycling dither pattern. The integration time for program \#4265 is $12\,000\,{\rm s}$, with a 9-point small cycling dither.
No off-center background observations or leakage calibration exposures were taken. Instead, the background is measured from background pixels within the $3\arcsec \times 3\arcsec$ IFU field of view.
The final $3\sigma$ flux RMS (average over the different targets) of our observations is $6.6\times10^{-21}\,{\rm erg\,s^{-1}\,cm^{-1}\,\AA^{-1}}$ for G235M and $1.3\times10^{-21}\,{\rm erg\,s^{-1}\,cm^{-1}\,\AA^{-1}}$ for G395M/G395H \citep[the flux per target varies slightly per target, for details see,][]{fujimoto25}.

The data were reduced using the STScI pipeline (version 1.16.0) with the CRDS context \texttt{jwst\_1298.pmap} to a final pixel scale of $0.1\arcsec/{\rm px}$. We followed the procedure developed by the ERS TEMPLATES team \citep[\#1355; PIs Rigby \& Vieira, see also][]{rigby23,welch23,birkin23}. Several additional steps to improve the final data quality were implemented, including improved background subtraction, stripe removal, and rescaling of the error cubes \citep[][]{fujimoto25}.
The cubes were subsequently astrometrically aligned to the JWST COSMOS-Web images, which are referenced to Gaia DR3 \citep{gaia23}, see also \citet{franco25}. To this end, first NIRCam F277W and F444W pseudo continuum images were created by convolving the cubes (pixel by pixel) with the corresponding filter transmission curves. The centroid positions of the collapsed cubes (obtained by a Gaussian fit) are then compared to the corresponding JWST/NIRCam images. We found offsets on the order of $0.1-0.3\arcsec$ between the NIRCam image and collapsed cube, which we correct for. These offsets are within the expected pointing uncertainty of JWST. The final astrometric precision is comparable with that of COSMOS-Web ({\it i.e.} a few $10\,{\rm mas}$).
In addition, we checked the absolute and relative calibration of both grism configurations. This was achieved by convolving the respective spectra (spatially integrated over an aperture of $0.5\arcsec$ diameter) with the transmission curves of the NIRCam F277W and F444W filters. This synthetic broad-band photometry was then compared to the photometry measured on the F277W and F444W COSMOS-Web images in the same aperture. We find that the absolute flux calibration is better than $30\%$ with a $1\sigma$ scatter of $\sim10\%$. The relative calibration of the two grism configurations is better than $5\%$.

The stacked and spatially integrated spectrum of all the $18$ targets is shown in Figure~\ref{fig:specstack}. The spectra of the individual targets are shown in Figure~\ref{fig:eachgalaxy} in Appendix~\ref{app:eachgalaxy} (see Section~\ref{sec:measurements} for details). All the strong optical emission lines are detected as well as some fainter lines such as [\ion{Ne}{3}]$_{3868}$, \ion{He}{1} ($3889\,{\rm \AA}$, $5875\,{\rm \AA}$), H$\epsilon$, H$\delta$, H$\gamma$, [\ion{O}{1}]$_{6300}$, [\ion{Ar}{3}]$_{7138}$, and auroral oxygen lines (\oiii$_{4363}$ and the \oii$_{7322,32}$ doublet). These are diagnostic lines sensitive to the ionization parameter ($\log{(U)}$), gas-metallicity, nebular dust attenuation ($\ebmvn$), electron density ($n_e$), and in some cases electron temperatures ($T_e$) which can be used to derive $T_e$-based gas-phase metallicities.

\subsection{ALMA Sub-mm Observations}\label{sec:dataalma}

All $18$ targets are part of the ALPINE survey, a large cycle 5 ALMA program (\#2017.1.00428.L, PI: Le F\`evre) targeting the \cplus~emission in $118$ galaxies at $z=4-6$. A survey overview is provided by \citet{lefevre20} with the data reduction presented by \citet{bethermin20}. The ancillary data and the measurement of the physical properties of the ALPINE galaxies is described in \citet{faisst20b}.

The ALPINE-CRISTAL-JWST sample has additional follow-up ALMA observations at $\lesssim0.3\arcsec$, to complement the ALPINE-ALMA data at higher spatial resolutions.
Out of the $18$ targets, $15$ have follow-up observations by the ALMA-CRISTAL survey \citep[\#2021.1.00280.L, PI: Herrera-Camus,][]{herreracamus25}. The sensitivity achieved by CRISTAL is comparable to ALPINE but at a $3-4\times$ higher angular resolution and $2-7\times$ longer integration times.
Out of the remaining three galaxies, two ({\it DC-873756} and {\it VC-5101218326}) were observed by the program \#2019.1.00226.S (PI: Ibar) and one ({\it VC-5110377875}) by the program \#2022.1.01118.S (PI: B\'ethermin). For the former, the medium ($0.3\arcsec$) and high-resolution ($0.15\arcsec$) visibilities were combined. The data from these two additional programs are discussed in more detail in \citet{devereaux24} and \citet{bethermin23}, respectively.
In all these cases, the data are reduced using the standard ALMA reduction pipeline \texttt{CASA} \citep{casateam22}, combining the high-resolution data with the lower resolution observations from the ALPINE survey.
This combination provide an excellent coverage from small compact to more extended diffuse sub-mm emission.

\subsection{JWST NIRCam and MIRI Observations}\label{sec:datajwstimage}

The JWST/NIRCam and MIRI observations provide rest-frame optical imaging at unprecedented resolution. These imaging data are crucial for resolved SED studies \citep[see][]{li24}.

All except one target are covered by the JWST cycle 1 GO treasury program {\it COSMOS-Web} \citep[\#1727, PIs: Kartaltepe \& Casey;][]{casey22}. This program provides observations in four NIRCam filters (F115W, F150W, F277W, and F444W) and one MIRI filter (F770W). Due to the smaller field of view of MIRI and the adopted survey strategy, only $7$ out of the $18$ galaxies have MIRI coverage.
We use the official data products distributed by the COSMOS-Web team, which include additional updates on background subtraction and artifact removal compared to the standard STScI pipeline. Details of the NIRCam and MIRI data reductions are described by \citet[][]{franco25} and \citet{harish25}, respectively. A catalog of sources is presented in \citet{shuntov25}.
In addition, four galaxies were covered by the JWST cycle 1 GO treasury program {\it PRIMER} (\#1837, PI: Dunlop, reduced by the COSMOS-Web team) providing (in addition to COSMOS-Web) imaging in the NIRCam filters F090W, F115W, F150W, F200W, F277W, F356W, F444W, and F410M as well as in MIRI F770W and F1800W.
All data are aligned to the Gaia astrometric reference frame, consistent with the JWST/NIRCam IFU cubes (Section~\ref{sec:dataifu}).

We refer to Table~\ref{tab:obsdata} for a full list of JWST data and to Table~\ref{tab:obsdataext} for a full list of program numbers. The JWST imaging data is also shown in Figure~\ref{fig:eachgalaxy} in Appendix~\ref{app:eachgalaxy} for each of our targets.

\subsection{Other Ancillary Data and Planned Observations}\label{sec:dataother}

All targets reside in the COSMOS field \citep{scoville07} providing HST/ACS F814W imaging data \citep{koekemoer07} as well as a wealth of data from ground based facilities. See \citet{weaver22} for an overview of these data. In addition, the galaxies are covered by several HST WFC3/IR pointings including filters F105W, F110W, F125W, F140W, and F160W. These provide crucial rest-frame UV and optical wavelength coverage offset from the central wavelengths of the JWST filters.
The programs are listed in Table~\ref{tab:obsdataext} and include {\it 3D-HST} \citep[][]{brammer12,momcheva16}, {\it COSMOS-DASH} \citep{mowla19,cutler22}, {\it COSMOS-DASH-3D} \citep{mowla22}, and the CANDELS-HST survey \citep{grogin11,koekemoer11}.

The ALPINE-CRISTAL-JWST survey targets will also be part of several future observations by HST and JWST.
Five out of the $18$ galaxies will be covered by the {\it ORCHIDS} cycle 3 JWST program in NIRSpec/IFU G395H/F290LP (\#5974, PI: Aravena) with the goal to understand better the \cplus~halos surrounding these galaxies with deep spectroscopy.
In total $12$ galaxies will be included in the {\it COSMOS-3D} cycle 3 JWST/NIRCam (grism) survey (\#5893, PI: Kakiichi) providing additional imaging observations in F115W/F200W/F356W (NIRCam) and F1000W/F2100W (MIRI) as well as slitless spectroscopy in F444W.
Finally, $17$ galaxies will be included in the {\it CLUTCH} cycle 32 HST multi-cycle program (\#17802, PI: Kartaltepe), providing deeper ACS/F814W observations together with new ACS/F435W, ACS/F606W, and WFC3/F098W imaging (note that the galaxies will be undetected in the bluer bands provided by {\it CLUTCH} due to their high redshifts).

\subsection{SED-based Stellar Masses and SFRs}\label{sec:physical}

Various physical measurements were inherited from previous works (see Table~\ref{tab:physical}). Specifically, stellar masses and SFRs were derived from SED fitting including the photometry from ground-based observatories and Spitzer from the COSMOS2020 catalog \citep{weaver22} and JWST imaging from the COSMOS-Web program. 
Here we adopt the measurements derived using \texttt{CIGALE} \citep{boquien19,burgarella05} and described in detail in \citet{mitsuhashi24} and \citet{herreracamus25}. We note that these results are consistent with \citet{li24} who measure spatially resolved stellar masses and SFRs using \texttt{MAGPHYS} \citep{dacunha08,dacunha15} including the ALMA continuum data, and \citet{lines25} who use \texttt{Bagpipes} \citep{carnall18,carnall19} not including the ALMA continuum data.

In addition to SED fitting, we computed the star formation rates from different observed indicators, spanning star formation timescales from $10-100\,{\rm Myrs}$. These indicators include \cplus, \halpha, \oii, and the UV$+$far-IR luminosity.
For the \cplus-derived SFRs, we used the parameterization by \citet{schaerer20} (including UV$+$IR, with $a=7.37$ and $b=0.83$).
From the dust-corrected \halpha~and \oii~luminosities, we derived SFRs using the \citet{kennicutt98} parameterizations\footnote{We correct from a Salpeter to Chabrier IMF by multiplying with a factor of $0.61$ \citep[e.g.,][]{madau14}.} and the nebular dust attenuation measurements described in Section~\ref{sec:measuredust}.
For targets detected in the ALPINE$+$CRISTAL ALMA continuum, we computed the total SFR measurement based on the sum of the SFRs derived from UV and far-IR continuum ($\rm SFR_{UV+IR}$). The former is derived from rest-frame UV photometry of HST and ground-based imaging \citep[see][]{faisst20b} and the latter is the total far-IR luminosity derived from the ALMA $158\,{\rm \mu m}$ continuum by using the appropriate conversion factors \citep{bethermin20,mitsuhashi24}. In both cases, the luminosity is converted to SFR using the \citet{kennicutt98} parameterizations. 
Table~\ref{tab:physical} lists the different SFR measurements. The SFR indicators are compared in Section~\ref{sec:measuresfr}.

\begin{figure*}[t!]
\centering
\includegraphics[angle=0,width=\textwidth]{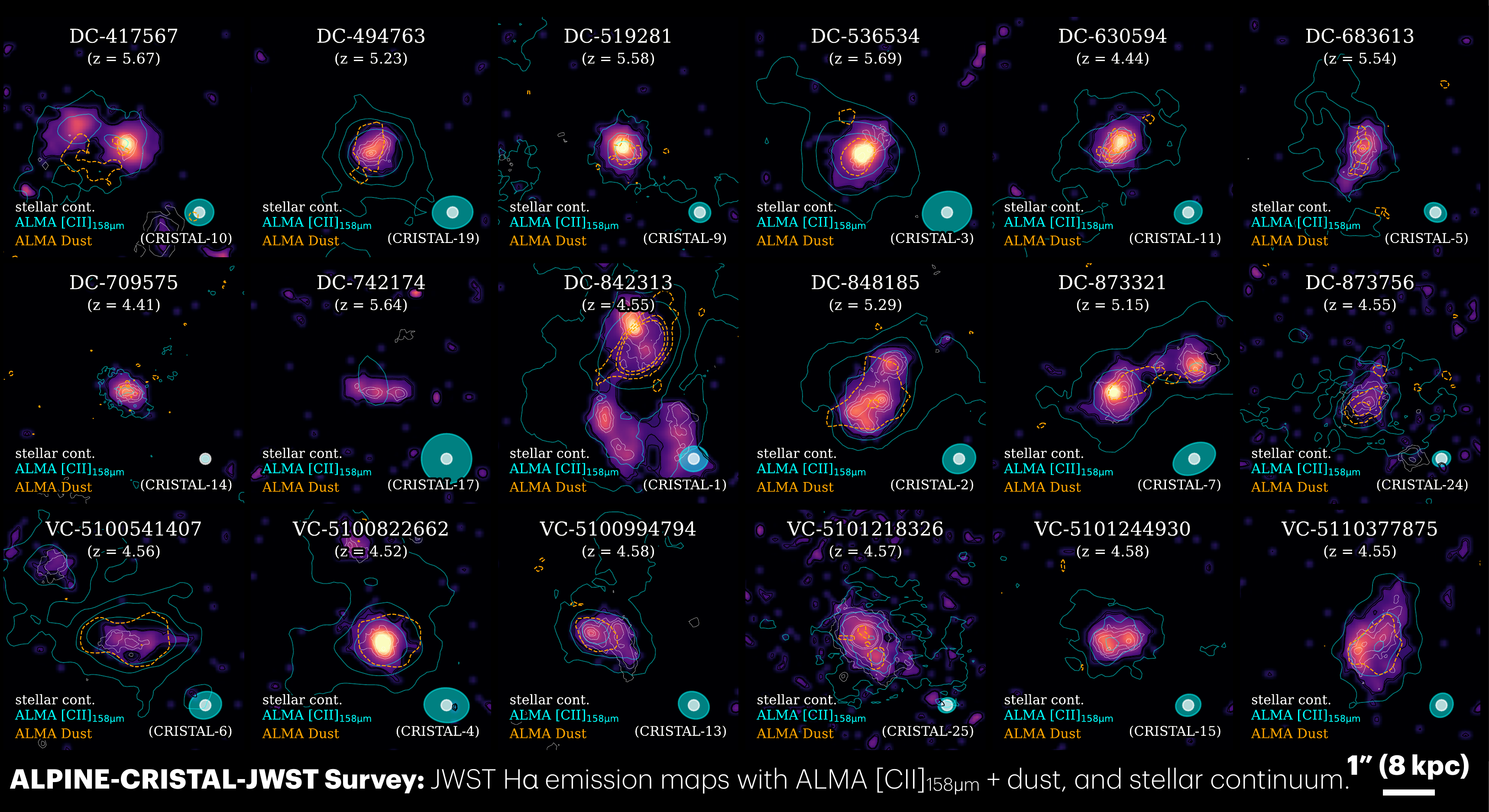}\vspace{-0mm}
\caption{\halpha~emission line maps derived from the NIRSpec/IFU observations for the $18$ galaxies in the ALPINE-CRISTAL-JWST sample. The line maps are cut off at $5\sigma$. The white contours show the rest-frame optical light (JWST/NIRCam F277W and HST WFC3/F160W for {\it DC-417567}) and cyan contours show \cplus~emission from the ALMA-CRISTAL program.
The orange-dashed contours show the ALMA $158\,{\rm \mu m}$ continuum dust emission.
The contours show 5, 10, 20, 40, and 50$\sigma$ levels for JWST and 3, 10, 20$\sigma$ levels for ALMA data. The ALMA (cyan) and NIRSpec (white) beam/PSF are indicated in all panels.
\label{fig:collageHa}}
\end{figure*}

\section{JWST/NIRSpec IFU Measurements}\label{sec:measurements}

In this section, we outline basic measurements of diagnostic lines and related properties derived from the NIRSpec IFU spectra. We focus here on spatially integrated measurements and emission line maps. Resolved measurements will be explored in future works.

\subsection{Propagation of Measurement Uncertainties}

Because there are significant cross-correlations and dependencies between the parameters derived in the following, it is not straightforward to propagate uncertainties through these calculations. To provide realistic uncertainties on the following measurements, we therefore adopted an extensive end-to-end Monte-Carlo sampling method to properly propagate and estimate the uncertainties of all the quantities reported below.
To do so, we created $200$ different NIRSpec/IFU cube realizations by perturbing each pixel according to its uncertainty derived from the rescaled IFU error cubes, which include all observational effects and sky background \citep{fujimoto25}.
This method generally works well unless the errors are significantly correlated between the pixels. This is however not the case given that the PSF is comparable to the pixel size of the IFU data. Inter-pixel correlations are neglected here for simplicity. Each of the cube representations is then analyzed end-to-end, resulting in $200$ different measurements. The final measurements are obtained as the medians and their $1\sigma$ percentiles.

\begin{figure*}[t!]
\centering
\includegraphics[angle=0,width=\textwidth]{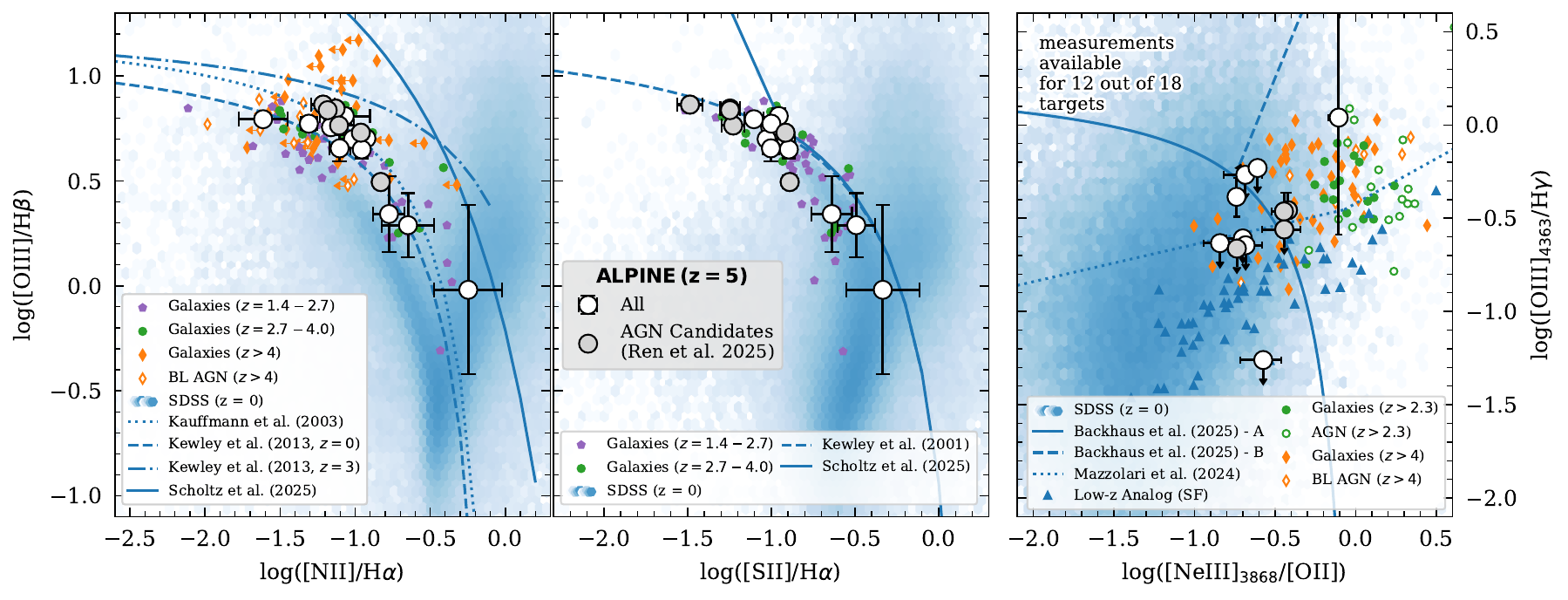}\vspace{-3mm}
\caption{The $18$ ALPINE-CRISTAL-JWST targets (large circles with type 1 AGN from \citet{ren25} indicated in gray) on different line ratio diagnostic diagrams.
{\it Left:} \Ntwo~BPT diagram. The AGN region is to the right of the dividing lines from \citet{kauffmann03} (dotted), \citep{kewley13} (dashed for $z=0$ and dot-dashed for $z=3$, note that we do not show the parameterization for $z>3$ as that would be an extrapolation to the study of \citet{kewley13}), and \citet{scholtz25} \citep[solid; derived from][]{feltre16,gutkin16}. Observations from the literature at $z=0$ \citep[blue cloud, SDSS;][]{abazajian09}, at $z=1.4-2.7$ (purple) and $z=2.7-4.0$ (green) from the AURORA survey \citep{shapley25}, and at $z>4$ \citep[orange diamonds;][]{harikane23,nakajima23} are also shown. The target with the largest uncertainties and a \Ntwo~close to unity is {\it DC-873756}, potentially a BPT-selected AGN (see discussion in text).
{\it Middle:} \Stwo~line ratio diagram. The AGN region is to the right of the dividing lines from \citet{kewley01} (dashed), and \citet{scholtz25} \citep[solid; derived from][]{nakajima22}. Observations from the literature at $z=0$ \citep[blue cloud, SDSS;][]{abazajian09} and at $z=1.4-2.7$ (purple) and $z=2.7-4.0$ (green) from the AURORA survey \citep{shapley25} are shown.
{\it Right:} High-ionization auroral [\ion{Ne}{3}]$_{3868}$/[\ion{O}{2}] vs. [\ion{O}{3}]$_{4363}$/H$\gamma$ diagram. Only shown for 12 galaxies with robust [\ion{Ne}{3}]$_{3868}$ and [\ion{O}{3}]$_{4363}$ measurements. The AGN region is above the dividing line from \citet{mazzolari24} (dotted) or to the right of the combined dividing lines by \citet{backhaus25} (dashed and solid). The region in the upper left above the dashed and solid line may contain star-forming galaxies and AGN \citep[see][]{backhaus25}. The target with the largest uncertainty and [\ion{Ne}{3}]$_{3868}$/[\ion{O}{2}] close to unity is {\it DC-742174}, a metal-poor compact multi-component system.
\label{fig:bpt}
}
\end{figure*}

\subsection{Emission Line Maps and Integrated Spectra}\label{sec:emlinmaps}

We first create spatially integrated spectra for each of the $18$ targets from their NIRSpec cubes. For the spatial integration, we define a mask encompassing emission of the sum of the \oii, \oiii, \halpha, and \hbeta~line maps. Creating these maps for a given target involves an iterative process: First, we spatially integrated the spectrum in a $1\arcsec$ aperture around the target coordinates (center of mass measured on NIRCam images) to derive a first-guess integrated spectrum. From that spectrum we measured the full-width-at-half-maximum (FWHM) of the emission lines (see Section~\ref{sec:linefitting}). We then created maps of each of the emission lines by integrating the cubes across three FWHMs ($\sim 3\times340\,{\rm km\,s^{-1}}$) along the spectral dimension. To derive the final masks, we stacked these emission line maps and selected pixels for which the total flux is detected at an ${\rm SNR}>5$.
Note that because different emission lines are covered by the two gratings G235M and G395M, we created two such masks for each target.
Using those masks, we then derived the spatially integrated spectrum for each grating by summing up the spectra in each pixel in the mask.
In a final step, the two one-dimensional spectra of the two gratings are combined, using median stacking over the wavelength region where they overlap. We note that generally the SNR ratio is higher in G395M in the overlapping region.
Figure~\ref{fig:collageHa} shows the \halpha~maps together with the JWST NIRCam imaging and ALMA data. 
Figure~\ref{fig:eachgalaxy} show the G235M$+$G395M (or G235M$+$G395H in the case of {\it DC-842313}) combined spectrum as well as emission line maps for each target individually.

\begin{figure*}[t!]
\centering
\includegraphics[angle=0,width=\textwidth]{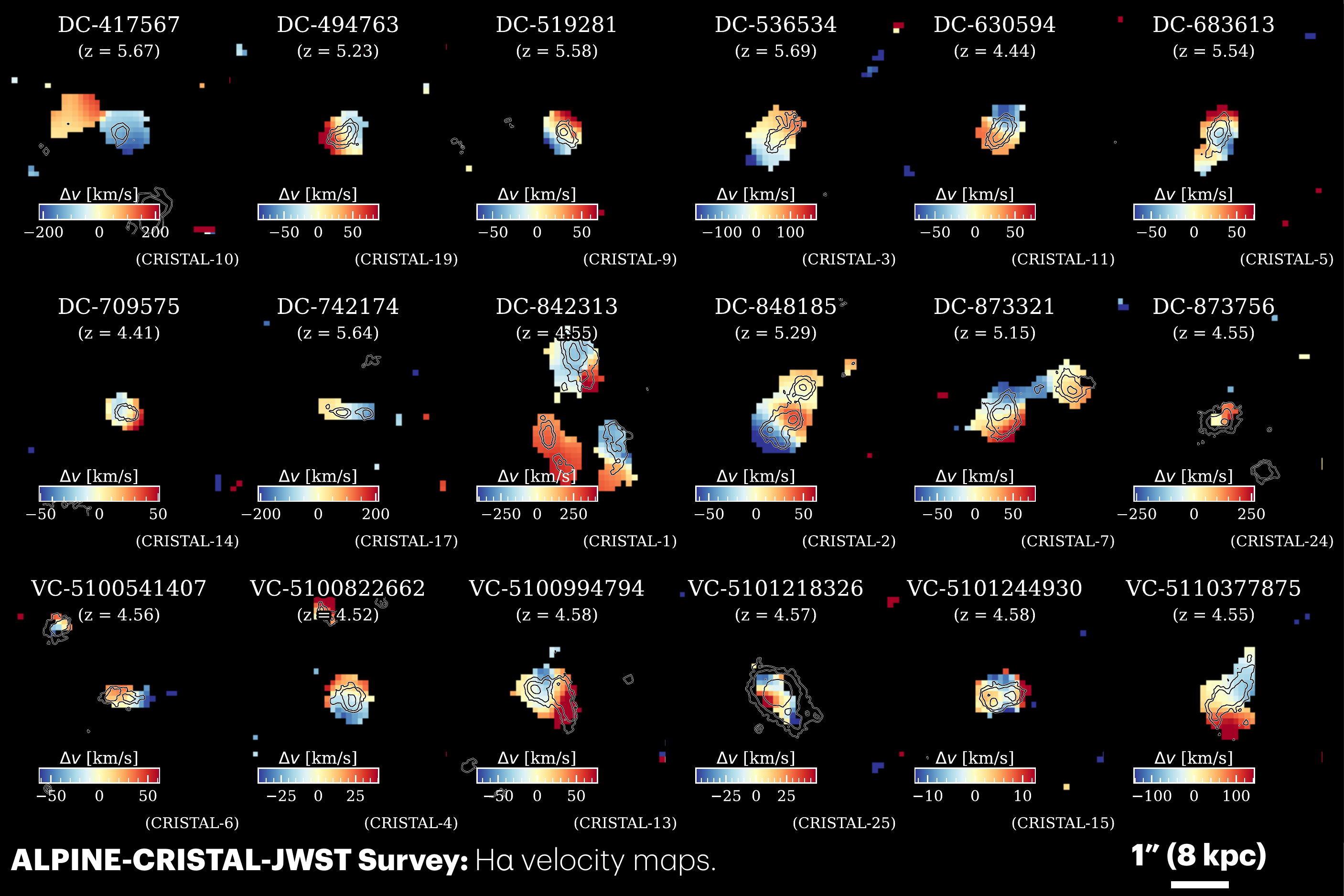}\vspace{-0mm}
\caption{Collage of the \halpha~velocity maps of the $18$ targets in the ALPINE-CRISTAL-JWST sample. The stellar continuum from the COSMOS-Web JWST/NIRCam F277W imaging is shown with contours (5, 10, and 30$\sigma$).
\label{fig:velocitymaps}}
\end{figure*}

\subsection{Emission Line Fitting and Line Diagnostic Diagrams (``BPT diagrams'')}\label{sec:linefitting}

Having measured the spatially integrated spectra of each of the targets, we then fit the total fluxes of bright optical lines including some fainter auroral lines. The line measurements are summarized in Table~\ref{tab:linefluxes} and the process is described below.

We simultaneously fit lines within seven different emission line complexes: 
{\it (i)} the {\it spectrally} unresolved \oii$_{\rm 3727}$~doublet;
{\it (ii)} the [\ion{Ne}{3}]$_{\rm 3868}$~Neon line;
{\it (iii)} \hgamma~and the [\ion{O}{3}]$_{\rm 4363}$ auroral line;
{\it (iv)} \hbeta, [\ion{O}{3}]$_{\rm 4959}$, and [\ion{O}{3}]$_{\rm 5007}$;
{\it (v)} [\ion{N}{2}]$_{\rm 6548}$, \halpha, and [\ion{N}{2}]$_{\rm 6585}$;
{\it (vi)} the [\ion{S}{2}]$_{\rm 6718}$ and [\ion{S}{2}]$_{\rm 6732}$ doublet; and
{\it (vii)} the [\ion{O}{2}]$_{\rm 7322}$ and [\ion{O}{2}]$_{\rm 7332}$ auroral lines.
With this split, we can deblend lines that are close by and at the same time treat the complexes independently to account for changes in the noise and line spread function.
The line flux ratios of the \nii~lines are fixed to three \citep{osterbrock06}. We assume the same FWHM for line doublets, however, not for different line species in different line complexes. This is because the line spread function may change across the wavelength range covered by the gratings. Fitting for the FWHM and central wavelengths between different complexes mitigates this effect. In the end, we found that the results are consistent within  $\sim10\%$ in FWHM (except broad lines, see below).
Here, we fit each line (except \halpha) with a single Gaussian with variable integrated flux, central wavelength, and FWHM (or $\sigma$) using the Python \texttt{lmfit} package\footnote{\url{https://lmfit.github.io/lmfit-py/}}. The continuum is subtracted by masking the line and fitting first-order polynomial.
For \halpha~we fit a double Gaussian representing a broad and narrow line component for cases in which the reduced $\chi^2$ is significantly smaller (by a factor of two, ranging between $0.5$ and $2.0$) compared to a single Gaussian fit. We found that a double Gaussian is a better fit for seven out of 18 targets. For four out of these seven targets, the broad component contributes significantly ($>30\%$) to the total \halpha~flux (see Table~\ref{tab:linefluxes}). These targets are {\it DC-519281}, {\it DC-536534}, {\it DC-842313}, and {\it DC-848185} (shown to have strong outflows by \citet{davies25} and may have a type 1 AGN in the north-west component).
All of these were also selected by \citet{ren25} as type 1 AGN candidates based on their broad \halpha~lines.
For the remaining three targets, the broad \halpha~line may be dominated by outflows. This will be analyzed in a subsequent paper.

Figure~\ref{fig:bpt} shows the targets on three different line ratio diagnostic diagrams (type 1 AGN candidates from \citet{ren25} are indicated in gray --- only narrow-line flux is considered). On each diagram, we show common separation lines between AGN and star-forming galaxies \citep{kewley01,kauffmann03,kewley13,mazzolari24,scholtz25,backhaus25} as well as measurements from the literature at low and high redshifts \citep{shapley25,harikane23,nakajima23}.
We find that none of the targets are robustly identified as AGN based on these line diagnostic diagrams. However, they do reside close to the AGN vs. star-forming galaxy dividing lines, in agreement with other measurements from the literature (see figure caption) and expectations from lower redshift analogs \citep[see][]{Faisst18}.
We note that the separation lines by \citet{kewley13} and \citet{scholtz25} in the classic BPT diagram (left panel in Figure~\ref{fig:bpt}) take into account the stronger ionization in high-redshift galaxies compared to galaxies at lower redshifts \citep[e.g.,][]{kewley01}. In other words, the separation of star forming galaxies and AGN includes the harder radiation in star-forming early galaxies. The fact that the ALPINE-CRISTAL galaxies are well below these dividing lines suggests that they might be evolved enough to resemble more lower redshift galaxies.

Two galaxies are worth mentioning in more detail:

{\it DC-873756} is the only galaxy that lies in the AGN region (based on the \citealt{kauffmann03} and \citealt{kewley13} parameterizations) of the classical \Ntwo~vs. [\ion{O}{3}]$_{5007}$/\hbeta~BPT diagram (however, it is not an outlier in the other diagrams or in the \citealt{scholtz25} parameterization). It is not identified as type 1 AGN.
{\it DC-873756} is a peculiar galaxy, showing a compact core with a strong dust continuum emission offset from the diffuse stellar continuum. Its spectrum shows a \Ntwo~ratio close to unity but due to the dust obscuration, the bluer emission lines are weak, which affects its reliability on the other two diagnostic diagrams (see Figure~\ref{fig:eachgalaxy}).
Taken at face value, this galaxy could be an AGN candidate, but deeper observations are needed for confirmation.

{\it DC-742174}, a relatively compact system with at least three components. It is an outlier on the [\ion{Ne}{3}]$_{3868}$/[\ion{O}{2}] vs. [\ion{O}{3}]$_{4363}$/H$\gamma$ diagram, showing a logarithmic [\ion{Ne}{3}]$_{3868}$/[\ion{O}{2}] line ratio close to unity. This system is not an outlier on the other diagrams and also not a type 1 AGN candidate based on the work by \citet{ren25}. The increased [\ion{Ne}{3}]$_{3868}$ could therefore originate from a low metallicity due to the monotonic anticorrelation of [\ion{Ne}{3}]$_{3868}$ with oxygen abundance \citep{nagao06,perezmontero07,shi07}. In fact, the [\ion{Ne}{3}]$_{3868}$/[\ion{O}{2}] ratio of {\it DC-742174} is in the upper range of the distribution of MOSDEF galaxies \citep{jeong20,shapley17} if compared to a similar mass of $\rm \log(M_*/M_\odot) = 9.6$. Its low metallicity is indeed also suggested by the strong line method (see below), finding a $\rm 12+log(O/H) \sim 7.84$, and also consistent with a weak \cplus~emission (Figure~\ref{fig:eachgalaxy}). Another possibility (likely related to low metallicity) could be a harder radiation field, which is also found in high equivalent-width emission line galaxies \citep[e.g.,][]{khostovan24}.

\subsection{Velocity Maps}

With the present NIRSpec/IFU observations we are in a unique position to study for the first time the kinematics of the ionized gas in these post-EoR galaxies.
To showcase the potential of the ALPINE-CRISTAL-JWST survey in this regard, we show in Figure~\ref{fig:velocitymaps} the ionized gas velocity maps, which are based on the \halpha~fluxes for each of the targets. We first apply the emission line based mask (see Section~\ref{sec:emlinmaps}) to the cube and then fit the \halpha~centroid in each spatial pixel using the same methods as described in Section~\ref{sec:linefitting}.
Note that the optical velocity resolution ($50-70\,{\rm km\,s^{-1}}$) is worse than obtained with \cplus~with ALMA \citep[$10-20\,{\rm km\,s^{-1}}$;][]{jones21,romano21,telikova24,herreracamus22,posses24}. A collage of these \halpha-based velocity maps for each of the 18 targets is shown in Figure~\ref{fig:velocitymaps} (in units of $\rm km\,s^{-1}$). Detailed \cplus-based velocity maps will be presented in \citet{leelilian25}.
Simulations suggest that the inferred $V/\sigma$ (rotation vs. dispersion) ratio may vary significantly depending on which line tracer ({\it i.e.} ionized or neutral gas) is used \citep[e.g.,][]{kohandel20,kohandel24}. Observational results are still limited due to the lack of simultaneous observations of ionized and neutral gas tracers.
Using a single line tracer could bias the interpretation of galaxy dynamics, making multi-wavelength data needed for a  comprehensive physical picture of galaxy dynamics.
On the other hand, investigating the velocity structure of cold and hot gas disks may tell us about disk formation.
To first order, we find a similar velocity structure between \halpha~and \cplus, showing some ordered rotation as well as mergers \citep[e.g.,][]{danhaive25,wisnioski25,leelilian25}. The fraction of disk-dominated systems is around $40-50\%$ as measured by \cplus~\citep{jones21,romano21,telikova24} and \halpha~\citep{telikova24,leelilian25} for the ALPINE/CRISTAL sample. A more detailed comparison between the ionized gas (\halpha) and the cold ISM gas (as traced by \cplus) will be presented in a forthcoming paper.

\subsection{Nebular Dust attenuation}\label{sec:measuredust}
The nebular dust attenuation differs from the stellar attenuation due to the configuration of clouds in the ISM \citep[e.g.,][]{calzetti01,reddy15,salim20,reddy25}. The differential dust attenuation\footnote{The differential dust attenuation between stars and nebular lines is defined as $f_{\rm dust}={E_{\rm s}(B-V)\,/\,E_{\rm n}(B-V)}$.} between nebular and stellar light is therefore an important quantity for many science applications \citep[see discussion by][]{faisst19}. With JWST spectroscopic observations, we can now derive the nebular dust attenuation of our targets through the \halpha/\hbeta~Balmer decrement via
\begin{equation}
    \ebmvn = \frac{2.5}{\left( k_\lambda({\rm H}\beta) - k_\lambda({\rm H}\alpha) \right)} \,\log(R/R_0),
    \label{eq:balmerdec}
\end{equation}
where $k_\lambda({\rm H\alpha})$ and $k_\lambda({\rm H\beta})$ are the reddening curve values at the wavelength of \halpha~and \hbeta, respectively. $R$ and $R_0$ are the observed and intrinsic $\frac{{\rm H}\alpha}{{\rm H}\beta}$ flux ratio, respectively.
From that, any observed emission line flux ($F^{\rm line}_o$) can be translated into an intrinsic line flux ($F^{\rm line}_i$) by correcting for the presence of galaxy-intrinsic dust via
\begin{equation}
    F^{\rm line}_i(\lambda) = F^{\rm line}_o(\lambda) \times 10^{0.4\,{\rm E_n(B-V)}\,k(\lambda)}.
\end{equation}

We assumed a reddening curve $k(\lambda)$ similar to local starburst galaxies \citep{calzetti00}, motivated by the fact that the ALPINE galaxies are relatively metal enriched (see Section~\ref{sec:measuremetallicity}), even though some evolution of the dust properties are suggested by works measuring the dust attenuation curve \citep{markov25,fisher25}.
Note that the intrinsic flux ratio $R_0$ varies on the assumption of electron temperature, density, and recombination type (case A and B). Specifically, case B is valid for a medium that is optically thick where recombination to the ground state is ignored as the ionizing photons are absorbed locally (``on-the-spot'' approximation). Vice versa, case A is expected in an optically thin cloud \citep[see discussion in][]{nebrin23}.

\begin{figure}[t!]
\centering
\includegraphics[angle=0,width=\columnwidth]{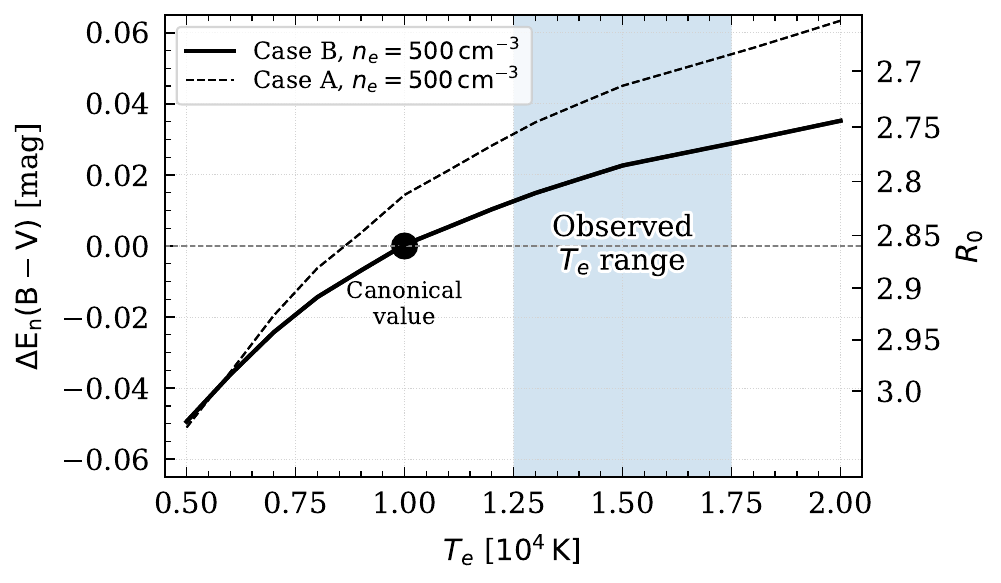}\vspace{-3mm}
\caption{Dependence of nebular dust attenuation (left $y$- axis) derived from the $\frac{{\rm H}\alpha}{{\rm H}\beta}$ Balmer decrement on electron temperature ($T_e$) and recombination type (case A: dashed line; case B: solid line) using \texttt{PyNeb}. The right $y$-axes shows the $R_0$ value with the canonical value of $2.86$ (for case B and $T_e = 10^4\,{\rm K}$ highlighted). Note that the dependence on electron density ($n_e$) is insignificant. The blue area shows the observed $T_e$ range measured for ALPINE-CRISTAL-JWST galaxies with auroral line detections \citep[see][]{faisst25b}. 
\label{fig:caseabtest}}
\end{figure}

Usually, a canonical value of $R_0 = 2.86$ is assumed \citep{cardelli89}, which is valid for case B recombination and $T_e = 10^4\,{\rm K}$. In Figure~\ref{fig:caseabtest}, we show how $R_0$, and the corresponding difference in $\ebmvn$, change with respect to this canonical values as a function of $T_e$ for both case A and B recombination. The curves are derived using the \texttt{PyNeb} Python package \citep{luridiana15} with the \citet{storey95} atomic database and for $n_e = 500\,{\rm cm^{-3}}$ \citep[consistent with observations of the ALPINE galaxies;][]{faisst25b,vanderhoof22,veraldi25}.
In the following, we quote the $\ebmvn$~values (and use this as dust correction) for the canonical emission line dust correction (case B, $T_e = 10^4\,{\rm K}$) for simplicity and for comparison to other works. However, we caution that the quoted $\ebmvn$~values could be underestimated by up to $0.04-0.05\,{\rm mag}$ (although smaller than measurement errors) due to {\it lower} $R_0$ values assuming the observed $T_e$ range (blue area in Figure~\ref{fig:caseabtest}, see also \citet{faisst25b} and \citet{delooze25} where $T_e$ is derived from fainter auroral lines). Lower values can also be achieved through Balmer self-absorption such as suggested in low-mass strong line emitters \citep[][]{scarlata24,pirzkal23,yang17,atek09}.
The final values of $\ebmvn$ of the galaxies in our sample are listed in Table~\ref{tab:physical}.

\begin{figure}[t!]
\centering
\includegraphics[angle=0,width=\columnwidth]{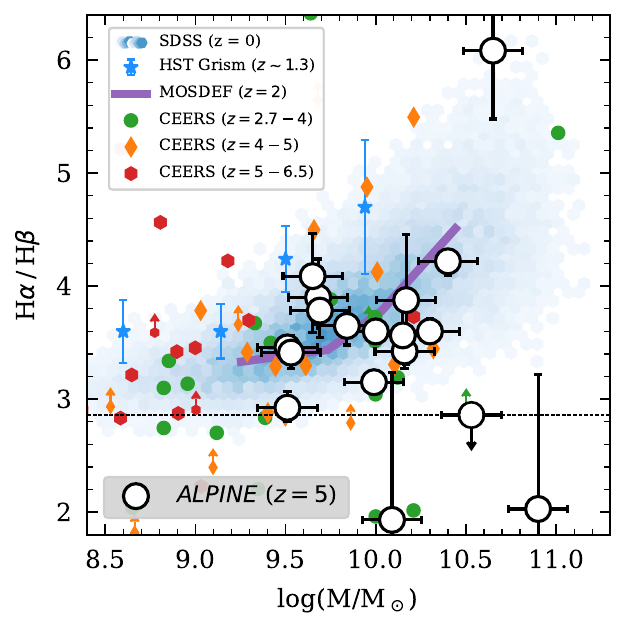}\vspace{-3mm}
\caption{
Relation between \halpha/\hbeta~total line flux ratios and stellar mass. The ALPINE-CRISTAL-JWST $z\sim5$ observations are shown as large symbols together with data at $z\sim0$ \citep[SDSS, total stellar masses, blue background;][]{abazajian09}, at $z\sim1.3$ \citep[HST grism, blue stars;][]{battisti22}, at $z\sim2$ \citep[MOSDEF, purple line;][]{shapley22}, and at $z>2.7$ \citep[CEERS, other symbols, see][]{shapley23}. The expected \halpha/\hbeta~line ratio for case B recombination is shown as dashed horizontal line. 
\label{fig:dustdecrement}}
\end{figure}

We also compared the $\ebmvn$~values to the stellar dust attenuation values, $\ebmvs$, derived from SED fitting \citep{tsujita25}. We find a differential dust attenuation ($f_{\rm dust}$), which is consistent with the value measured for local starburst galaxies \citep[$f_{\rm dust}= 0.44$;][]{calzetti00}. However, we note that this is a spatial integrated average and locally the differential dust attenuation may vary. This will be discussed in detail by \citet{tsujita25}.

Figure~\ref{fig:dustdecrement} suggests that the relation between the \halpha/\hbeta~line ratio and stellar mass is not evolving significantly with cosmic time since $z=5$. We find a similar relation to that found by the MOSDEF survey \citep[$z\sim2$;][]{reddy15,shapley22} and the $z\sim0$ galaxy population from SDSS \citep[using the total masses;][]{abazajian09}. However, we find generally lower values compared to the study by \citet{battisti22}, focusing on $z=1.3$ galaxies observed with HST grism. Note that using SDSS fiber stellar masses (instead of total masses) would move the $z=0$ cloud to $\sim 0.4\,{\rm dex}$ lower stellar masses, making it consistent with the $z=1.3$ measurements from \citet{battisti22}.
Literature studies at similar redshifts of $z=4-6$ \citep{shapley23} are consistent with our measurements when extrapolating the relation from their lower stellar masses to the higher stellar masses probed by the ALPINE-CRISTAL-JWST survey sample.
For two galaxies ({\it VC-5100541407} and {\it VC-5101218326}) we find \halpha/\hbeta~line ratios that are below the canonical case B value of $2.86$. This could mean either hotter temperatures (Figure~\ref{fig:caseabtest}) or that case B is not valid in these cases \citep[see discussion in][]{scarlata24}. However, we caution to draw such conclusions as the measurement uncertainties are significant and the values are consistent at $1\sigma$-level with the case B value.
We note that there are also several CEERS galaxies that show \halpha/\hbeta~values smaller than the canonical value. \citet{shapley23} argue that the majority of these cases are due to remaining systematics in the NIRSpec grating-to-grating flux calibration.
The upper limit (hatched circle) is {\it DC-873756}, a potential BPT-selected AGN with uncertain \halpha~measurement and weak \hbeta~flux due to significant dust attenuation in the central core as suggested by the ALMA dust continuum detection.

\begin{figure}[t!]
\centering
\includegraphics[angle=0,width=\columnwidth]{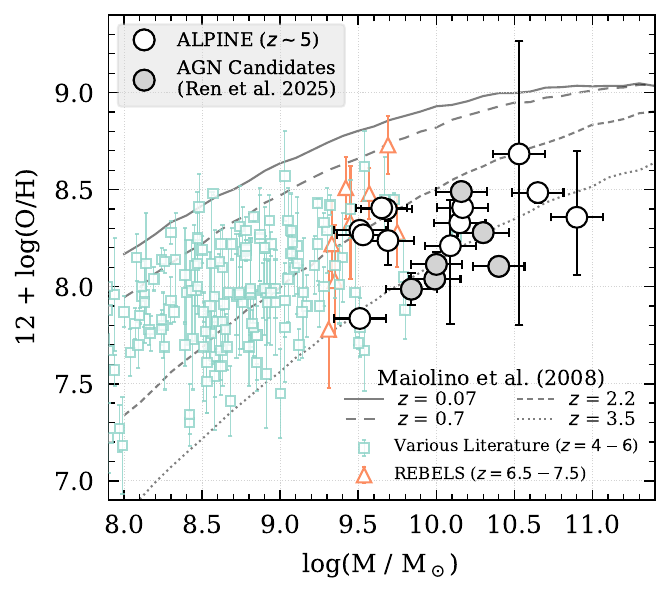}\vspace{-3mm}
\caption{The stellar mass vs. metallicity relation of the $z\sim5$ ALPINE-CRISTAL-JWST sample. Other observations with JWST at $z=4-6$ (aqua open squares; see text for references) and at $z=6.5-7.5$ \citep[gold triangles; from the REBELS-IFU sample,][]{rowland25} are also shown. The lines show the parameterizations by \citet{maiolino08} at $z\lesssim3.5$. The gray large symbols show ALPIEN-CRISTAL-JWST type 1 AGN candidates by \citep{ren25}.
\label{fig:mzr}}
\end{figure}

\begin{figure*}[t!]
\centering
\includegraphics[angle=0,width=\textwidth]{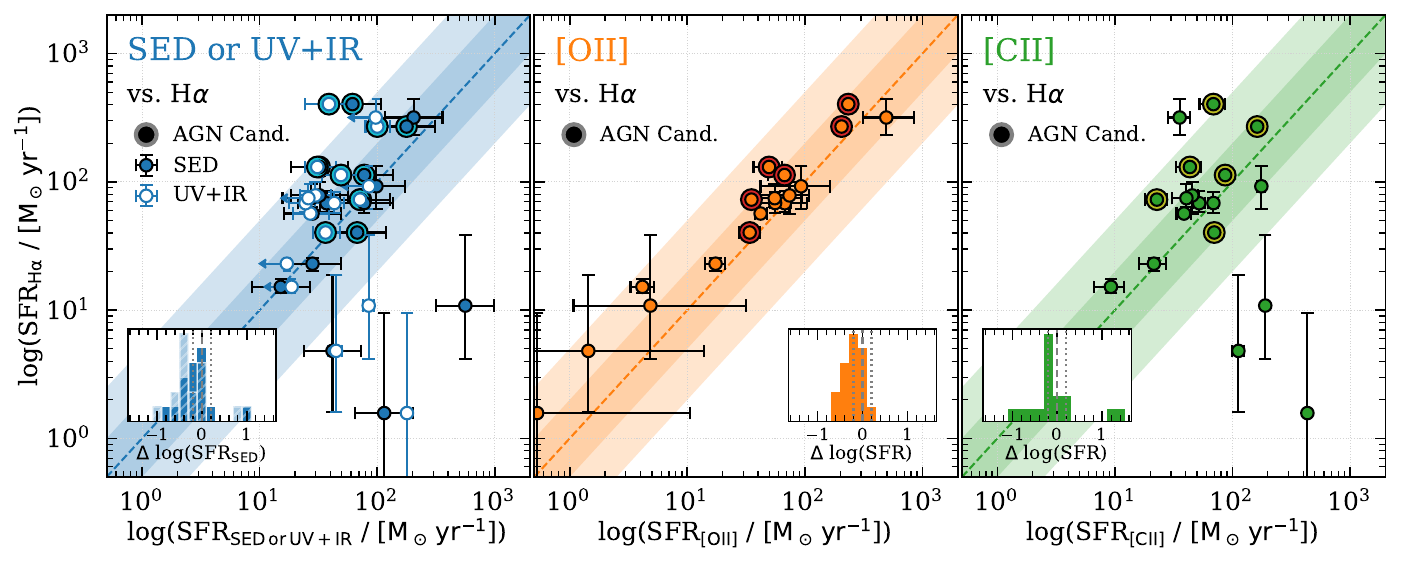}\vspace{-3mm}
\caption{Comparison of different star formation indicators to \halpha~(SED and UV$+$IR-based, left; \oii-based, middle; \cplus-based, right). Limiting UV$+$IR-based SFRs are given for upper limits in $L_{\rm IR}$ (left panel, empty symbols). The dark (light) bands show a factor of two (five) deviation from the 1-to-1 relation. The scatter between SED and \cplus-based SFR compared to \halpha~SFRs may be due to the different SFR timescales probed by these indicators. The inset shows histograms of the logarithmic difference in SFRs (negative values meaning higher \halpha~SFR). The dashed line indicates the median, while the dotted lines show the $1\sigma$ range. Type 1 AGN candidates from \citet{ren25} are indicated by larger symbols (see also Section~\ref{sec:sampleselection}). Note that these are consistent with the expected relations.
\label{fig:sfrcomparison}}
\end{figure*}

\subsection{Strong-Line Metallicity Measurement}\label{sec:measuremetallicity}

We measured the gas-phase metallicity, defined as $Z_{\rm neb} \equiv 12+\log({\rm O/H})$, of our galaxies using the strong optical emission lines detected for all individual galaxies. In \citet{faisst25b}, we will investigate in more detail $T_e$-based metallicities and the comparison of different strong-line calibrations.

Here, we measured the spatially integrated metal content of the galaxies using strong optical lines such as \oii, \hbeta, and \oiii~using the calibration revisited by \citet{sanders24}.
Specifically, we derived a best-fit metallicity for each galaxy by jointly fitting the line ratios 
$R_{\rm 23}\equiv$ \Rtwothree, 
$R_{\rm 2}\equiv$ \Rtwo,
$O_{\rm 32}\equiv$ \Othreetwo, and 
$R_{\rm 3}\equiv$ \Rthree~to the \citet{sanders24} parameterizations.
All emission line ratios were corrected for dust attenuation assuming the \citet{calzetti00} reddening curve and $\ebmvn$ as derived in Section~\ref{sec:measuredust}. The measured emission line ratios for each galaxy are listed in Table~\ref{tab:lineratios}.
We first used the parameterization from \citet{sanders24} and derive a $\chi^2(Z_{\rm neb})$ value for a grid of metallicities for each of the line ratios. To obtain the final best-fit metallicity and uncertainties, we combined the $\chi^2$ results and computed the $16-50-84$ percentiles.
The final strong line metallicity measurements are listed in Table~\ref{tab:physical}.
We note that using all the above line ratios could lead to overfitting (as they are not independent of each other).
We tested the impact of this on our final results by refitting the line ratios only using the combinations $R_{\rm 2}+O_{\rm 32}$ or $R_{\rm 3}+O_{\rm 32}$. We found that the metallicity measurements differ by less than $0.05\,{\rm dex}$ in $\rm 12+\log(O/H)$, an overall difference of $<1\sigma$.

Figure~\ref{fig:mzr} shows the stellar mass vs. metallicity relation for our ALPINE-CRISTAL-JWST survey targets. The measured abundances range from $7.7-8.5$ in $\rm 12+\log(O/H)$, corresponding to $10-70\%$ of solar metallicity\footnote{We assume here a solar value of $12+\log({\rm O/H}) = 8.69$ \citep{allendeprieto01}, however, note that this value could be higher \citep{bergemann21}.}.
We compared our measurements at $z\sim5$ to observations at other redshifts. These include the mass-metallicity parameterization from \citet{maiolino08} \citep[based on][]{nagao06} at $z<3.5$, measurements from the REBELS-IFU sample at $z=6.5-7.5$ \citep{rowland25}, and various other observations with JWST from the literature at $z=4-6$ \citep[including][]{marszewski24,curti23,nakajima23,morishita24,sarkar25}. Note that the metallicity measurements at $z>6.5$ are mainly based on \oiii~and \oii, while the ones derived for our sample are based on a combination of all strong optical lines. 
Overall, we find that the ALPINE-CRISTAL-JWST galaxies, which lie at the higher mass end of the galaxy distribution at these redshifts, are $\sim 0.5\,{\rm dex}$ more metal enriched than galaxies at the same redshifts at lower stellar masses of $\log(M_*/{\rm M_\odot}) < 9$. 
However, we do not find a significant evolution in metal abundances between $z=3.5$ and $7.5$ at a given stellar mass, emphasizing that early galaxies must already be significantly metal enriched. The detailed comparison of these findings to analytical and cosmological models will be presented in \citet{faisst25b}. In addition, spatially resolved metallicity measurements will be presented in \citet{fujimoto25} as well as other future works.

\subsection{Comparison of Different Star Formation Indicators}\label{sec:measuresfr}

Figure~\ref{fig:sfrcomparison} shows a comparison between the different SFR indicators discussed in Section~\ref{sec:physical}.
To first order, we find a good agreement between \halpha-derived SFRs and SFRs based on SED fitting and UV$+$FIR luminosity (left panel), \oii~emission (middle panel), and \cplus~\citep[right panel; using the calibration by][]{schaerer20}.
To second order, we find an increased scatter with an amplitude of a factor of five or more between \cplus~or continuum-based/SED-based SFR and \halpha~or \oii-based SFRs. The scatter is towards {\it higher} \halpha~or \oii~SFRs.
This systematic scatter can be explained naturally by a bursty star formation, which is common in star-forming galaxies at high redshifts \citep[e.g.,][]{faisst19,mehta23,navarrocarrera24,endsley24,sun25}. For example, \citet{faisst19} measured similar discrepancies between \halpha~and SED-based SFR measurements for a large sample of $z=4.5$ galaxies. (Although note that in this case \halpha~was measured from Spitzer [$3.6\,{\rm \mu m}$]--[$4.5\,{\rm \mu m}$] colors with an unknown $f_{\rm dust}$, which caused significant uncertainties.) 
\halpha~is tied to ionized gas around star forming regions and correlates directly with star formation on timescales of $\sim 10\,{\rm Myrs}$ \citep[e.g.,][]{kennicutt98}. \cplus~is expected to be emitted in various ISM phases, hence a correlation with star formation similar to the UV or far-IR continuum is expected \citep{herreracamus15,delooze11}. For the dependence of the \halpha~vs. \cplus~SFR relation with burstiness, see also the theoretical work by \citet{veraldi25}.
We find three outliers ({\it DC-873756}, {\it VC-5100541407}, and {\it VC-5101218326}) with significantly larger uncertainties in \halpha. 
The first one is the potential BPT-detected AGN (Figure~\ref{fig:bpt}). The other two seem to be significantly dust obscured according to their ALMA far-IR continuum. Consequently, they show a lower SNR in \oii, \hbeta, and \oiii~line emission, which increases the uncertainty in the nebular dust correction measurement, which in turn leads to an underestimation of their dust-corrected \halpha~and \oii-based SFRs.

\begin{figure}[t!]
\centering
\includegraphics[angle=0,width=0.45\textwidth]{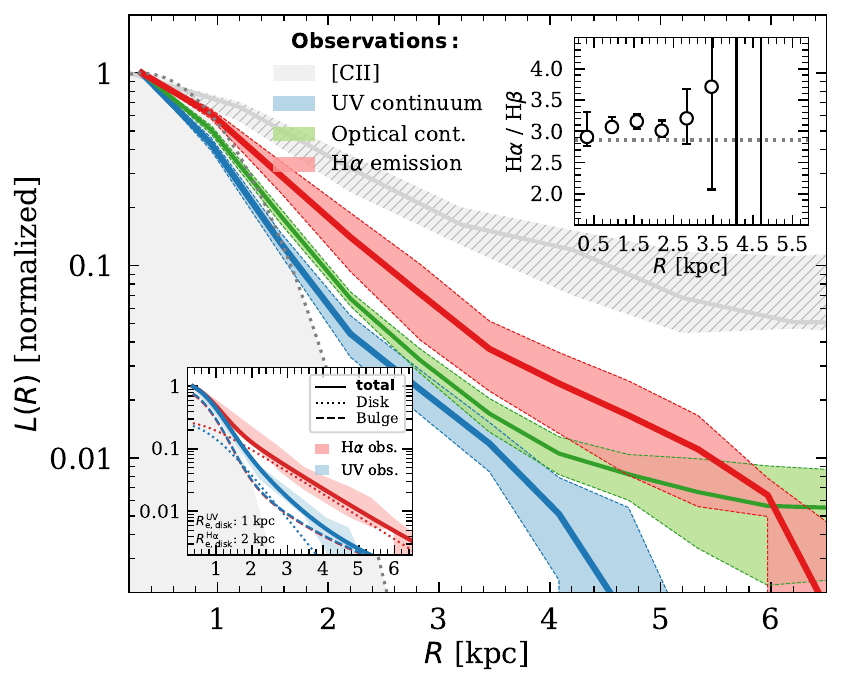}\vspace{-3mm}
\caption{
Average \cplus~(gray, hatched), UV continuum (blue), optical continuum (green), and \halpha~(red) radial profiles (thick lines with $1\sigma$ uncertainty). The profiles are normalized to their peak intensity. Only isolated and compact galaxies have been used (see text). The gray region shows the IFU spatial PSF profile. The upper right inset shows the average radial profile of the \halpha/\hbeta~flux ratio, suggesting a relatively constant radial dust attenuation. The lower left inset shows the decomposition of the UV (light blue) and \halpha~(light red) radial profiles in a disk ($n_{\rm Sersic} = 1$, dotted) and bulge ($n_{\rm Sersic}=4$, dashed) component. The bulge components have the same half-light radius, while $R_{\rm e,disk}^{\rm H\alpha} \sim 2\times R_{\rm e,disk}^{\rm UV}$.
\label{fig:resolvedexample}
}
\end{figure}

\subsection{Extended H$\alpha$ Emission at $z=5$}\label{sec:extendedHa}

In addition to the study of spatially integrated properties outlined in this paper, JWST's IFU observations open the door for new cutting-edge studies of the enrichment of the CGM around high-redshift galaxies. So far, extended halos of carbon-enriched warm gas have been traced by \cplus~emission out to as far as $10\,{\rm kpc}$ \citep[e.g.,][]{fujimoto19}. Only now with NIRSpec/IFU data, we are able to trace the ionized gas structure out to similar radii. This allows unique comparisons of the extended UV (ionized by young stars), \halpha~(ionized gas), and \cplus~(warm gas) and their comparison to simulations to inform feedback prescriptions and the abundance of satellite galaxies to ultimately study the origin of these halos \citep[e.g.,][]{ginolfi20,pizzati20,pizzati23,schimek24,romano24,elgueta24,birkin25,ikeda25}.

Figure~\ref{fig:resolvedexample} shows the stacked UV, \halpha, and \cplus~radial profiles obtained from a sub-sample of isolated, non-merging galaxies from the ALPINE-CRISTAL-JWST sample ({\it DC-494763}, {\it DC-519281}, {\it DC-630594}, {\it DC-709575}, {\it VC-5100822662}, and {\it VC-5101218326}).
A two-component fit (inset) confirms \halpha~disk components of two times larger effective radii compared to continuum emission. This suggests the extent of the ionized medium into the CGM, either due to surrounding star forming regions (e.g., star-forming satellites) or ionization of gas from the young stellar population in the central galaxy.
We can also see that the \cplus~emission is more extended that the \halpha~emission (and significantly more than the UV emission) suggesting maintained \cplus~halos far out into the CGM. This is consistent with a relatively enriched CGM at similar levels of the ISM \citep{wang25} and a comparison with simulations suggests that outflows may play a role more significant than satellite galaxies in this enrichment process \citep{lunjunliu25}.

\section{Summary and Future Work}\label{sec:conclusions}

In this paper, we have introduced the ALPINE-CRISTAL-JWST survey, a benchmark survey to study the spatially resolved properties of the ISM and CGM of typical main-sequence $z=4-6$ galaxies at $\rm \log(M_*/M_\odot) > 9.5$.
The unique combination of HST, JWST, and ALMA enables for the first time the observations of stars, dust, and gas at UV through far-IR wavelengths at comparable spatial resolution of $\sim 1\,{\rm kpc}$. Only with such a multi-wavelength exploration we can progress our understanding of the intertwined physical processes in galaxies at early times. 
Located right at the post-EoR cosmic time, the ALPINE-CRISTAL-JWST survey provides the necessary link between primordial galaxy evolution during the EoR ($z>6$) and mature galaxy evolution at cosmic noon ($z=2-3$). 
Furthermore, it complements studies of JWST-detected galaxies at lower stellar masses. For example, we show that the metal content of galaxies across all masses evolves little between $z=5$ and cosmic noon. This is discussed in more detail in a forthcoming paper \citep{faisst25b}.

The ALPINE-CRISTAL-JWST survey helps answering many open questions regarding the ISM and CGM properties of post-EoR galaxies, including how and where stars are formed, how their ISM is built up, and how the CGM is enriched. The survey also sheds light on the contribution of faint AGN to the typical galaxy population at these redshifts.

In addition to the spatial integrated properties outlined in this paper, JWST's IFU enables new resolved observations of the rest-frame optical light, which, combined with imaging and ALMA observations on similar spatial scales, will lead surely to new scientific discoveries.

This includes the measurement of the star formation burstiness in different parts of the galaxies through the spatially resolved \halpha/UV continuum emission ratio \citep{hadi25}, reflecting bursts of star formations \citep[e.g.,][]{faisst19}.
Another interesting avenue of research may be extended \halpha~halos and their connection to UV and \cplus~emission (see Section~\ref{sec:extendedHa}). The study of the multi-phase gas (ionized, warm, neutral) in the halos of these galaxies provide unique insights into the processes of feedback (stellar and AGN), outflow, and the existence of satellite galaxies in enriching the CGM of these galaxies. The comparison to state-of-the-art simulations \citet{lunjunliu25} will enable us to gain a comprehensive picture of these processes and their impact on early galaxy formation and evolution.

In addition to these interesting avenues, the ALPINE-CRISTAL-JWST survey will enable future science investigations capitalizing on its unique combination of the spectral and spatial resolution and multi-wavelength coverage of different cutting-edge facilities:

\begin{enumerate}
    \item[{\bf I.}] The integrated stellar mass vs. metallicity relation and its comparison to state-of-the-art model predictions will be further discussed in \citet{faisst25b}.
    
    \item[{\bf II.}] Resolved metallicity abundances and gradients as well as the spatially resolved relation between metallicity, SFR and stellar mass density will be discussed in the ALPINE-CRISTAL-JWST data reduction paper by \citet{fujimoto25} as well as \citet{leelilian25}.
    
    \item[{\bf III.}] The discovery of low-luminosity type 1 (broad line) AGN and their black hole masses will be presented in \citet{ren25}.
    
    \item[{\bf IV.}] A comprehensive investigation of the difference in the chemical properties of the ISM and CGM will be presented in \citet{wang25}.
    
    \item[{\bf V.}] The study of the resolved stellar and nebular dust attenuation will be presented in \citet{tsujita25}.
    
    \item[{\bf VI.}] The properties of star-formation driven outflows and their role in producing the first massive quiescent galaxies will be discussed in \citet{davies25}.
    
    \item[{\bf VII.}] The metallicity dependence of the \cplus~vs. FIR luminosity ratio will be studied to understand \cplus~deficits \citep{herreracamus25b}.
    
    \item[{\bf VIII.}] A spatial study of star formation burstiness and comparison to theoretical models will be carried out in \citet{hadi25}.

    \item[{\bf IX.}] A detailed investigation of the (spatially resolved) relation between \cplus~and SFR will be carried out in \citet{accard25}, \citet{palla25}, and \citet{LiYuan25}.

    \item[{\bf X.}] Spatially resolved scaling relations including dust masses and dust temperatures will be studied in \citet{lopez25} and \citet{relano25}.
    
    \item[{\bf XI.}] The relation between N/O and O/H will be studied in \citet{delooze25}.
    
\end{enumerate}

Additional studies will focus on the spectroscopic properties of star forming clumps, constraining the IMF, the comparison of the kinematics of low and high ionization gas as traced by \cplus~and optical lines, the detailed spatially resolve star formation history modeling including spectroscopic data \citep[see also][]{li24}, the study of dust production mechanisms, and the characterization of spatial changes in the ionization parameter ($\xi_{\rm ion}$) that can inform the reionization of neutral hydrogen by galaxies in the EoR.

\begin{acknowledgments}
{\small
This work is based in part on observations made with the NASA/ESA/CSA \textit{James Webb} Space Telescope.
The JWST and HST data presented in this article were obtained from the Mikulski Archive for Space Telescopes (MAST) at the Space Telescope Science Institute.
These observations are associated with programs, HST-GO-13641 (\dataset[doi: 10.17909/xne1-7v26]{https://doi.org/10.17909/xne1-7v26}), JWST-GO-01727 (\dataset[doi: 10.17909/ph8h-qf05]{https://doi.org/10.17909/ph8h-qf05}), JWST-GO-03045 (\dataset[doi: 10.17909/cqds-qc81]{https://doi.org/10.17909/cqds-qc81}), and JWST-GO-04265 (\dataset[doi: 10.17909/wac6-9741]{https://doi.org/10.17909/wac6-9741}).
This research has made use of the NASA/IPAC Infrared Science Archive (IRSA), which is funded by the National Aeronautics and Space Administration and operated by the California Institute of Technology. The following dataset was used from IRSA: \dataset[doi: 10.26131/IRSA178]{https://doi.org/10.26131/IRSA178}.

Support for program JWST-GO-03045 was provided by NASA through a grant from the Space Telescope Science Institute, which is operated by the Association of Universities for Research in Astronomy, Inc., under NASA contract NAS 5-03127.
This work made use of Astropy:\footnote{http://www.astropy.org} a community-developed core Python package and an ecosystem of tools and resources for astronomy \citep{astropy13, astropy18, astropy22}.
This paper makes use of the following ALMA data: ADS/JAO.ALMA\#2017.1.00428.L, \#2019.1.00226.S, \#2022.1.01118.S, and \#2021.1.00280.L. ALMA is a partnership of ESO (representing its member states), NSF (USA) and NINS (Japan), together with NRC (Canada), MOST and ASIAA (Taiwan), and KASI (Republic of Korea), in cooperation with the Republic of Chile. The Joint ALMA Observatory is operated by ESO, AUI/NRAO and NAOJ.
M.A. is supported by FONDECYT grant number 1252054, and gratefully acknowledges support from ANID Basal Project FB210003 and ANID MILENIO NCN2024\_112.
A.N. and P.S. acknowledge support from the Narodowe Centrum Nauki (NCN), Poland, through the SONATA BIS grant UMO-2020/38/E/ST9/00077
M.B. gratefully acknowledges support from the ANID BASAL project FB210003 and from the FONDECYT regular grant 1211000.
S.F. acknowledges support from NASA through the NASA Hubble Fellowship grant HST-HF2-51505.001-A awarded by the Space Telescope Science Institute, which is operated by the Association of Universities for Research in Astronomy, Incorporated, under NASA contract NAS5-26555.
R.J.A was supported by FONDECYT grant number 1231718 and by the ANID BASAL project FB210003.
D.R. gratefully acknowledges support from the Collaborative Research Center 1601 (SFB 1601 sub-projects C1, C2, C3, and C6) funded by the Deutsche Forschungsgemeinschaft (DFG) --- 500700252.
R.A. acknowledges financial support from projects PID2023-147386NB-I00 and the Severo Ochoa grant CEX2021-001131-S funded by MCIN/AEI/10.13039/501100011033
K.T. acknowledges support from JSPS KAKENHI Grant No. 23K03466.
V.V. acknowledges support from the ANID BASAL project FB210003 and from ANID --- MILENIO --- NCN2024\_112.
M.S. was financially supported by Becas-ANID scholarship \#21221511, and also acknowledges support from ANID BASAL project FB210003.
M.P acknowledges financial support from the project ``LEGO --- Reconstructing the building blocks of the Galaxy by chemical tagging'' granted by the Italian MUR through contract PRIN2022LLP8TK\_001.
J. M. gratefully acknowledges support from ANID MILENIO NCN2024\_112
G.C.J acknowledges support by the Science and Technology Facilities Council (STFC), by the ERC through Advanced Grant 695671 ``QUENCH'', and by the UKRI Frontier Research grant RISEandFALL.
R.L.D is supported by the Australian Research Council through the Discovery Early Career Researcher Award (DECRA) Fellowship DE240100136 funded by the Australian Government.
E.dC acknowledges support from the Australian Research Council through project DP240100589.
E.I. gratefully acknowledge financial support from ANID - MILENIO - NCN2024\_112 and ANID FONDECYT Regular 1221846.
A.F. is partly supported by the ERC Advanced Grant INTERSTELLAR H2020/740120, and by grant NSF PHY-2309135 to the Kavli Institute for Theoretical Physics.
M.R. acknowledges support from project PID2023-150178NB-I00 financed by MCIU/AEI/10.13039/501100011033, and by FEDER, UE. 
LV acknowledges support from the INAF Minigrant ``RISE: Resolving the ISM and Star formation in the Epoch of Reionization'' (PI: Vallini, Ob. Fu. 1.05.24.07.01)
This work was supported by the French government through the France 2030 investment plan managed by the National Research Agency (ANR), as part of the Initiative of Excellence of Universit\'e Cote d'Azur under reference number ANR-15-IDEX-01.
We acknowledge the Lorentz Center for giving us the opportunity and infrastructure to organize a successful and very stimulating workshop, which discussions helped to realize this work. 
}
\end{acknowledgments}

%

\vspace{5mm}
\facilities{IRSA, HST(ACS), HST(WFC3/IR), JWST(NIRCam), JWST(NIRSpec), ALMA}


\software{\texttt{Astropy} \citep{astropy13,astropy18,astropy22},
          \texttt{PyNeb} \citep{luridiana15},
          \texttt{CIGALE} \citep{boquien19,burgarella05},
          \texttt{Bagpipes} \citep{carnall18,carnall19},
          \texttt{MAGPHYS} \citep{dacunha08,dacunha15}
          }



\clearpage
\newpage
\appendix
\setcounter{figure}{0}
\renewcommand\thefigure{\thesection.\arabic{figure}}  
\setcounter{table}{0}
\renewcommand{\thetable}{\thesection.\arabic{table}}

\section{Tables}\label{sec:tables}
The following tables describe a summary of the targets and available data (Table~\ref{tab:obsdataext}), a summary of physical measurements (Table~\ref{tab:physical}), and a list of measured emission line ratios (Table~\ref{tab:lineratios}) and line fluxes (Table~\ref{tab:linefluxes}).

\newpage
\begin{table*}
\scriptsize
\centering
\setlength{\tabcolsep}{3pt}\vspace{-3mm}
\caption{Summary of targets and available data.}
\label{tab:obsdataext}
\begin{tabular}{c c c C{2cm} C{2cm} c C{3.5cm}  C{6.5cm}} 
\hline \hline
\multicolumn{2}{c}{Name} && \multicolumn{2}{c}{JWST} && HST & ALMA \\ \cline{1-2} \cline{4-5}
ALPINE & CRISTAL &&  Spec. & Imaging && Spec./Imag. & \\ \hline
DC-417567 & CRISTAL-10  && {\bf 3045}, 5974, 6480 & -- && 9822, 16443, 13641 & 2012.1.00523.S, 2017.1.00428.L, 2021.1.00280.L, 2019.1.00459.S\\ \hline
DC-494763 & CRISTAL-20 && {\bf 3045}, 5893 & 1727 && 9822, 13657, 16259, 16443, 16684, 17802 & 2017.1.00428.L, 2021.1.00280.L, 2023.1.00180.L\\ \hline
DC-519281 & CRISTAL-09 && {\bf 3045}, 5974, 5893 & 1727 && 9822, 12578, 13669, 14114, 16443, 17802 & 2017.1.00428.L, 2021.1.00280.L\\ \hline
DC-536534 & CRISTAL-03 && {\bf 3045}, 5893 & 1727 && 9822, 12578, 13641, 16259, 16443 16684, 17802 & 2012.1.00523.S, 2017.1.00428.L, 2021.1.00280.L, 2023.1.00180.L\\ \hline
DC-630594 & CRISTAL-11 && {\bf 3045}, 5893, 6368 & 1727, 1837, 1840, 2321, 2514 && 9822, 12440, 12328, 15100, 15647, 17802  & 2013.1.01292.S, 2015.1.00379.S, 2017.1.00428.L, 2021.1.00280.L, 2021.1.00225.S\\ \hline
DC-683613 & CRISTAL-05 && {\bf 3045}, 5893, 5974 & 1727, 1837, 5893 && 9822/9999, 12328, 13641, 13657, 15100, 16443, 17802 & 2012.1.00523.S, 2017.1.00428.L, 2018.1.01359.S, 2019.1.00459.S, 2021.1.00280.L, 2021.1.00705.S, 2021.1.00225.S, 2023.1.00180.L
\\ \hline
DC-709575 & CRISTAL-14 && {\bf 3045}, 5893 & 1727, 5893 && 9822, 13294, 13641, 13868, 16259, 16443, 17802 & 2017.1.00428.L, 2021.1.00280.L, 2021.1.00225.S\\ \hline
DC-742174 & CRISTAL-17 && {\bf 3045}, 5545, 5893, 6368 & 1727, 1837 && 9822/9999, 12328, 12440, 12461, 15647, 16443, 17802 & 2017.1.00428.L, 2021.1.00280.L, 2021.1.00225.S, 2023.1.00180.L\\ \hline
DC-842313 & CRISTAL-01 && {\bf 3045}, 4265 & 1727, 3954  && 9822, 13641, 13294, 14114, 14719, 16443, 17802  & 2012.1.00978.S, 2016.1.00171.S, 2016.1.00478.S,  2017.1.00428.L, 2019.1.01587.S, 2021.1.00280.L, 2022.1.00863.S\\ \hline
DC-848185 & CRISTAL-02 && {\bf 3045} & 1727, 2417 && 9822, 12328, 13641, 13384, 14114, 16443, 17802 & 2011.0.00964.S, 2012.1.00523.S, 2013.1.01258.S, 2015.1.00928.S, 2015.1.00388.S, 2016.1.01149.S, 2017.1.00428.L, 2018.1.00348.S, 2019.1.00459.S, 2021.1.00280.L, 2021.1.00705.S\\ \hline
DC-873321 & CRISTAL-07 && {\bf 3045}, 5974 & 1727 && 9822, 13641, 16443, 17802 & 2012.1.00523.S, 2017.1.00428.L, 2021.1.00280.L, 2022.1.00863.S\\ \hline
DC-873756 & CRISTAL-24 && {\bf 3045}, 5974 & 1727 && 9822, 13641, 16259, 16443, 17802 & 2017.1.00428.L, 2019.1.00226.S, 2024.1.01401.S\\ \hline
VC-5100541407 & CRISTAL-06 && {\bf 3045}, 6480 & 1727 && 9822, 12578, 14495, 14596, 16443, 17802 & 2017.1.00428.L, 2021.1.00280.L\\ \hline
VC-5100822662 & CRISTAL-04 && {\bf 3045}, 5893 & 1727 && 9822, 12578, 13669, 14114, 16443, 17802 & 2017.1.00428.L, 2021.1.00280.L\\ \hline
VC-5100994794 & CRISTAL-13 && {\bf 3045}, 5893, 6368 & 1727, 1837 && 9822, 12328, 12440, 15647, 17802 & 2017.1.00428.L, 2021.1.00225.S, 2021.1.00280.L\\ \hline
VC-510128326 & CRISTAL-25 && {\bf 3045}, 5893 & 1727 && 9822, 15692, 16259, 16443, 17802 & 2015.1.00379.S, 2016.1.01546.S, 2017.1.00428.L, 2019.1.00226.S\\ \hline
VC-5101244930 & CRISTAL-15 && {\bf 3045}, 5893 & 1727, 1837 && 9822, 12328, 12440, 15100, 16443, 17802 & 2013.1.01292.S, 2017.1.00428.L, 2021.1.00280.L, 2021.1.00225.S\\ \hline
VC-5110377875 & -- && {\bf 3045}, 5893 & 1727 && 9822, 16259, 16443, 17802 & 2015.1.00379.S, 2017.1.00428.L, 2022.1.01118.S\\ \hline
\end{tabular}
\tablecomments{\footnotesize
\underline{Proposal IDs for JWST:}
\href{https://www.stsci.edu/jwst/science-execution/program-information?id=1727}{1727} (PI: Kartaltepe, NIRCam F115W/F150W/F277W/F444W, MIRI F770W),
\href{https://www.stsci.edu/jwst/science-execution/program-information?id=1837}{1837} (PI: Dunlop, NIRCam F090W/F115W/F200W/F277W/F356W/F444W, MIRI F770W/F1800W),
\href{https://www.stsci.edu/jwst/science-execution/program-information?id=1840}{1840} (PI: Alvarez-Marquez, NIRCam F115W/F150W/F200W/F250W/F335W/F444W),
\href{https://www.stsci.edu/jwst/science-execution/program-information?id=2321}{2321} (PI: Best, NIRCam F212N/F200W/F444W/F470N),
\href{https://www.stsci.edu/jwst/science-execution/program-information?id=2417}{2417} (PI: Riechers, NIRCam F200W/F356W/F444W, MIRI F1000W/F1500W/F2100W),
\href{https://www.stsci.edu/jwst/science-execution/program-information?id=2514}{2514} (PI: Williams, NIRCam F115W/F150W/F200W/F277W/F356W/F444W),
\href{https://www.stsci.edu/jwst/science-execution/program-information?id=3045}{3045} (PI: Faisst, NIRSpec/IFU G235M/G395M),
\href{https://www.stsci.edu/jwst/science-execution/program-information?id=3954}{3954} (PI: Lelli, MIRI F770W),
\href{https://www.stsci.edu/jwst/science-execution/program-information?id=5545}{5545} (PI: Barrufet, NIRSpec Prism\footnote{\label{fn:msa}MSA may not include source.}),
\href{https://www.stsci.edu/jwst/science-execution/program-information?id=5893}{5893} (PI: Kakiichi, NIRCam/Grism F115W/F200W/F356W/F444W, MIRI F1000W/F2100W),
\href{https://www.stsci.edu/jwst/science-execution/program-information?id=5974}{5974} (PI: Aravena, NIRSpec/IFU G395H),
\href{https://www.stsci.edu/jwst/science-execution/program-information?id=6245}{6245} (PI: Gonzalez-Lopez, NIRCam F200W/F356W, NIRSpec G235M/G395H),
\href{https://www.stsci.edu/jwst/science-execution/program-information?id=6368}{6368} (PI: Dickinson, NIRSpec/MSA Prism$^{\rm \ref{fn:msa}}$),
\href{https://www.stsci.edu/jwst/science-execution/program-information?id=6480}{6480} (PI: Schouws, NIRCam/Grism, NIRCam F070W/F115W/F200W/F356W/F444W)
$\star$~\underline{Proposal IDs for HST:}
\href{https://www.stsci.edu/hst-program-info/program/?program=9822}{9822}/\href{https://www.stsci.edu/hst-program-info/program/?program=9999}{9999} (PI: Scoville, ACS F814W),
\href{https://www.stsci.edu/hst-program-info/program/?program=12328}{12328} (PI: van Dokkum, ACS F814W, WFC3 F140W/G141/G800L),
\href{https://www.stsci.edu/hst-program-info/program/?program=12440}{12440} (PI: Faber, ACS F814W/F606W, WFC3 F125W/F160W),
\href{https://www.stsci.edu/hst-program-info/program/?program=12461}{12461} (PI: Riess, ACS F606W/F814W),
\href{https://www.stsci.edu/hst-program-info/program/?program=12578}{12578} (PI: Forster-Schreiber WFC3 F110W/F160W),
\href{https://www.stsci.edu/hst-program-info/program/?program=13294}{13294} (PI: Karim, ACS F814W),
\href{https://www.stsci.edu/hst-program-info/program/?program=13384}{13384} (PI: Riechers, ACS F606W, WFC3 F125W/F160W),
\href{https://www.stsci.edu/hst-program-info/program/?program=13641}{13641} (PI: Capak, WFC3 F105W/F125W/F160W),
\href{https://www.stsci.edu/hst-program-info/program/?program=13657}{13657} (PI: Kartaltepe, WFC3 F160W),
\href{https://www.stsci.edu/hst-program-info/program/?program=13669}{13669} (PI: Carollo WFC3 F438W),
\href{https://www.stsci.edu/hst-program-info/program/?program=13868}{13868} (PI: Kocevski, WFC3 F160W),
\href{https://www.stsci.edu/hst-program-info/program/?program=14114}{14114} (PI: van Dokkum, WFC3 F160W),
\href{https://www.stsci.edu/hst-program-info/program/?program=14495}{14495} (PI: Sobral, WFC3 F140W/G141),
\href{https://www.stsci.edu/hst-program-info/program/?program=14596}{14596} (PI: Fan, WFC3 F110W/F160W),
\href{https://www.stsci.edu/hst-program-info/program/?program=14719}{14719} (PI: Best, WFC3 F606W/F140W),
\href{https://www.stsci.edu/hst-program-info/program/?program=15100}{15100} (PI: Cooke, ACS F435W),
\href{https://www.stsci.edu/hst-program-info/program/?program=15647}{15647} (PI: Teplitz, ACS F435W, WFC3 F275W),
\href{https://www.stsci.edu/hst-program-info/program/?program=15692}{15692} (PI: Faisst, WFC3 F105W/F160W),
\href{https://www.stsci.edu/hst-program-info/program/?program=16259}{16259}/\href{https://www.stsci.edu/hst-program-info/program/?program=16443}{16443} (PI: Momcheva, WFC3 F160W/G141),
\href{https://www.stsci.edu/hst-program-info/program/?program=16684}{16684} (PI: Lemaux, WFC3 F160W/G141),
\href{https://www.stsci.edu/hst-program-info/program/?program=17802}{17802} (PI: Kartaltepe, ACS F435W/F606W, WFC3 F098W)
$\star$~\underline{Proposal IDs for ALMA:}
\href{https://almascience.nrao.edu/aq/?projectCode=2017.1.00428.L}{2011.0.00964.S} (PI: Riechers, Band 7), 
\href{https://almascience.nrao.edu/aq/?projectCode=2012.1.00523.S}{2012.1.00523.S} (PI: Capak, Band 7), 
\href{https://almascience.nrao.edu/aq/?projectCode=2012.1.00978.S}{2012.1.00978.S} (PI: Karim, Band 7),
\href{https://almascience.nrao.edu/aq/?projectCode=2013.1.01258.S}{2013.1.01258.S} (PI: Riechers, Band 7), 
\href{https://almascience.nrao.edu/aq/?projectCode=2013.1.01292.S}{2013.1.01292.S} (PI: Leiton, Band 7), 
\href{https://almascience.nrao.edu/aq/?projectCode=2015.1.00379.S}{2015.1.00379.S} (PI: Schinnerer, Band 6),
\href{https://almascience.nrao.edu/aq/?projectCode=2015.1.00388.S}{2015.1.00388.S} (PI: Lu, Band 6), 
\href{https://almascience.nrao.edu/aq/?projectCode=2015.1.00928.S}{2015.1.00928.S} (PI: Pavesi, Band 6), 
\href{https://almascience.nrao.edu/aq/?projectCode=2016.1.00171.S}{2016.1.00171.S} (PI: Daddi, Band 3), 
\href{https://almascience.nrao.edu/aq/?projectCode=2016.1.00478.S}{2016.1.00478.S} (PI: Miettinen, Band 7), 
\href{https://almascience.nrao.edu/aq/?projectCode=2016.1.01149.S}{2016.1.01149.S} (PI: Keating, Band 3), 
\href{https://almascience.nrao.edu/aq/?projectCode=2016.1.01546.S}{2016.1.01546.S} (PI: Cassata, Band 3), 
\href{https://almascience.nrao.edu/aq/?projectCode=2017.1.00428.L}{2017.1.00428.L} (ALPINE; PI: Le F\`efre, Band 7), 
\href{https://almascience.nrao.edu/aq/?projectCode=2018.1.00231.S}{2018.1.00231.S} (MORA; PI: Casey, Band 4), 
\href{https://almascience.nrao.edu/aq/?projectCode=2018.1.00348.S}{2018.1.00348.S} (PI: Faisst, Band 8), 
\href{https://almascience.nrao.edu/aq/?projectCode=2018.1.01359.S}{2018.1.01359.S} (PI: Aravena, Band 7),
\href{https://almascience.nrao.edu/aq/?projectCode=2019.1.00226.S}{2019.1.00226.S} (PI: Ibar, Band 7), 
\href{https://almascience.nrao.edu/aq/?projectCode=2019.1.00459.S}{2019.1.00459.S} (PI: Scoville, Band 4), 
\href{https://almascience.nrao.edu/aq/?projectCode=2019.1.01587.S}{2019.1.01587.S} (TRICEPS; PI: Lelli, Band 7), 
\href{https://almascience.nrao.edu/aq/?projectCode=2021.1.00225.S}{2021.1.00225.S} (exMORA; PI: Casey, Band 4), 
\href{https://almascience.nrao.edu/aq/?projectCode=2021.1.00280.L}{2021.1.00280.L} (CRISTAL; PI: Herrera-Camus, Band 7), 
\href{https://almascience.nrao.edu/aq/?projectCode=2021.1.00705.S}{2021.1.00705.S} (PI: Cooper, Band 4), 
\href{https://almascience.nrao.edu/aq/?projectCode=2022.1.00863.S}{2022.1.00863.S} (PI: Hodge, Band 3),
\href{https://almascience.nrao.edu/aq/?projectCode=2022.1.01118.S}{2022.1.01118.S} (PI: B\'ethermin, Band 7), 
\href{https://almascience.nrao.edu/aq/?projectCode=2023.1.00180.L}{2023.1.00180.L} (CHAMPS; PI: Faisst, Band 6), 
\href{https://almascience.nrao.edu/aq/?projectCode=2024.1.01401.S}{2024.1.01401.S} (PI: Herrera-Camus, Band 9)
}
\end{table*}
\newpage

\newpage
\begin{table*}
\centering
\footnotesize
\setlength{\tabcolsep}{3pt}
\caption{Summary of physical measurements.}
\label{tab:physical}
\begin{tabular}{ccccccccccc}
\hline \hline
Name & $\rm M_*$$^{\triangledown}$ & SFR$_{\rm SED}$$^{\triangledown}$ & SFR$_{\rm UV+IR}$ & SFR$_{\rm [CII]}$ & SFR$_{\rm H\alpha}$ & SFR$_{\rm [OII]}$ & Age & E$_{\rm s}$(B-V)$^\flat$ & E$_{\rm n}$(B-V) & 12+log(O/H)$^{\dagger}$\\
 & $(\rm 10^9\,M_\odot)$ & $(\rm M_\odot\,yr^{-1})$ & $(\rm M_\odot\,yr^{-1})$ & $(\rm M_\odot\,yr^{-1})$ & $(\rm M_\odot\,yr^{-1})$ & $(\rm M_\odot\,yr^{-1})$ & (Myr) & (mag) & (mag) &  \\ \hline
DC-417567 & $9.8^{+10.2}_{-5}$ & $72^{+42}_{-27}$ & $72^{+11}_{-10}$ & $23^{+5}_{-5}$ & $73^{+4}_{-4}$ & $35^{+4}_{-4}$ & $105^{+100}_{-56}$ & $0.16^{+0.04}_{-0.04}$ & $0.08^{+0.02}_{-0.02}$ & $8.04^{+0.02}_{-0.02}$ \\
DC-494763 & $3.2^{+4.2}_{-1.8}$ & $28^{+36}_{-16}$ & $27^{+21}_{-9}$ & $39^{+5}_{-5}$ & $57^{+4}_{-4}$ & $42^{+5}_{-4}$ & $175^{+352}_{-119}$ & $0.07^{+0.01}_{-0.01}$ & $0.16^{+0.02}_{-0.02}$ & $8.29^{+0.02}_{-0.02}$ \\
DC-519281 & $6.9^{+10.1}_{-4.1}$ & $32^{+35}_{-17}$ & $31^{+17}_{-7}$ & $43^{+11}_{-10}$ & $131^{+11}_{-20}$ & $50^{+15}_{-13}$ & $324^{+331}_{-194}$ & $0.11^{+0.01}_{-0.01}$ & $0.21^{+0.02}_{-0.04}$ & $7.99^{+0.08}_{-0.08}$ \\
DC-536534 & $25.1^{+23.9}_{-12.2}$ & $62^{+64}_{-31}$ & $35^{+22}_{-9}$ & $69^{+17}_{-16}$ & $404^{+30}_{-35}$ & $234^{+37}_{-28}$ & $456^{+275}_{-224}$ & $0.43^{+0.02}_{-0.02}$ & $0.33^{+0.02}_{-0.03}$ & $8.1^{+0.04}_{-0.05}$ \\
DC-630594 & $4.8^{+5.4}_{-2.5}$ & $37^{+39}_{-19}$ & $25^{+11}_{-8}$ & $52^{+6}_{-6}$ & $68^{+16}_{-10}$ & $68^{+37}_{-17}$ & $186^{+227}_{-88}$ & $0.09^{+0.01}_{-0.01}$ & $0.26^{+0.07}_{-0.05}$ & $8.4^{+0.04}_{-0.04}$ \\
DC-683613 & $14.5^{+17.9}_{-8}$ & $68^{+67}_{-34}$ & $38^{+16}_{-11}$ & $70^{+7}_{-7}$ & $40^{+4}_{-5}$ & $34^{+8}_{-7}$ & $274^{+271}_{-159}$ & $0.07^{+0.01}_{-0.01}$ & $0.15^{+0.02}_{-0.04}$ & $8.49^{+0.03}_{-0.05}$ \\
DC-709575 & $3.4^{+4.7}_{-2}$ & $28^{+39}_{-16}$ & $>17$ & $21^{+6}_{-5}$ & $23^{+3}_{-3}$ & $17^{+3}_{-3}$ & $190^{+302}_{-95}$ & $0.06^{+0.01}_{-0.01}$ & $0.15^{+0.04}_{-0.04}$ & $8.27^{+0.02}_{-0.03}$ \\
DC-742174 & $3.2^{+4.9}_{-1.9}$ & $15^{+16}_{-8}$ & $>19$ & $9^{+3}_{-3}$ & $15^{+2}_{-2}$ & $4^{+1}_{-1}$ & $354^{+338}_{-215}$ & $0.09^{+0.01}_{-0.01}$ & $0.02^{+0.04}_{-0.04}$ & $7.84^{+0.05}_{-0.05}$ \\
DC-842313 & $44.7^{+96.6}_{-30.5}$ & $204^{+918}_{-167}$ & $>97$ & $36^{+8}_{-7}$ & $318^{+124}_{-85}$ & $493^{+357}_{-179}$ & $394^{+250}_{-221}$ & $0.25^{+0.02}_{-0.02}$ & $0.64^{+0.11}_{-0.09}$ & $8.48^{+0.03}_{-0.05}$ \\
DC-848185 & $20^{+18.1}_{-9.5}$ & $178^{+290}_{-110}$ & $93^{+31}_{-18}$ & $162^{+12}_{-11}$ & $270^{+24}_{-15}$ & $205^{+35}_{-26}$ & $98^{+52}_{-48}$ & $0.15^{+0.01}_{-0.01}$ & $0.2^{+0.03}_{-0.02}$ & $8.28^{+0.02}_{-0.03}$ \\
DC-873321 & $10^{+11.4}_{-5.3}$ & $78^{+64}_{-35}$ & $47^{+25}_{-14}$ & $87^{+12}_{-12}$ & $113^{+7}_{-10}$ & $67^{+12}_{-18}$ & $112^{+84}_{-63}$ & $0.13^{+0.01}_{-0.01}$ & $0.2^{+0.02}_{-0.03}$ & $8.11^{+0.06}_{-0.07}$ \\
DC-873756 & $33.9^{+6.9}_{-5.7}$ & $115^{+76}_{-46}$ & $175^{+14}_{-10}$ & $434^{+19}_{-19}$ & $2^{+8}_{-1}$ & $1^{+10}_{-0}$ & $550^{+317}_{-169}$ & $0.43^{+0.02}_{-0.02}$ & $>-0.14$ & $8.68^{+0.58}_{-0.88}$ \\
VC-5100541407 & $12.3^{+12.2}_{-6.1}$ & $42^{+50}_{-23}$ & $45^{+9}_{-6}$ & $112^{+14}_{-13}$ & $5^{+14}_{-3}$ & $1^{+12}_{-1}$ & $400^{+427}_{-239}$ & $0.1^{+0.02}_{-0.02}$ & $0.11^{+0}_{-0.79}$ & $8.21^{+0.24}_{-0.4}$ \\
VC-5100822662 & $14.1^{+13.4}_{-6.9}$ & $78^{+48}_{-30}$ & $42^{+8}_{-7}$ & $69^{+7}_{-7}$ & $68^{+15}_{-11}$ & $56^{+21}_{-14}$ & $237^{+196}_{-122}$ & $0.06^{+0.01}_{-0.01}$ & $0.19^{+0.05}_{-0.05}$ & $8.33^{+0.04}_{-0.04}$ \\
VC-5100994794 & $4.5^{+5.3}_{-2.4}$ & $32^{+51}_{-20}$ & $30^{+10}_{-8}$ & $45^{+5}_{-5}$ & $78^{+22}_{-22}$ & $74^{+35}_{-33}$ & $193^{+252}_{-95}$ & $0.09^{+0.01}_{-0.01}$ & $0.3^{+0.08}_{-0.11}$ & $8.41^{+0.04}_{-0.05}$ \\
VC-5101218326 & $79.4^{+86.5}_{-41.4}$ & $562^{+534}_{-274}$ & $83^{+14}_{-7}$ & $190^{+10}_{-10}$ & $11^{+28}_{-7}$ & $5^{+27}_{-4}$ & $646^{+97}_{-275}$ & $0.16^{+0.01}_{-0.01}$ & $0.1^{+0}_{-0.69}$ & $8.36^{+0.34}_{-0.3}$ \\
VC-5101244930 & $4.9^{+5.6}_{-2.6}$ & $28^{+20}_{-12}$ & $>26$ & $41^{+11}_{-10}$ & $75^{+22}_{-11}$ & $55^{+38}_{-19}$ & $183^{+365}_{-88}$ & $0.03^{+0}_{-0}$ & $0.24^{+0.08}_{-0.06}$ & $8.24^{+0.1}_{-0.13}$ \\
VC-5110377875 & $14.7^{+6.8}_{-4.7}$ & $99^{+75}_{-43}$ & $>85$ & $176^{+13}_{-13}$ & $92^{+40}_{-31}$ & $93^{+70}_{-51}$ & $161^{+127}_{-108}$ & $0.11^{+0.01}_{-0.01}$ & $0.26^{+0.12}_{-0.14}$ & $8.41^{+0.08}_{-0.09}$ \\ \hline
\end{tabular}
\tablecomments{
$^{\triangledown}$ From \citet{herreracamus25} based on \texttt{CIGALE} fits in \citet{mitsuhashi24}.\\
$^\flat$ From \citet{tsujita25}.\\
$^{\dagger}$ Uses the \citet{sanders24} strong-line calibration.\\
}
\end{table*}
\newpage



\newpage
\begin{table*}
\centering
\setlength{\tabcolsep}{3pt}
\caption{Summary of measured (dust corrected) line ratios.}
\label{tab:lineratios}
\begin{tabular}{c c c c c c c c} 
\hline \hline
Name & $\log(R_{\rm 23})$ & $\log(R_{\rm 3})$ & $\log(R_{\rm 2})$ & $\log(O_{\rm 32})$ & $\log(N_{\rm 2})$ & $\log(S_{\rm 2})$ & $\log(S_{\rm R})$ \\ \hline
DC-417567 & $1.03^{+0.01}_{-0.01}$ & $0.97^{+0.01}_{-0.01}$ & $0.14^{+0.02}_{-0.03}$ & $0.71^{+0.02}_{-0.02}$ & $-1.13^{+0.06}_{-0.03}$ & $-1.25^{+0.06}_{-0.06}$ & $0.11^{+0.10}_{-0.06}$ \\
DC-494763 & $0.95^{+0.01}_{-0.01}$ & $0.83^{+0.01}_{-0.01}$ & $0.33^{+0.03}_{-0.03}$ & $0.37^{+0.02}_{-0.03}$ & $-0.93^{+0.02}_{-0.03}$ & $-0.95^{+0.03}_{-0.05}$ & $-0.22^{+0.07}_{-0.07}$ \\
DC-519281 & $0.95^{+0.01}_{-0.02}$ & $0.89^{+0.01}_{-0.01}$ & $0.04^{+0.09}_{-0.09}$ & $0.73^{+0.08}_{-0.09}$ & $-0.97^{+0.10}_{-0.05}$ & $-1.11^{+0.09}_{-0.12}$ & $-0.01^{+0.13}_{-0.11}$ \\
DC-536534 & $1.06^{+0.01}_{-0.01}$ & $0.99^{+0.01}_{-0.01}$ & $0.23^{+0.04}_{-0.06}$ & $0.64^{+0.05}_{-0.05}$ & $-1.01^{+0.08}_{-0.11}$ & $-1.29^{+0.13}_{-0.06}$ & $-0.20^{+0.12}_{-0.15}$ \\
DC-630594 & $0.99^{+0.05}_{-0.03}$ & $0.84^{+0.03}_{-0.03}$ & $0.46^{+0.07}_{-0.06}$ & $0.24^{+0.04}_{-0.04}$ & $-0.93^{+0.03}_{-0.01}$ & $-0.91^{+0.02}_{-0.02}$ & $-0.12^{+0.02}_{-0.03}$ \\
DC-683613 & $0.82^{+0.03}_{-0.03}$ & $0.62^{+0.02}_{-0.01}$ & $0.39^{+0.05}_{-0.05}$ & $0.10^{+0.05}_{-0.03}$ & $-0.76^{+0.04}_{-0.03}$ & $-0.83^{+0.07}_{-0.03}$ & $-0.05^{+0.05}_{-0.10}$ \\
DC-709575 & $0.99^{+0.02}_{-0.02}$ & $0.88^{+0.02}_{-0.02}$ & $0.34^{+0.03}_{-0.04}$ & $0.42^{+0.03}_{-0.02}$ & $-1.16^{+0.05}_{-0.04}$ & $-1.01^{+0.04}_{-0.05}$ & $-0.06^{+0.09}_{-0.04}$ \\
DC-742174 & $0.96^{+0.02}_{-0.02}$ & $0.92^{+0.02}_{-0.02}$ & $-0.11^{+0.06}_{-0.05}$ & $0.91^{+0.03}_{-0.05}$ & $-1.61^{+0.29}_{-0.04}$ & $-1.10^{+0.07}_{-0.04}$ & $-0.12^{+0.11}_{-0.15}$ \\
DC-842313 & $1.12^{+0.06}_{-0.04}$ & $0.94^{+0.03}_{-0.04}$ & $0.65^{+0.10}_{-0.08}$ & $0.14^{+0.07}_{-0.05}$ & $-0.75^{+0.04}_{-0.41}$ & $-0.64^{+0.03}_{-0.14}$ & $-0.13^{+0.02}_{-0.02}$ \\
DC-848185 & $0.97^{+0.02}_{-0.01}$ & $0.86^{+0.01}_{-0.01}$ & $0.34^{+0.03}_{-0.04}$ & $0.39^{+0.04}_{-0.02}$ & $-0.75^{+0.09}_{-0.08}$ & $-0.70^{+0.07}_{-0.08}$ & $-0.07^{+0.02}_{-0.03}$ \\
DC-873321 & $1.04^{+0.02}_{-0.02}$ & $0.96^{+0.01}_{-0.01}$ & $0.22^{+0.07}_{-0.09}$ & $0.62^{+0.08}_{-0.07}$ & $-1.08^{+0.04}_{-0.06}$ & $-1.16^{+0.06}_{-0.04}$ & $-0.04^{+0.04}_{-0.05}$ \\
DC-873756 & $0.54^{+0.44}_{-0.27}$ & $0.19^{+0.37}_{-0.21}$ & $-0.12^{+0.70}_{-0.44}$ & $-0.10^{+0.60}_{-0.72}$ & $-0.25^{+0.27}_{-0.18}$ & $-0.33^{+0.34}_{-0.10}$ & $-0.02^{+0.17}_{-0.16}$ \\
VC-5100541407 & $0.52^{+0.35}_{-0.13}$ & $0.42^{+0.23}_{-0.14}$ & $-0.02^{+0.40}_{-0.42}$ & $0.36^{+0.22}_{-0.21}$ & $-0.77^{+0.09}_{-0.11}$ & $-0.63^{+0.08}_{-0.16}$ & $-0.14^{+0.08}_{-0.25}$ \\
VC-5100822662 & $0.95^{+0.03}_{-0.04}$ & $0.82^{+0.02}_{-0.03}$ & $0.37^{+0.06}_{-0.06}$ & $0.33^{+0.05}_{-0.04}$ & $-0.97^{+0.07}_{-0.05}$ & $-0.98^{+0.09}_{-0.03}$ & $-0.09^{+0.03}_{-0.04}$ \\
VC-5100994794 & $0.94^{+0.03}_{-0.07}$ & $0.77^{+0.03}_{-0.05}$ & $0.43^{+0.07}_{-0.10}$ & $0.22^{+0.06}_{-0.05}$ & $-0.96^{+0.03}_{-0.02}$ & $-0.89^{+0.03}_{-0.02}$ & $0.03^{+0.03}_{-0.05}$ \\
VC-5101218326 & $0.64^{+0.19}_{-0.17}$ & $0.45^{+0.21}_{-0.11}$ & $0.07^{+0.37}_{-0.27}$ & $0.23^{+0.29}_{-0.36}$ & $-0.64^{+0.21}_{-0.16}$ & $-0.46^{+0.12}_{-0.17}$ & $0.01^{+0.13}_{-0.14}$ \\
VC-5101244930 & $1.01^{+0.06}_{-0.04}$ & $0.90^{+0.03}_{-0.03}$ & $0.32^{+0.13}_{-0.18}$ & $0.46^{+0.13}_{-0.12}$ & $-1.19^{+0.11}_{-0.06}$ & $-0.87^{+0.04}_{-0.05}$ & $-0.21^{+0.04}_{-0.05}$ \\
VC-5110377875 & $0.94^{+0.07}_{-0.10}$ & $0.78^{+0.05}_{-0.07}$ & $0.44^{+0.13}_{-0.15}$ & $0.23^{+0.09}_{-0.09}$ & $-1.09^{+0.06}_{-0.04}$ & $-0.98^{+0.05}_{-0.05}$ & $0.00^{+0.07}_{-0.07}$ \\ \hline
\end{tabular}
\tablecomments{
Definition of line ratios: $R_{\rm 23}\equiv$ \Rtwothree,
$R_{\rm 3}\equiv$ \Rthree,
$R_{\rm 2}\equiv$ \Rtwo,
$O_{\rm 32}\equiv$ \Othreetwo,
$N_{\rm 2}\equiv$ \Ntwo,
$S_{\rm 2}\equiv$ \Stwo,
$S_{\rm R}\equiv$ \SR.
Note that \oii~includes the doublet, which is unresolved.
}
\end{table*}
\newpage


\newpage
\begin{sidewaystable}
\centering
\scriptsize
\setlength{\tabcolsep}{3pt}
\caption{Observed emission line fluxes in $10^{-18}\,{\rm erg\,s^{-1}\,cm^{-2}}$.}
\label{tab:linefluxes}
\begin{tabular}{cccccccccccccccc} 
\hline \hline
Name & [\ion{O}{2}]$^\dagger$ & [\ion{Ne}{3}]$_{\rm 3868}$ & H$\gamma$ & [\ion{O}{3}]$_{\rm 4363}$ & H$\beta$ & [\ion{O}{3}]$_{\rm 4959}$ & [\ion{O}{3}]$_{\rm 5007}$ & [\ion{N}{2}]$_{\rm 6548}$ & \multicolumn{2}{c}{H$\alpha$} & [\ion{N}{2}]$_{\rm 6585}$ & [\ion{S}{2}]$_{\rm 6718}$ & [\ion{S}{2}]$_{\rm 6732}$ & [\ion{O}{2}]$_{\rm 7322}$ & [\ion{O}{2}]$_{\rm 7332}$ \\ \cline{10-11}
 &  & &  &  &  &  &  &  & (total) & (broad) &  &  &  &  &  \\ \hline
DC-417567 & $14.46^{+0.38}_{-0.46}$ & $5.53^{+0.36}_{-0.32}$ & $5.14^{+0.21}_{-0.27}$ & $1.82^{+0.42}_{-0.32}$ & $11.51^{+0.21}_{-0.26}$ & $26.93^{+0.25}_{-0.24}$ & $81.28^{+0.31}_{-0.29}$ & $0.90^{+0.13}_{-0.06}$ & $36.25^{+0.13}_{-0.36}$ & -- & $2.69^{+0.38}_{-0.17}$ & $0.90^{+0.16}_{-0.16}$ & $1.18^{+0.18}_{-0.16}$ & $0.59^{+0.29}_{-0.21}$ & $0.59^{+0.16}_{-0.15}$ \\
DC-494763 & $13.54^{+0.71}_{-0.30}$ & $2.76^{+0.41}_{-0.51}$ & $3.46^{+0.43}_{-0.31}$ & $0.88^{+0.50}_{-0.41}$ & $7.66^{+0.18}_{-0.18}$ & $13.43^{+0.19}_{-0.14}$ & $39.44^{+0.22}_{-0.24}$ & $1.02^{+0.05}_{-0.07}$ & $26.42^{+0.28}_{-0.24}$ & -- & $3.06^{+0.14}_{-0.21}$ & $1.87^{+0.13}_{-0.18}$ & $1.12^{+0.19}_{-0.15}$ & $<0.48$ & $0.32^{+0.13}_{-0.12}$ \\
DC-519281 & $11.02^{+2.38}_{-2.18}$ & $4.73^{+1.56}_{-1.82}$ & $3.99^{+0.93}_{-2.12}$ & $5.54^{+2.16}_{-2.43}$ & $12.55^{+0.34}_{-0.33}$ & $24.95^{+0.31}_{-0.37}$ & $74.96^{+0.45}_{-0.42}$ & $1.20^{+0.42}_{-0.14}$ & $45.87^{+0.80}_{-1.51}$ & $12.35^{+1.93}_{-0.97}$ & $3.61^{+1.25}_{-0.41}$ & $1.35^{+0.47}_{-0.26}$ & $1.32^{+0.33}_{-0.28}$ & $<1.19$ & $<0.85$ \\
DC-536534 & $24.97^{+2.44}_{-2.54}$ & $9.51^{+1.09}_{-1.12}$ & $11.03^{+0.81}_{-0.43}$ & $3.94^{+1.16}_{-1.08}$ & $22.02^{+0.41}_{-0.58}$ & $56.18^{+0.42}_{-0.62}$ & $167.58^{+0.62}_{-0.58}$ & $1.89^{+0.28}_{-0.36}$ & $92.32^{+1.63}_{-1.26}$ & $34.45^{+5.48}_{-3.10}$ & $5.68^{+0.83}_{-1.07}$ & $1.92^{+0.32}_{-0.34}$ & $1.23^{+0.33}_{-0.32}$ & $0.83^{+0.42}_{-0.26}$ & $0.75^{+0.51}_{-0.38}$ \\
DC-630594 & $18.55^{+0.64}_{-0.90}$ & $2.82^{+0.72}_{-0.60}$ & $3.32^{+0.47}_{-0.49}$ & $0.83^{+0.44}_{-0.43}$ & $8.81^{+0.55}_{-0.73}$ & $16.13^{+0.66}_{-0.41}$ & $45.68^{+0.66}_{-0.49}$ & $1.37^{+0.10}_{-0.05}$ & $34.31^{+0.22}_{-0.24}$ & -- & $4.10^{+0.29}_{-0.14}$ & $2.46^{+0.16}_{-0.16}$ & $1.84^{+0.13}_{-0.10}$ & $0.36^{+0.13}_{-0.18}$ & $0.72^{+0.26}_{-0.24}$ \\
DC-683613 & $10.31^{+0.97}_{-0.90}$ & $<2.03$ & $2.31^{+0.65}_{-0.57}$ & $1.70^{+0.84}_{-1.10}$ & $5.02^{+0.13}_{-0.24}$ & $5.34^{+0.10}_{-0.16}$ & $15.94^{+0.21}_{-0.15}$ & $0.85^{+0.07}_{-0.05}$ & $17.13^{+0.37}_{-0.65}$ & $2.80^{+1.02}_{-0.69}$ & $2.55^{+0.22}_{-0.16}$ & $1.21^{+0.13}_{-0.15}$ & $1.04^{+0.12}_{-0.13}$ & $<0.49$ & $0.42^{+0.15}_{-0.13}$ \\
DC-709575 & $8.97^{+0.24}_{-0.37}$ & $1.65^{+0.17}_{-0.15}$ & $1.79^{+0.22}_{-0.10}$ & $0.74^{+0.23}_{-0.13}$ & $4.89^{+0.16}_{-0.21}$ & $9.39^{+0.16}_{-0.15}$ & $28.37^{+0.24}_{-0.21}$ & $0.39^{+0.04}_{-0.04}$ & $16.67^{+0.17}_{-0.15}$ & -- & $1.16^{+0.12}_{-0.13}$ & $0.87^{+0.13}_{-0.12}$ & $0.78^{+0.08}_{-0.09}$ & $<0.26$ & $0.22^{+0.07}_{-0.06}$ \\
DC-742174 & $2.41^{+0.27}_{-0.22}$ & $1.87^{+0.18}_{-0.13}$ & $1.33^{+0.15}_{-0.17}$ & $<10.27$ & $3.19^{+0.15}_{-0.13}$ & $6.59^{+0.17}_{-0.11}$ & $19.90^{+0.20}_{-0.14}$ & $<0.23$ & $9.36^{+0.11}_{-0.18}$ & -- & $<0.69$ & $0.42^{+0.09}_{-0.09}$ & $0.33^{+0.06}_{-0.08}$ & $<0.18$ & $0.34^{+0.14}_{-0.09}$ \\
DC-842313 & $16.63^{+0.90}_{-1.28}$ & $3.62^{+1.32}_{-0.79}$ & $3.46^{+0.64}_{-0.67}$ & $1.79^{+0.48}_{-0.77}$ & $7.67^{+0.67}_{-0.77}$ & $17.54^{+0.58}_{-0.91}$ & $53.38^{+0.93}_{-1.26}$ & $1.29^{+0.12}_{-0.57}$ & $45.46^{+3.61}_{-0.53}$ & $22.25^{+1.86}_{-3.84}$ & $3.87^{+0.36}_{-1.72}$ & $3.13^{+0.11}_{-0.11}$ & $2.38^{+0.11}_{-0.14}$ & $0.50^{+0.39}_{-0.21}$ & $0.79^{+0.46}_{-0.14}$ \\
DC-848185 & $53.08^{+2.75}_{-2.72}$ & $9.91^{+2.44}_{-1.45}$ & $13.63^{+1.44}_{-2.36}$ & $2.62^{+1.93}_{-1.02}$ & $30.85^{+0.53}_{-0.96}$ & $56.71^{+0.66}_{-0.40}$ & $169.74^{+0.91}_{-0.68}$ & $4.05^{+0.12}_{-0.12}$ & $111.18^{+0.71}_{-1.02}$ & $44.77^{+9.87}_{-11.36}$ & $12.16^{+0.36}_{-0.37}$ & $7.51^{+0.37}_{-0.42}$ & $6.27^{+0.42}_{-0.30}$ & $1.60^{+0.30}_{-0.32}$ & $0.79^{+0.26}_{-0.11}$ \\
DC-873321 & $18.01^{+3.61}_{-2.64}$ & $6.52^{+1.59}_{-0.87}$ & $6.17^{+1.43}_{-1.07}$ & $<6.79$ & $13.72^{+0.37}_{-0.36}$ & $31.97^{+0.31}_{-0.24}$ & $96.64^{+0.51}_{-0.41}$ & $1.08^{+0.09}_{-0.08}$ & $49.37^{+0.35}_{-0.46}$ & $10.11^{+2.46}_{-3.62}$ & $3.25^{+0.26}_{-0.24}$ & $1.51^{+0.11}_{-0.20}$ & $1.36^{+0.14}_{-0.14}$ & $0.76^{+0.19}_{-0.17}$ & $0.25^{+0.16}_{-0.17}$ \\
DC-873756 & $18.75^{+12.14}_{-6.10}$ & $<18.61$ & $<16.98$ & $<15.99$ & $11.18^{+6.09}_{-5.88}$ & $<19.72$ & $10.10^{+4.83}_{-7.29}$ & $2.64^{+1.14}_{-0.72}$ & $<11.95$ & -- & $7.93^{+3.43}_{-2.15}$ & $4.13^{+1.58}_{-1.47}$ & $3.40^{+1.38}_{-1.04}$ & $<6.58$ & $<5.71$ \\
VC-5100541407 & $9.03^{+2.81}_{-2.93}$ & $<9.17$ & $<6.96$ & $<6.08$ & $6.92^{+3.59}_{-2.71}$ & $4.43^{+1.33}_{-1.35}$ & $14.54^{+1.96}_{-2.84}$ & $0.74^{+0.19}_{-0.15}$ & $13.61^{+1.08}_{-0.69}$ & -- & $2.22^{+0.56}_{-0.46}$ & $1.87^{+0.37}_{-0.54}$ & $1.22^{+0.37}_{-0.39}$ & $<1.08$ & $0.78^{+0.39}_{-0.29}$ \\
VC-5100822662 & $21.71^{+1.18}_{-1.18}$ & $4.54^{+1.24}_{-1.06}$ & $4.74^{+0.76}_{-0.38}$ & $1.11^{+0.51}_{-0.54}$ & $11.65^{+0.70}_{-0.61}$ & $19.09^{+0.99}_{-0.59}$ & $59.57^{+1.15}_{-0.81}$ & $1.32^{+0.23}_{-0.19}$ & $41.13^{+2.58}_{-0.92}$ & -- & $3.97^{+0.68}_{-0.57}$ & $2.23^{+0.09}_{-0.08}$ & $1.78^{+0.12}_{-0.15}$ & $0.25^{+0.08}_{-0.08}$ & $0.14^{+0.08}_{-0.08}$ \\
VC-5100994794 & $14.85^{+1.14}_{-0.98}$ & $4.10^{+1.32}_{-1.22}$ & $2.41^{+0.60}_{-0.48}$ & $<1.34$ & $7.97^{+1.10}_{-0.68}$ & $11.77^{+0.61}_{-0.74}$ & $36.88^{+0.92}_{-0.80}$ & $1.21^{+0.07}_{-0.05}$ & $32.59^{+0.21}_{-0.21}$ & -- & $3.63^{+0.21}_{-0.16}$ & $2.08^{+0.21}_{-0.12}$ & $2.22^{+0.08}_{-0.14}$ & $0.71^{+0.11}_{-0.18}$ & $0.58^{+0.19}_{-0.12}$ \\
VC-5101218326 & $23.93^{+5.92}_{-6.54}$ & $<19.23$ & $<16.31$ & $<13.56$ & $14.14^{+4.64}_{-5.46}$ & $10.39^{+3.47}_{-2.23}$ & $24.35^{+8.38}_{-5.44}$ & $2.20^{+0.86}_{-0.78}$ & $26.78^{+3.73}_{-2.42}$ & -- & $6.61^{+2.58}_{-2.34}$ & $3.57^{+2.54}_{-1.00}$ & $3.61^{+1.11}_{-0.55}$ & $<5.32$ & $<6.01$ \\
VC-5101244930 & $16.47^{+2.89}_{-5.07}$ & $4.20^{+1.17}_{-0.83}$ & $4.86^{+0.56}_{-0.76}$ & $<2.09$ & $10.02^{+0.78}_{-0.82}$ & $21.21^{+0.78}_{-0.85}$ & $61.55^{+1.15}_{-1.47}$ & $0.63^{+0.11}_{-0.06}$ & $38.33^{+0.21}_{-0.39}$ & $9.33^{+3.10}_{-1.89}$ & $1.88^{+0.34}_{-0.17}$ & $2.41^{+0.13}_{-0.17}$ & $1.49^{+0.15}_{-0.19}$ & $0.57^{+0.12}_{-0.12}$ & $0.52^{+0.19}_{-0.10}$ \\
VC-5110377875 & $23.83^{+2.07}_{-2.33}$ & $6.27^{+2.94}_{-2.57}$ & $4.87^{+1.56}_{-0.72}$ & $<11.26$ & $11.81^{+2.03}_{-1.53}$ & $16.78^{+1.50}_{-1.23}$ & $55.06^{+2.19}_{-2.48}$ & $1.17^{+0.21}_{-0.13}$ & $45.48^{+1.38}_{-0.68}$ & -- & $3.51^{+0.64}_{-0.39}$ & $2.32^{+0.27}_{-0.27}$ & $2.35^{+0.34}_{-0.31}$ & $0.77^{+0.27}_{-0.16}$ & $0.72^{+0.43}_{-0.31}$ \\ \hline
\end{tabular}
\tablecomments{
Limits are $3\sigma$.\\
$^\dagger$ The [\ion{O}{2}] doublet is not resolved.
}
\end{sidewaystable}
\newpage


\clearpage
\newpage

\section{Individual Galaxies}\label{app:eachgalaxy}
\begin{figure*}[b!]
\centering\vspace{-5mm}
\includegraphics[angle=0,width=\textwidth]{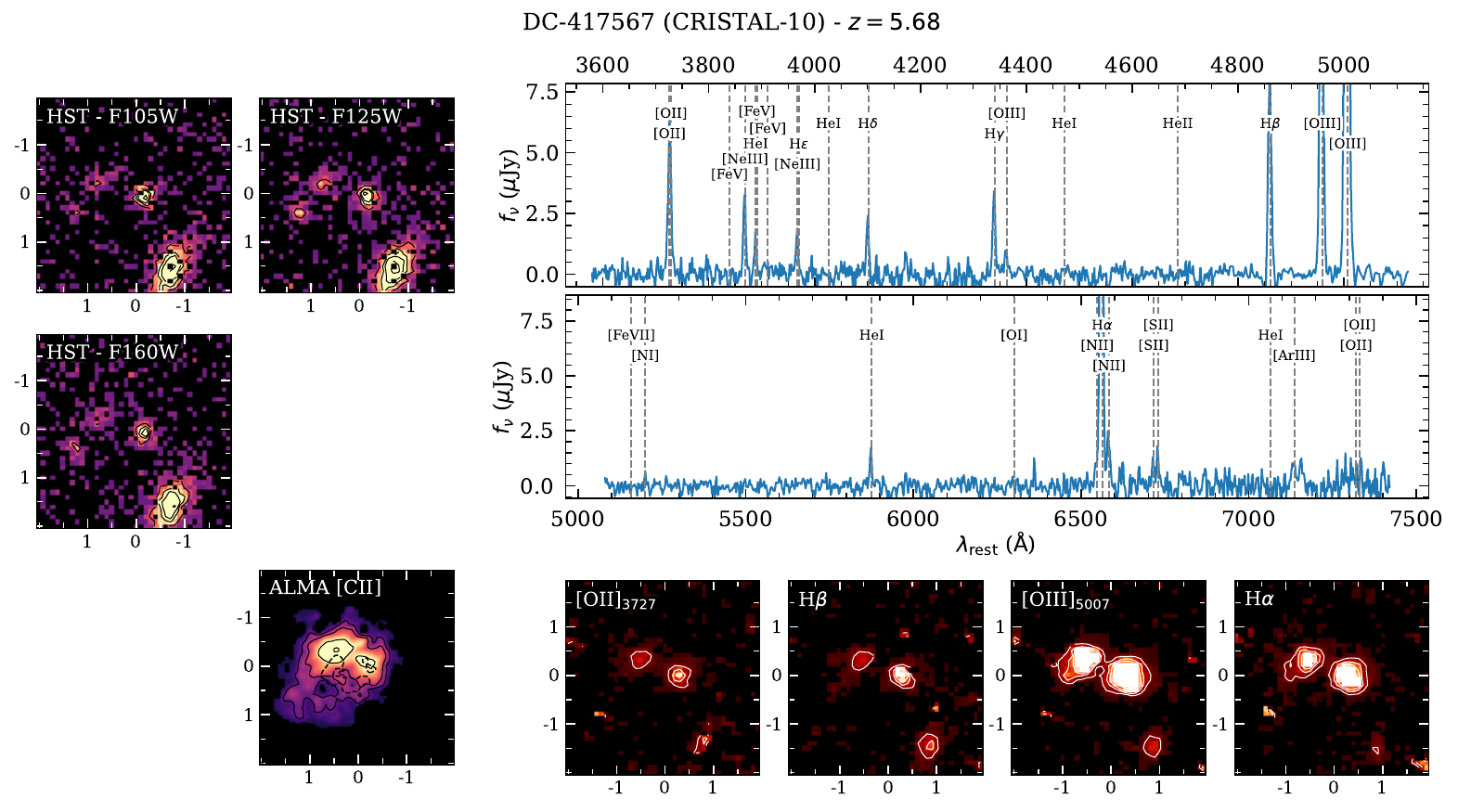}\\\vspace{-6mm}
\includegraphics[angle=0,width=\textwidth]{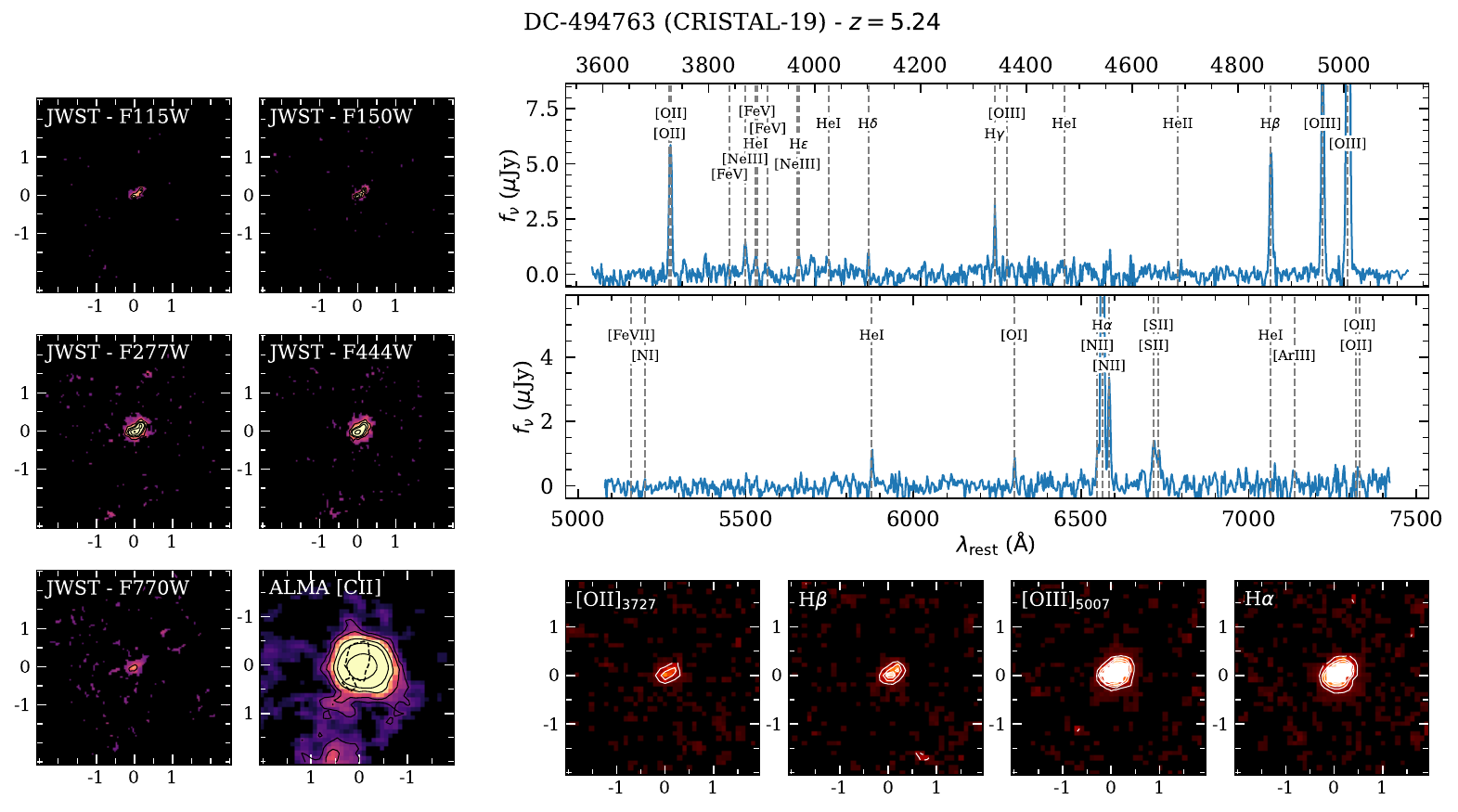}\vspace{-4mm}
\caption{Summary figures for targets {\it DC-417567} and {\it DC-494763}.
{\it Left six panels:} JWST/NIRCam and MIRI (if available) images (or HST for {\it DC-417567}) with corresponding contours (solid black) overlaid. The ALMA \cplus~moment map with \cplus~(solid black) and far-IR continuum (dashed black) emission contours overlaid.
{\it Lower right four panels:} Emission line maps (\oii, \hbeta, \oiii, and \halpha~from left to right).
{\it Large panels:} NIRSpec IFU integrated continuum-subtracted optical spectra with some emission line indicated.
All cutout images are $4\arcsec\times 4\arcsec$ in size and contours show 3, 5, 10, 15, 30, 50, and $100\sigma$ levels.
\label{fig:eachgalaxy}}
\end{figure*}

\begin{figure*}[t!]
\centering
\includegraphics[angle=0,width=\textwidth]{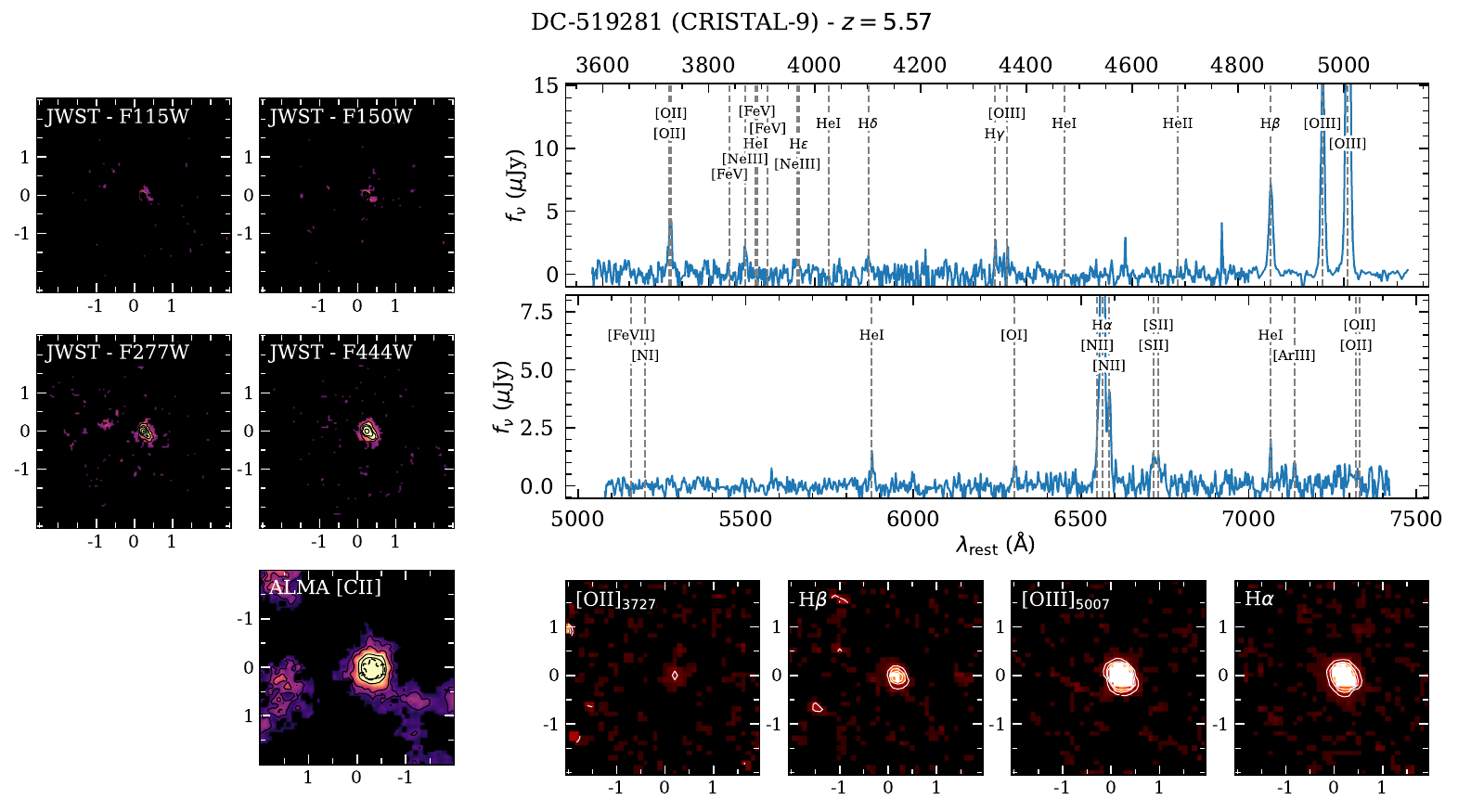}\\
\includegraphics[angle=0,width=\textwidth]{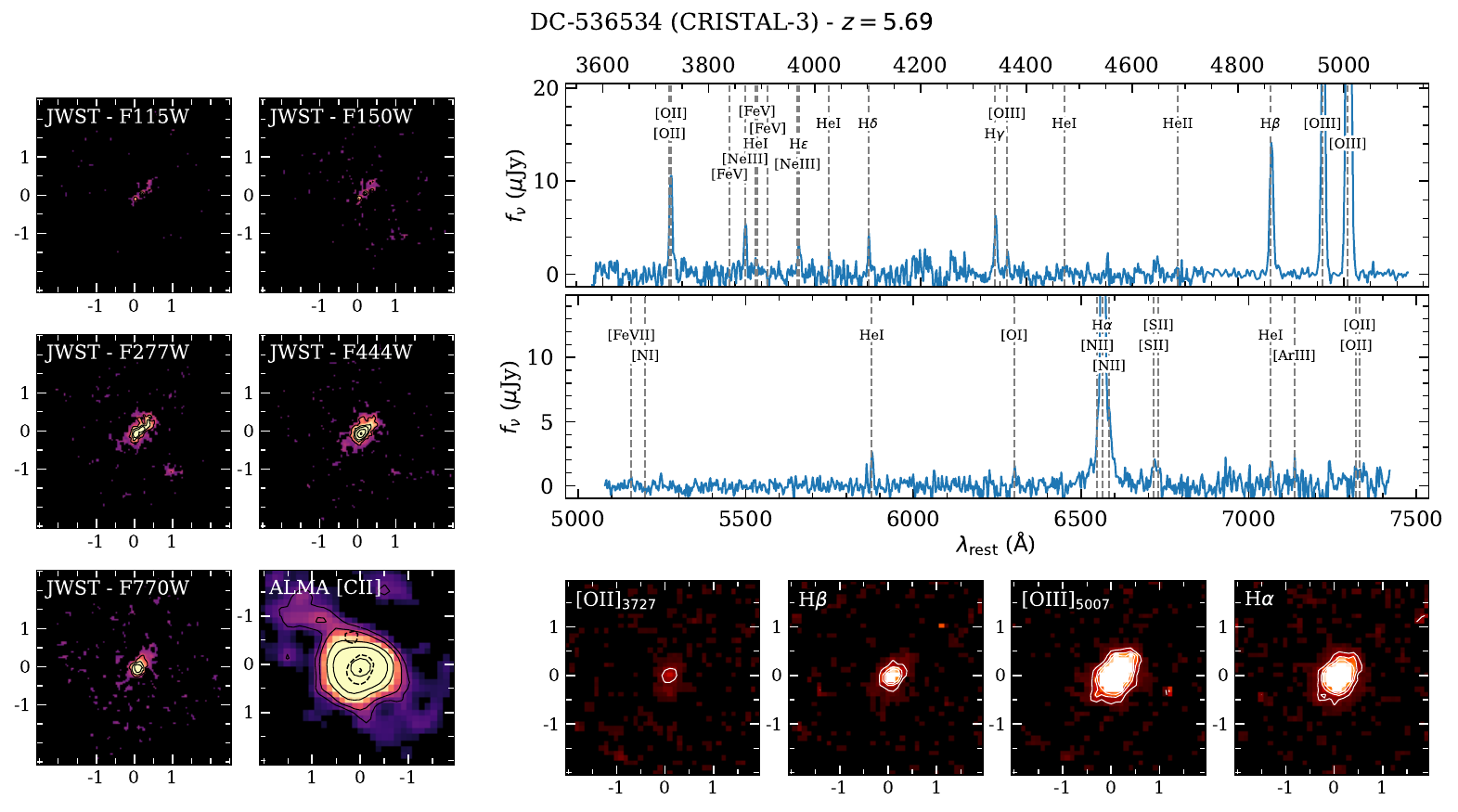}
\caption{Same as Figure~\ref{fig:eachgalaxy} but for targets {\it DC-519281} and {\it DC-536534}.
\label{fig:eachgalaxy2}}
\end{figure*}

\begin{figure*}[t!]
\centering
\includegraphics[angle=0,width=\textwidth]{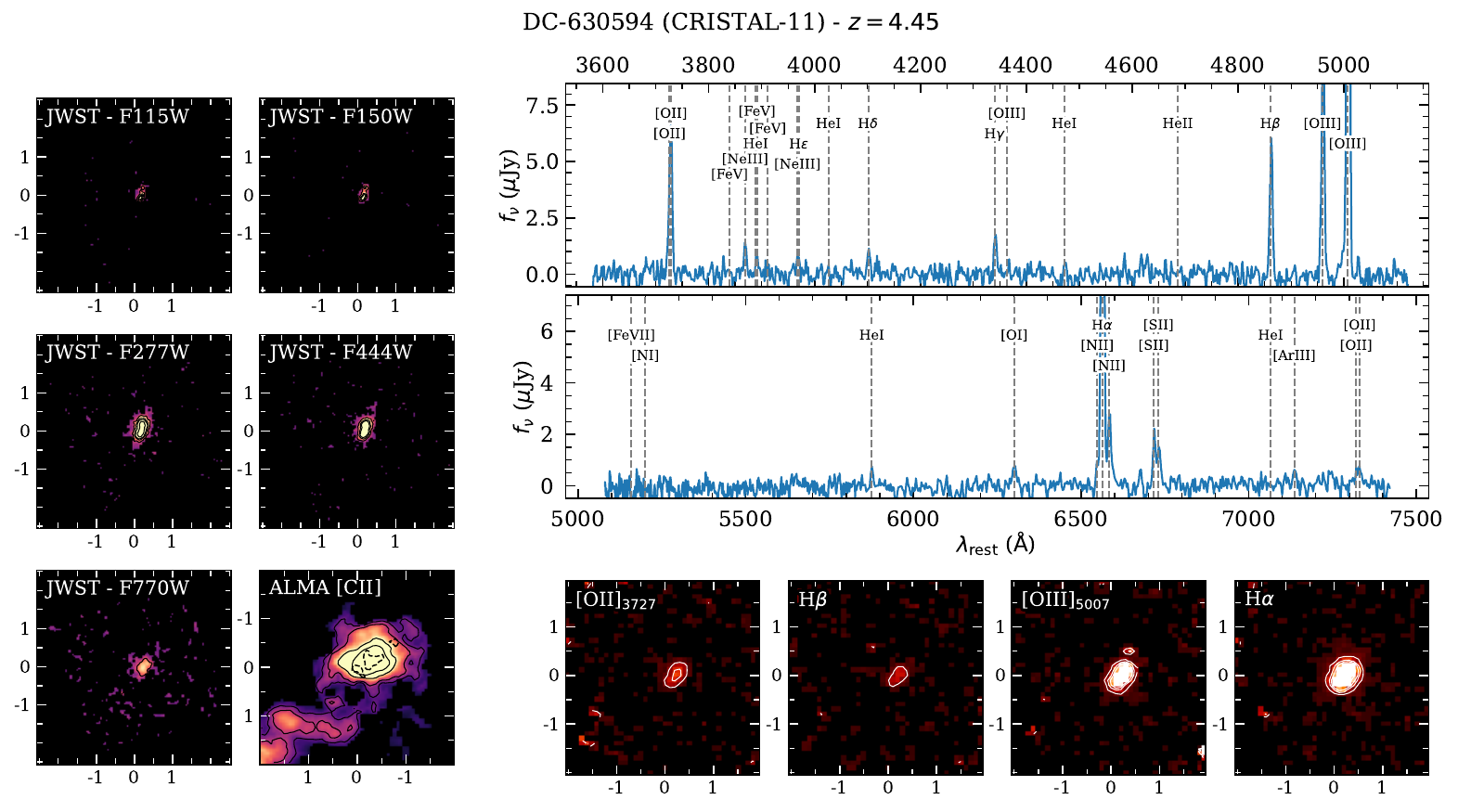}\\
\includegraphics[angle=0,width=\textwidth]{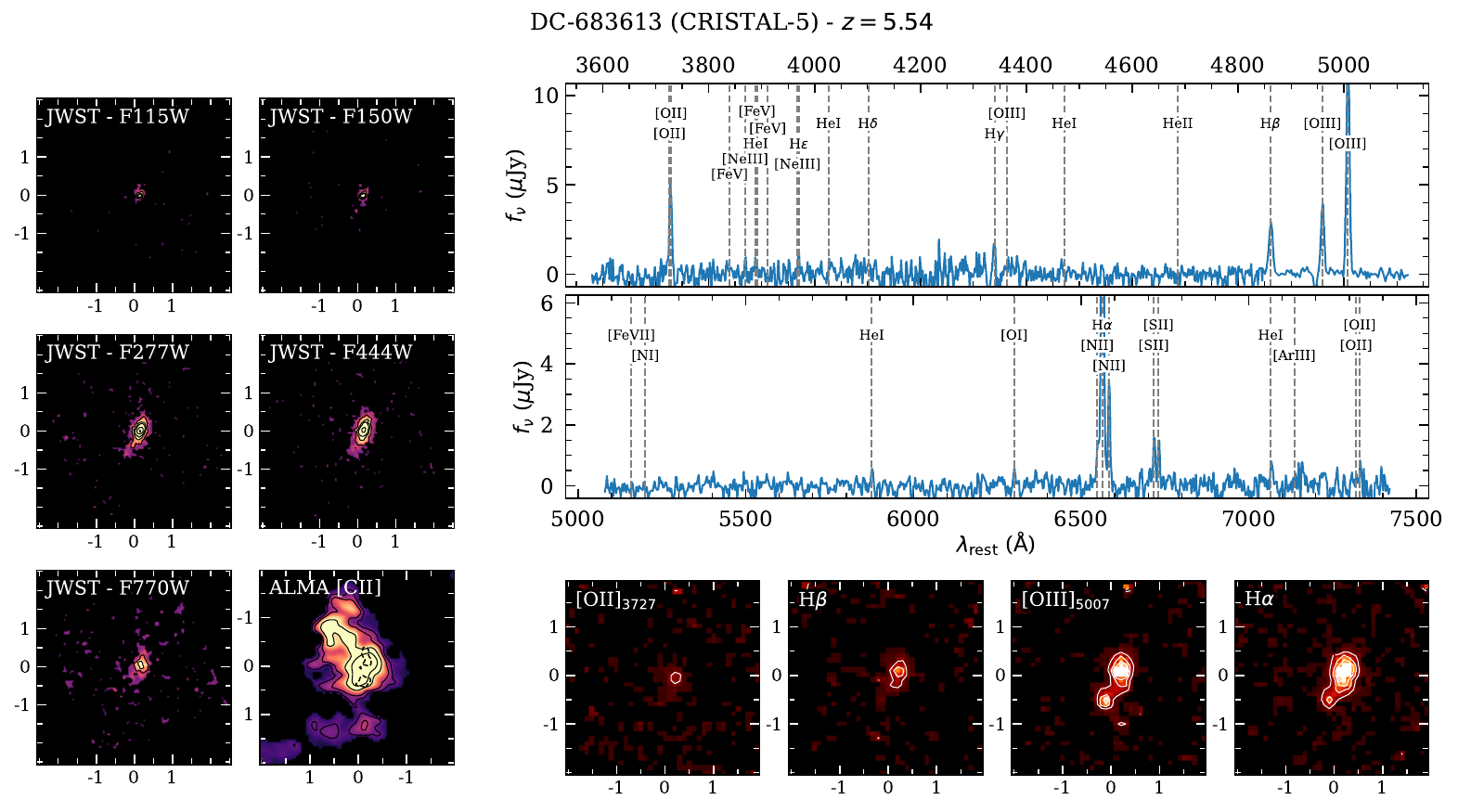}
\caption{Same as Figure~\ref{fig:eachgalaxy} but for targets {\it DC-630594} and {\it DC-683613}.
\label{fig:eachgalaxy3}}
\end{figure*}

\begin{figure*}[t!]
\centering
\includegraphics[angle=0,width=\textwidth]{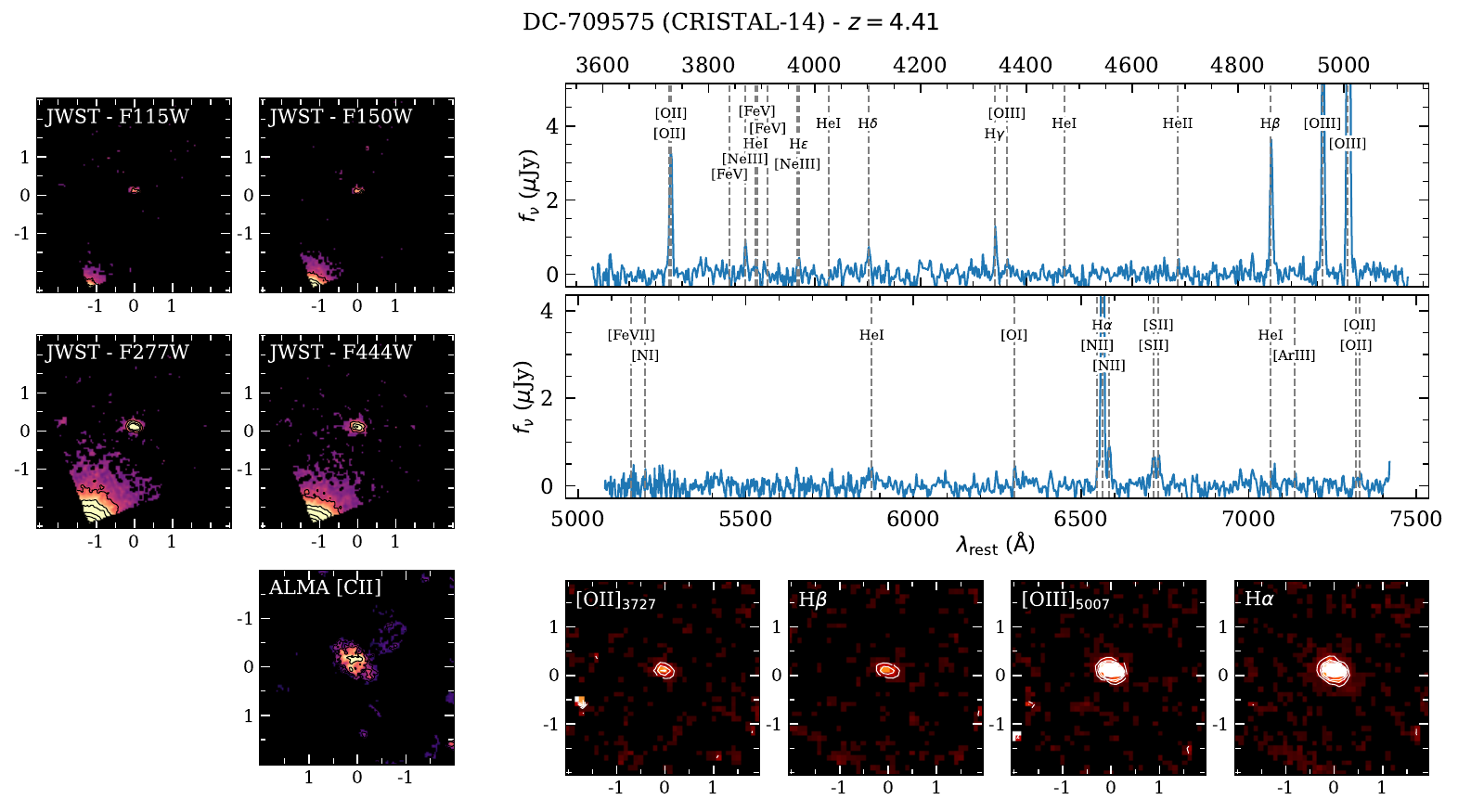}\\
\includegraphics[angle=0,width=\textwidth]{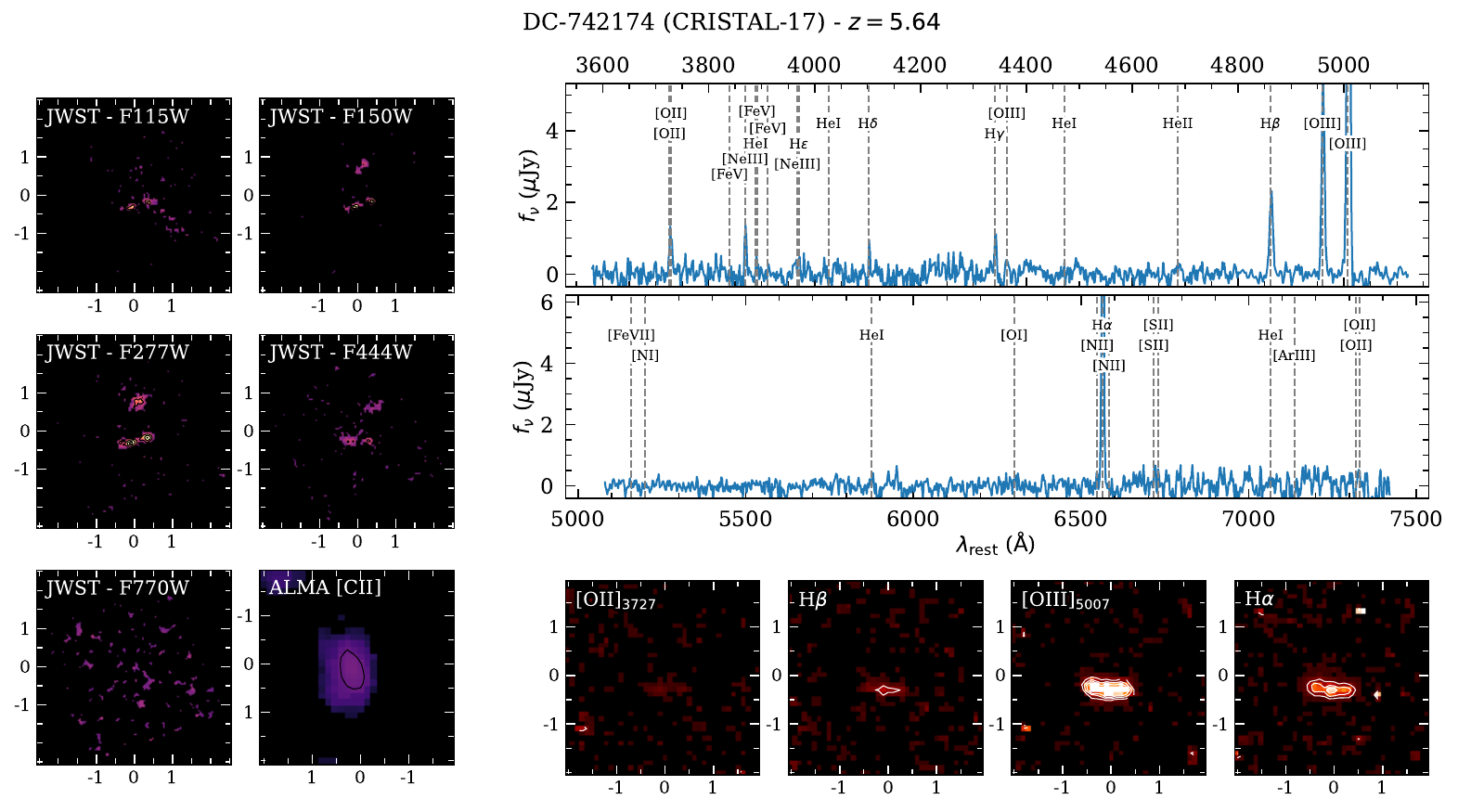}
\caption{Same as Figure~\ref{fig:eachgalaxy} but for targets {\it DC-709575} and {\it DC-742174}.
\label{fig:eachgalaxy4}}
\end{figure*}

\begin{figure*}[t!]
\centering
\includegraphics[angle=0,width=\textwidth]{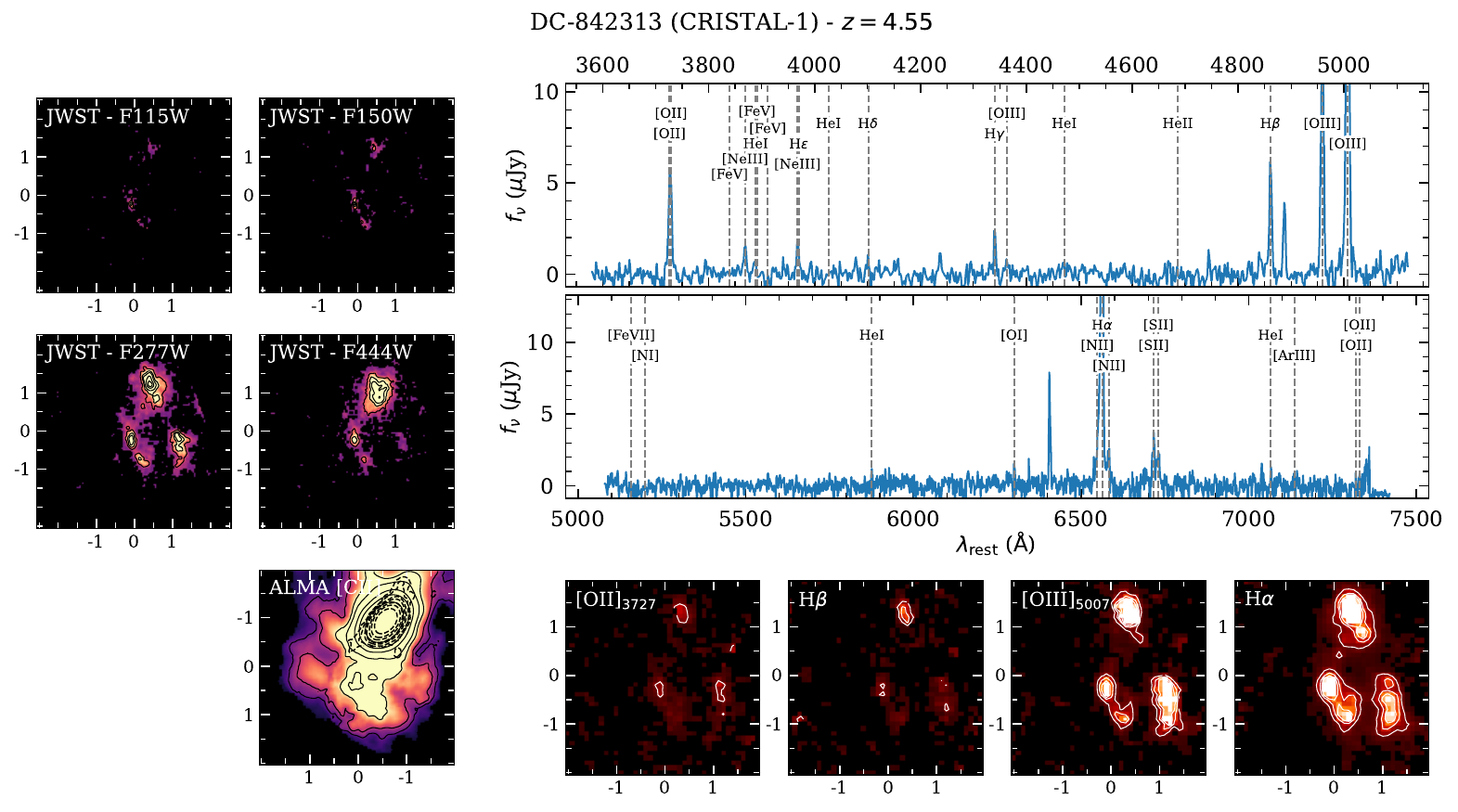}\\
\includegraphics[angle=0,width=\textwidth]{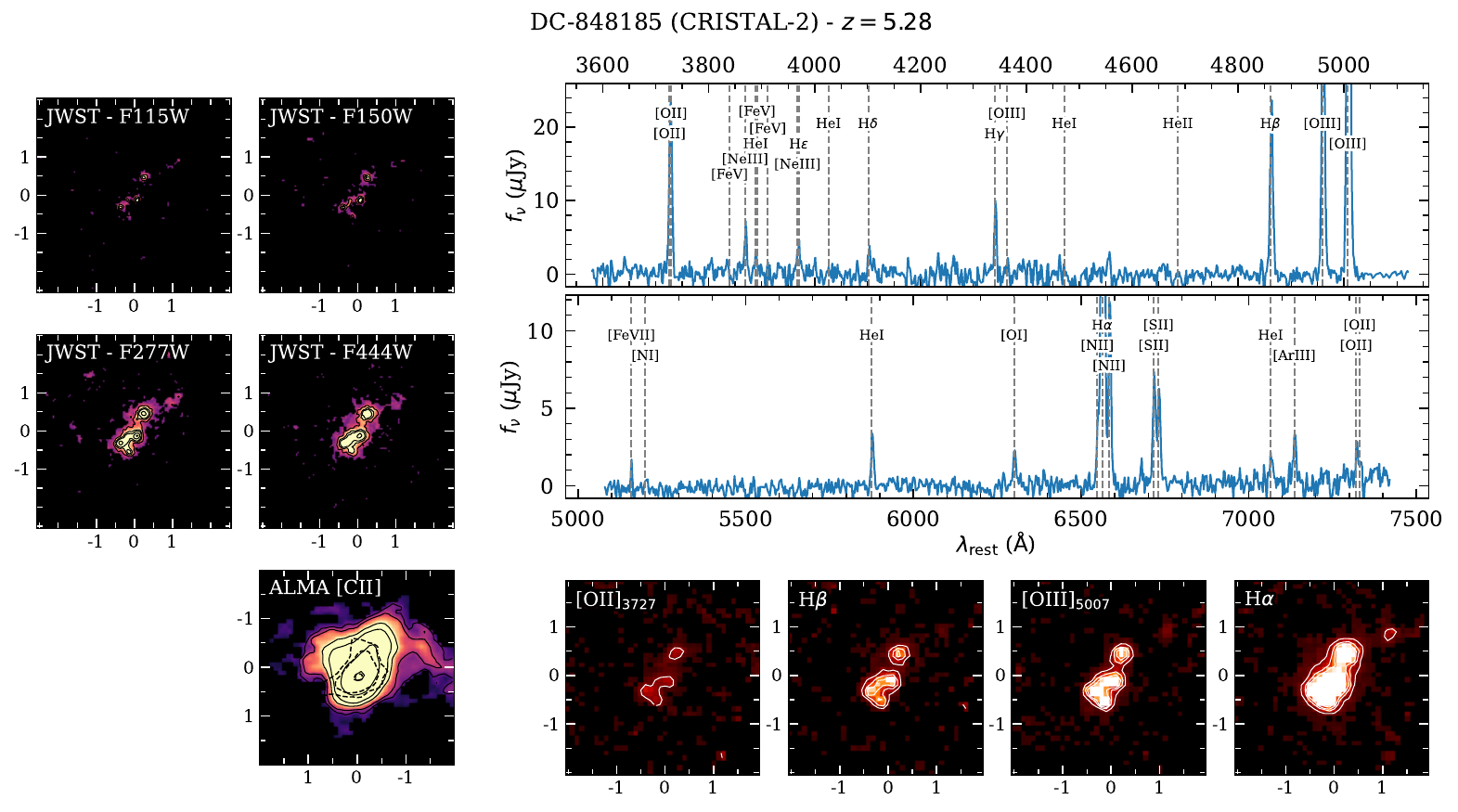}
\caption{Same as Figure~\ref{fig:eachgalaxy} but for targets {\it DC-842313} and {\it DC-848185}.
\label{fig:eachgalaxy5}}
\end{figure*}

\begin{figure*}[t!]
\centering
\includegraphics[angle=0,width=\textwidth]{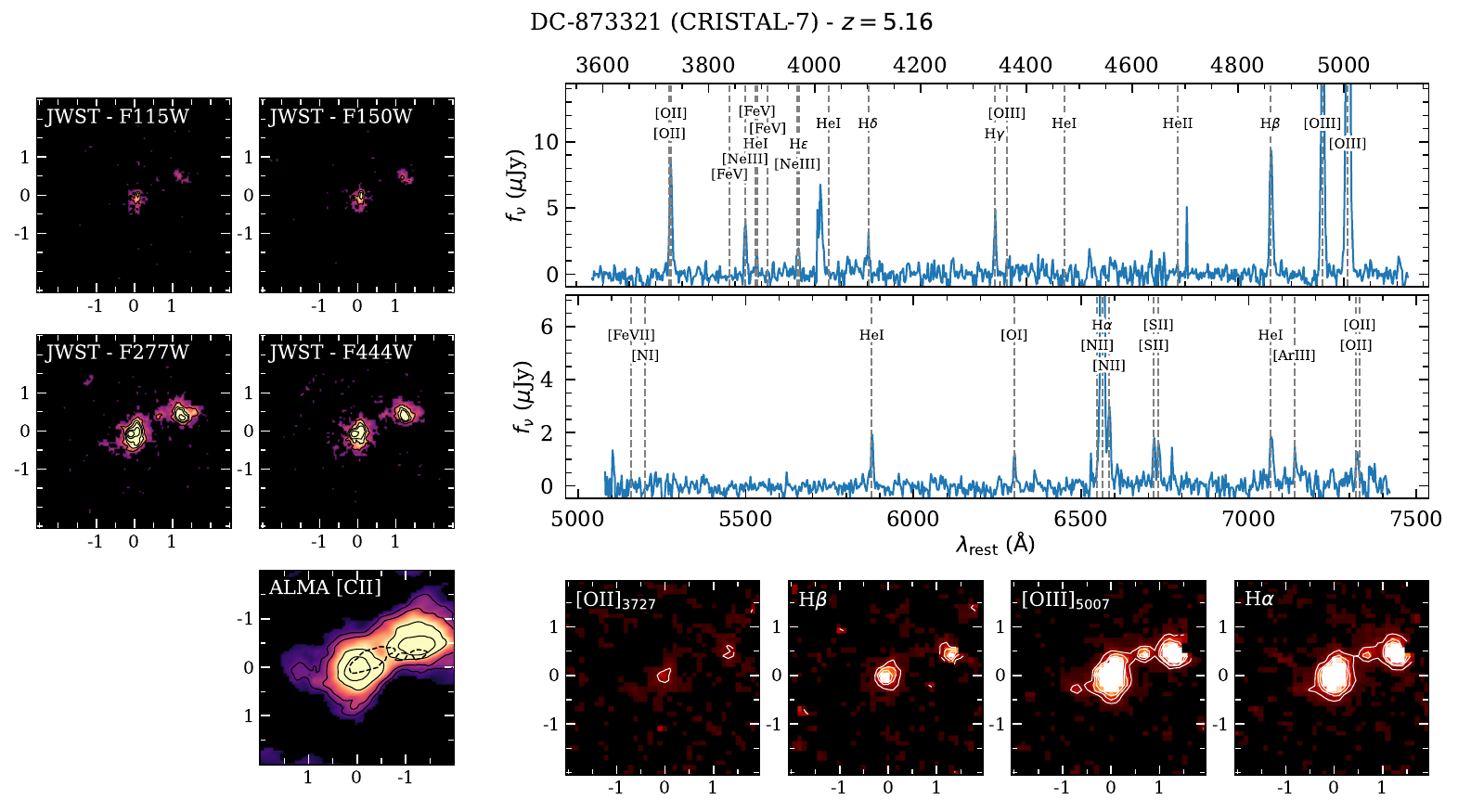}\\
\includegraphics[angle=0,width=\textwidth]{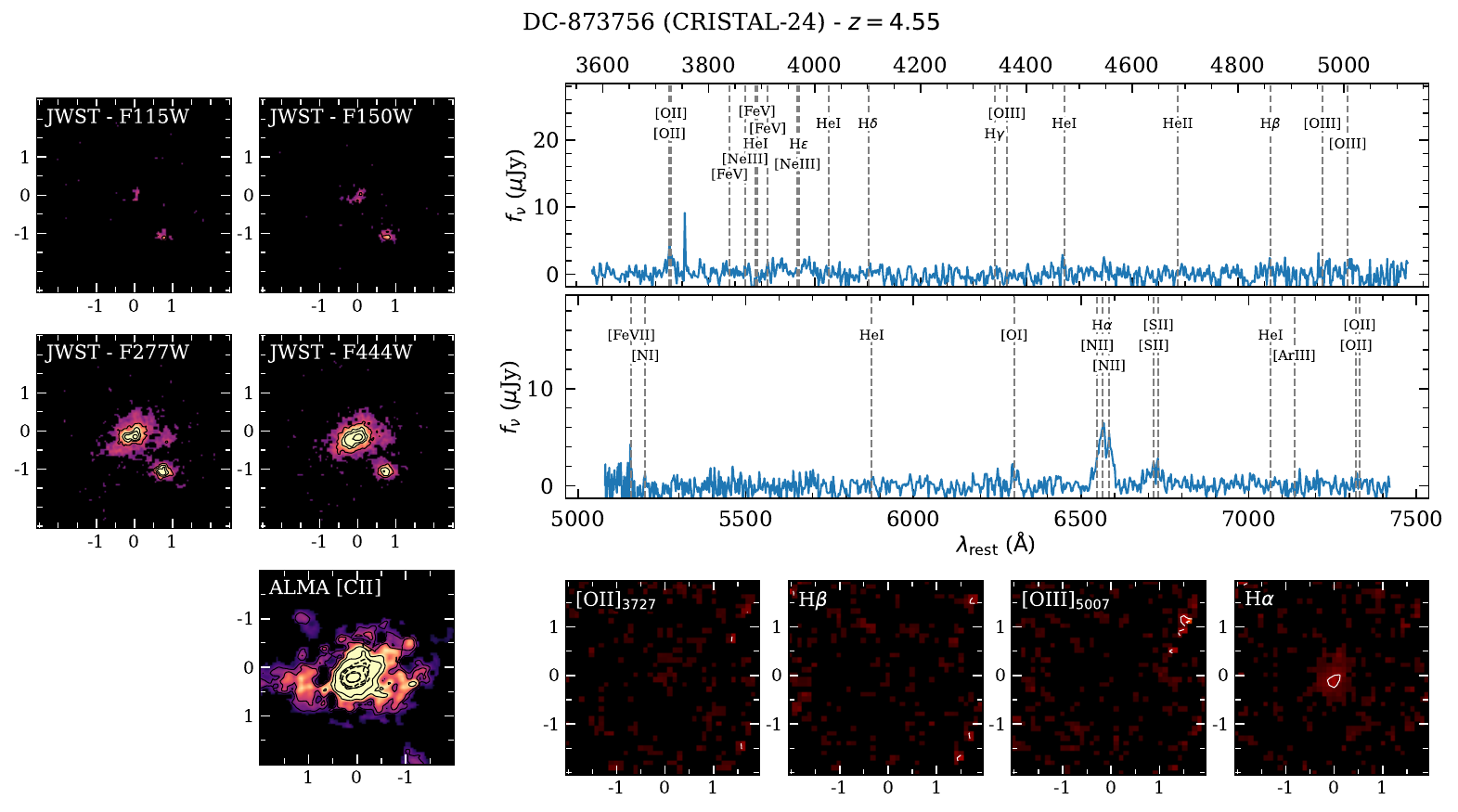}
\caption{Same as Figure~\ref{fig:eachgalaxy} but for targets {\it DC-873321} and {\it DC-873756}.
\label{fig:eachgalaxy6}}
\end{figure*}

\begin{figure*}[t!]
\centering
\includegraphics[angle=0,width=\textwidth]{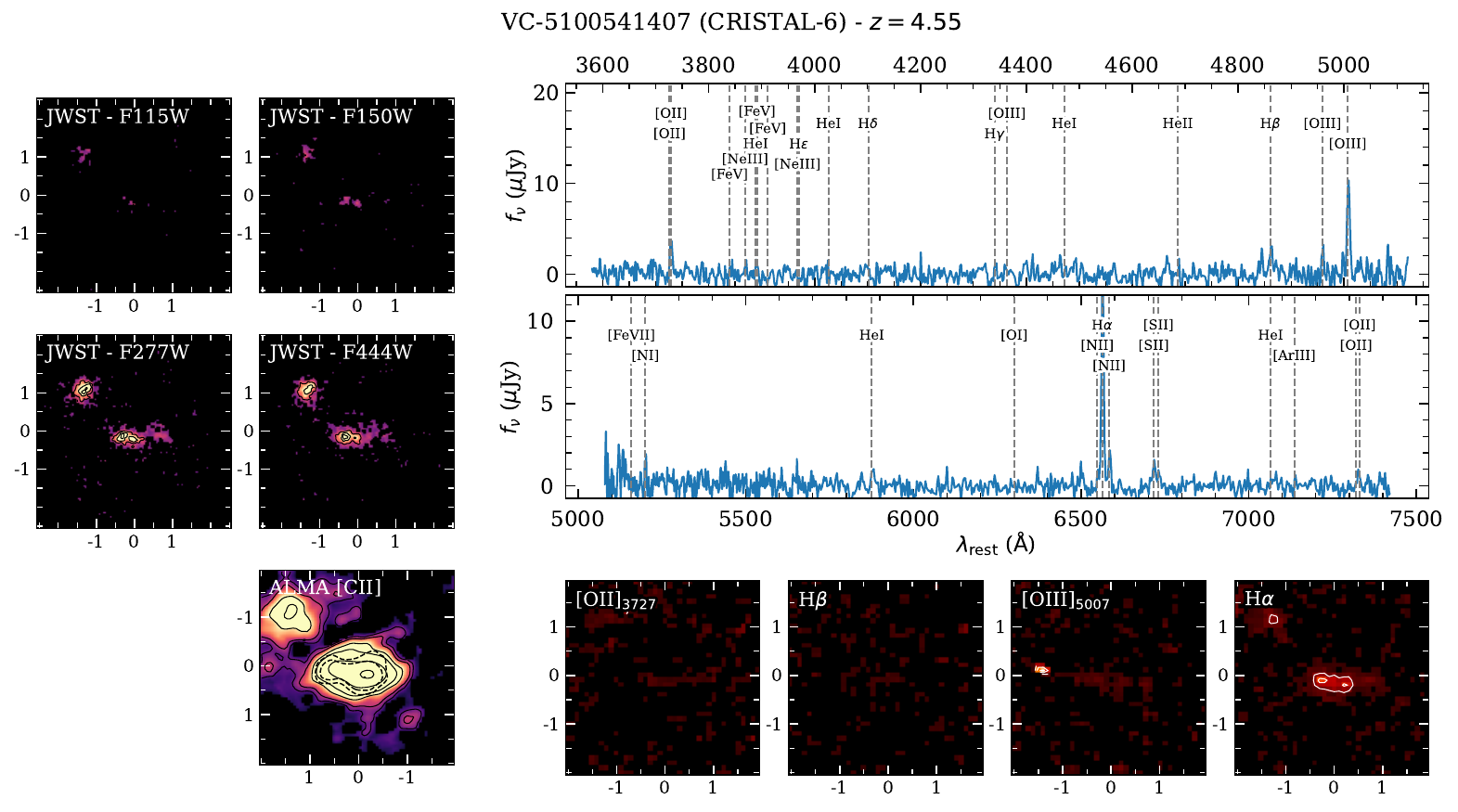}\\
\includegraphics[angle=0,width=\textwidth]{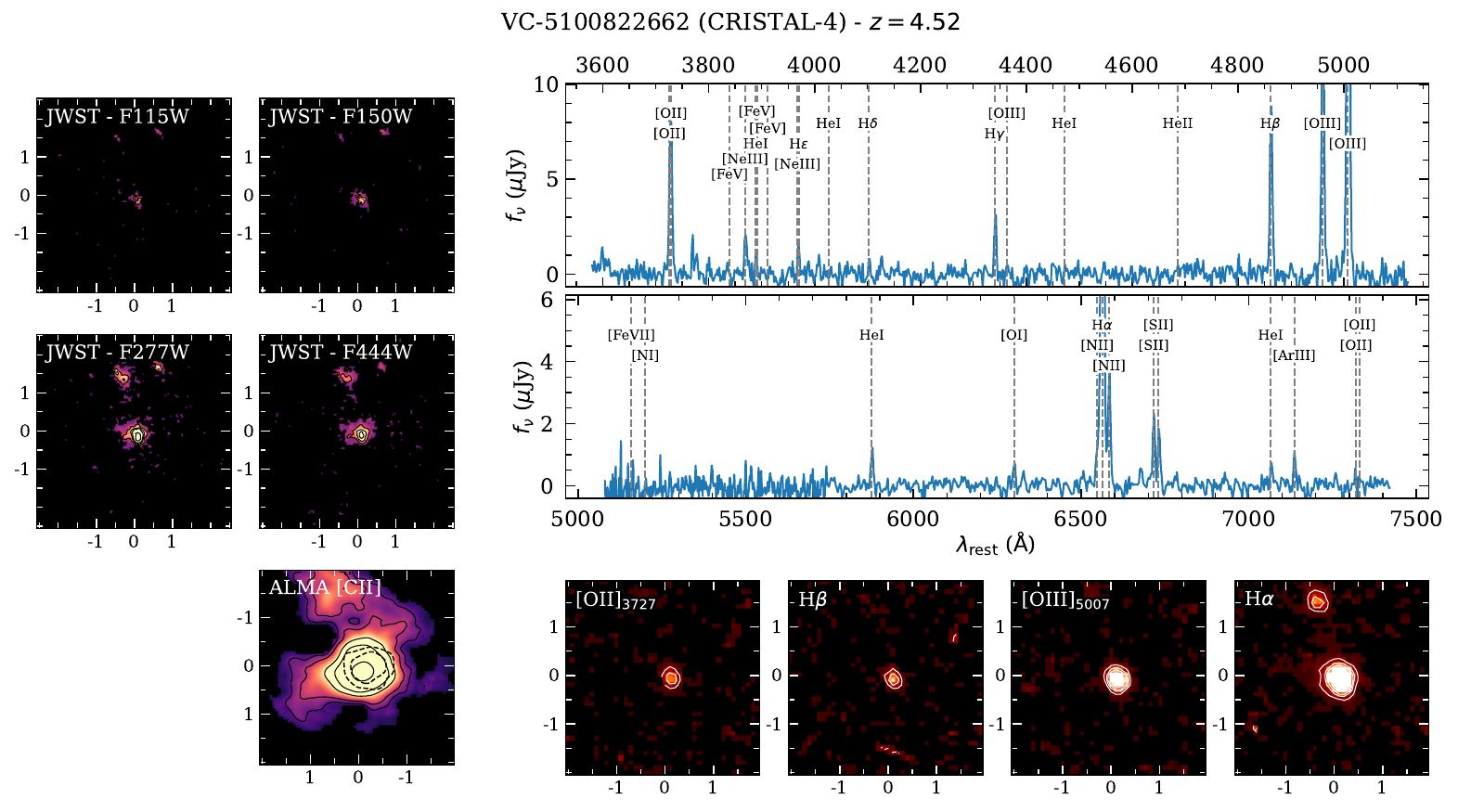}
\caption{Same as Figure~\ref{fig:eachgalaxy} but for targets {\it VC-5100541407} and {\it VC-5100822662}.
\label{fig:eachgalaxy7}}
\end{figure*}

\begin{figure*}[t!]
\centering
\includegraphics[angle=0,width=\textwidth]{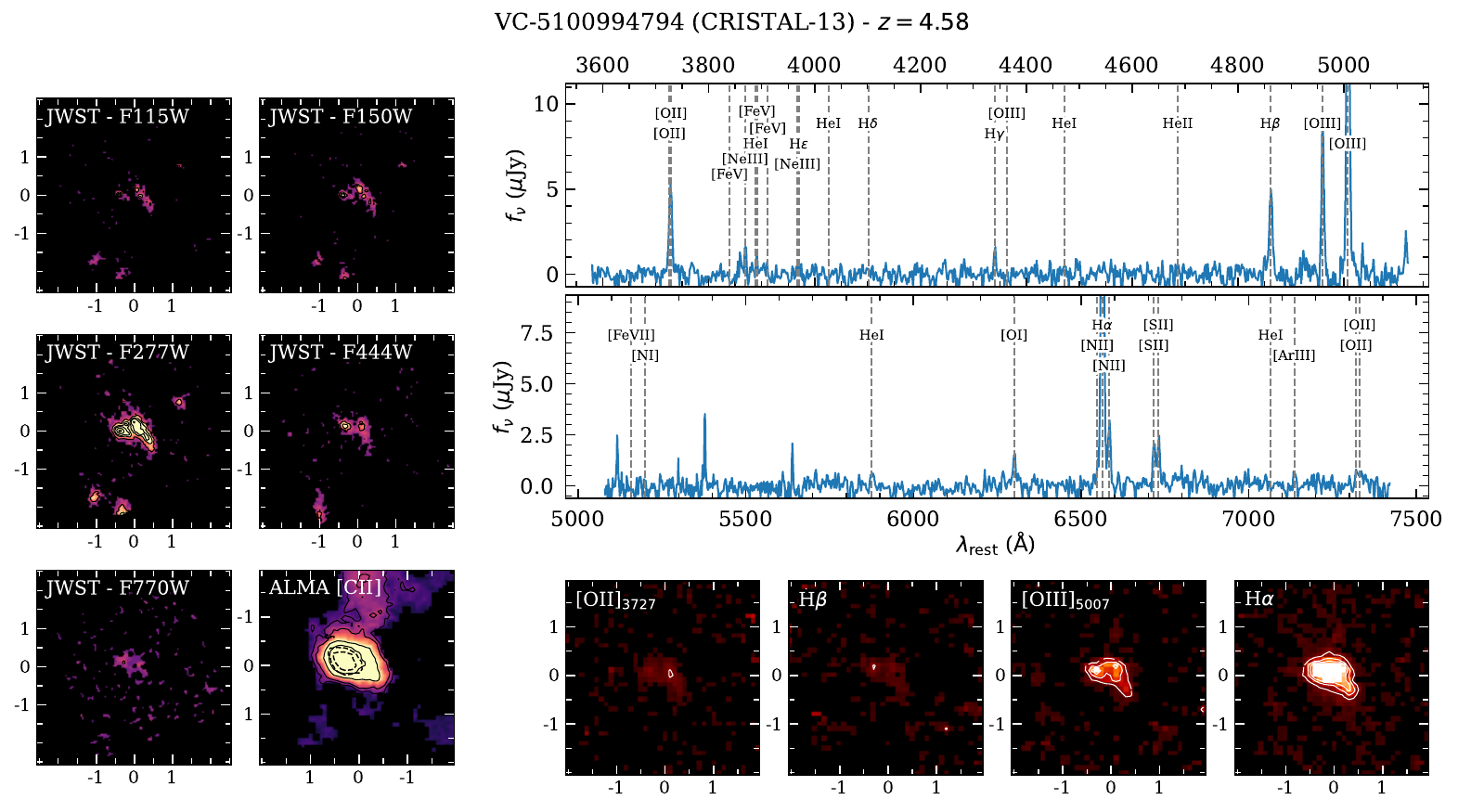}\\
\includegraphics[angle=0,width=\textwidth]{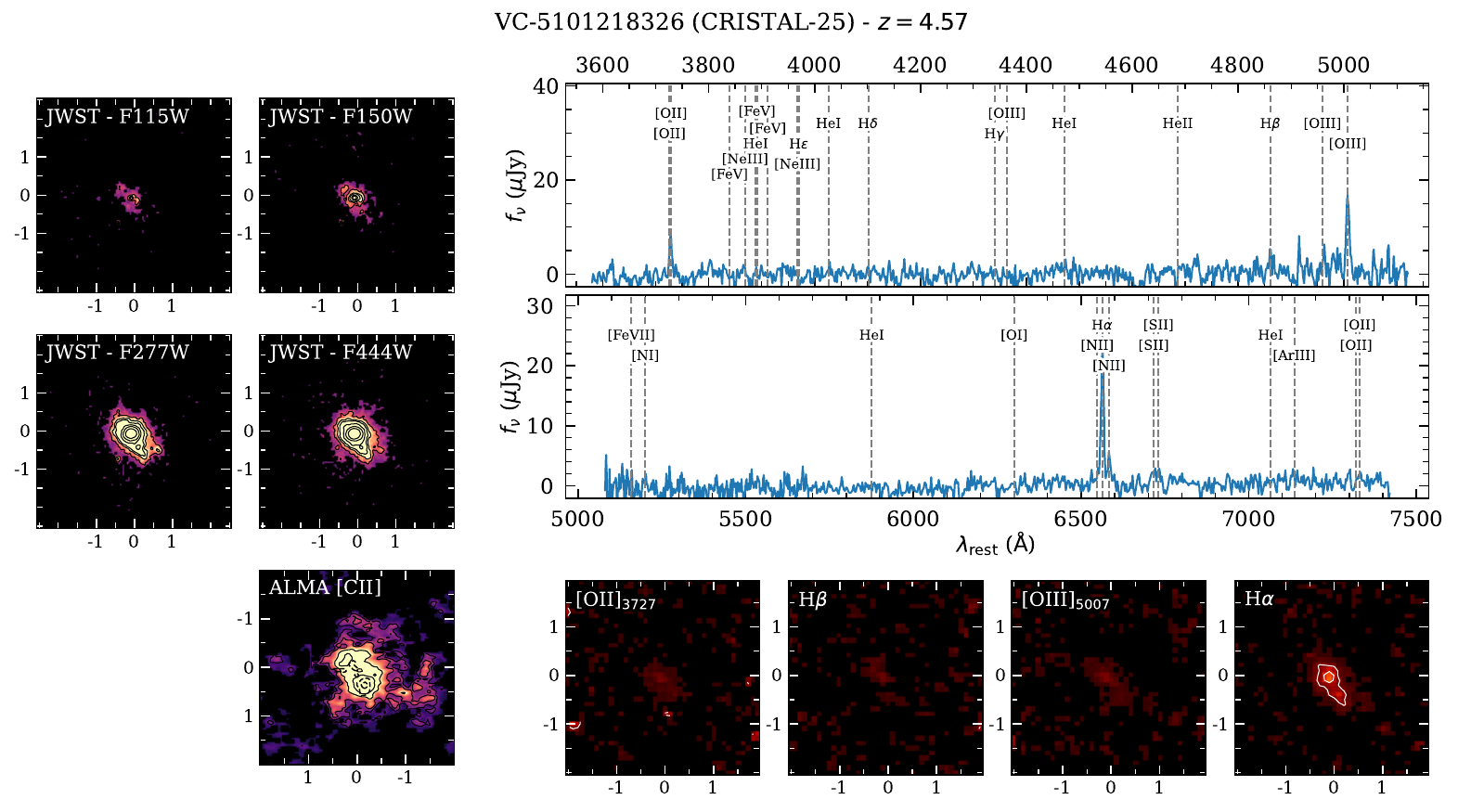}
\caption{Same as Figure~\ref{fig:eachgalaxy} but for targets {\it VC-5100994794} and {\it VC-5101218326}.
\label{fig:eachgalaxy8}}
\end{figure*}

\begin{figure*}[t!]
\centering
\includegraphics[angle=0,width=\textwidth]{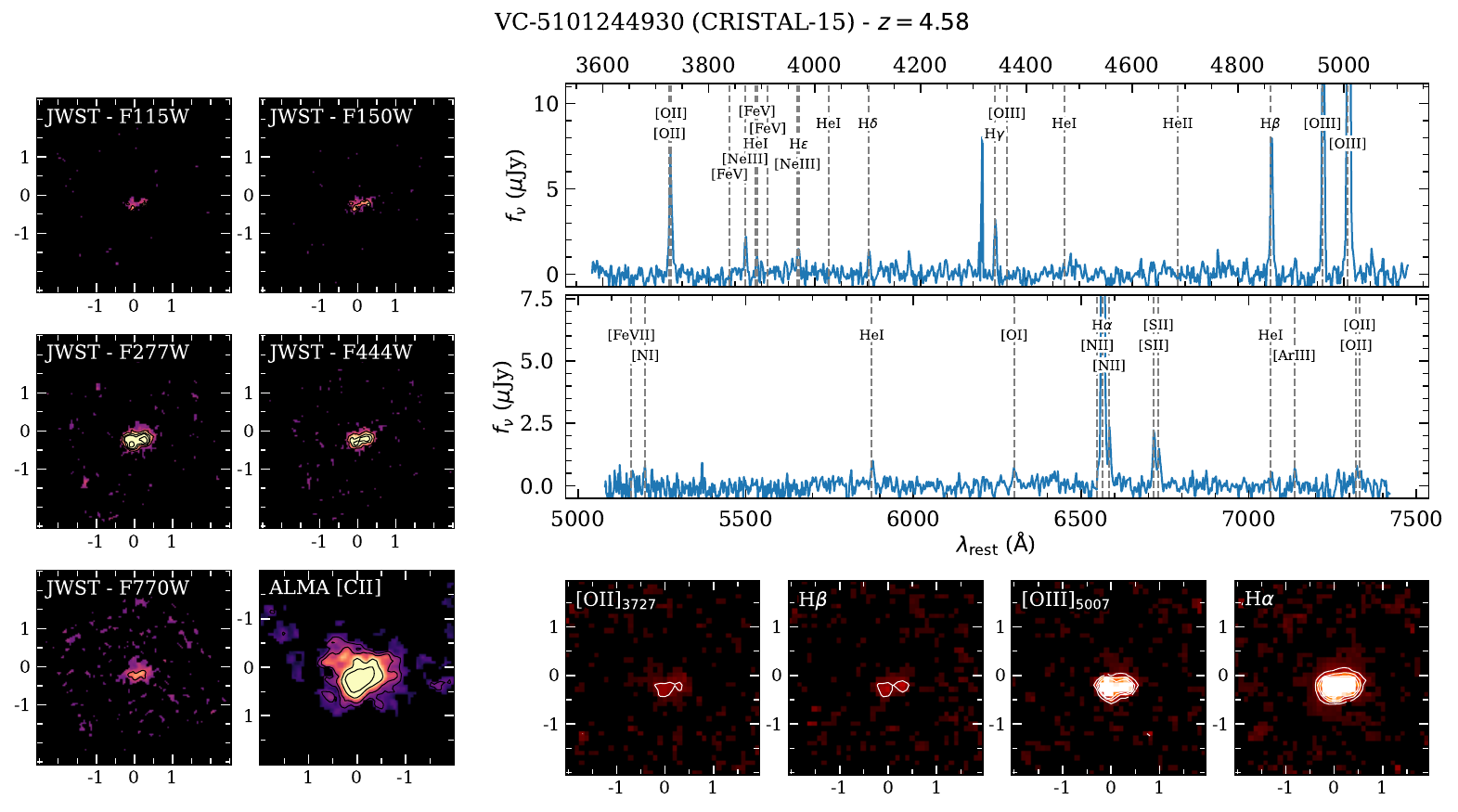}\\
\includegraphics[angle=0,width=\textwidth]{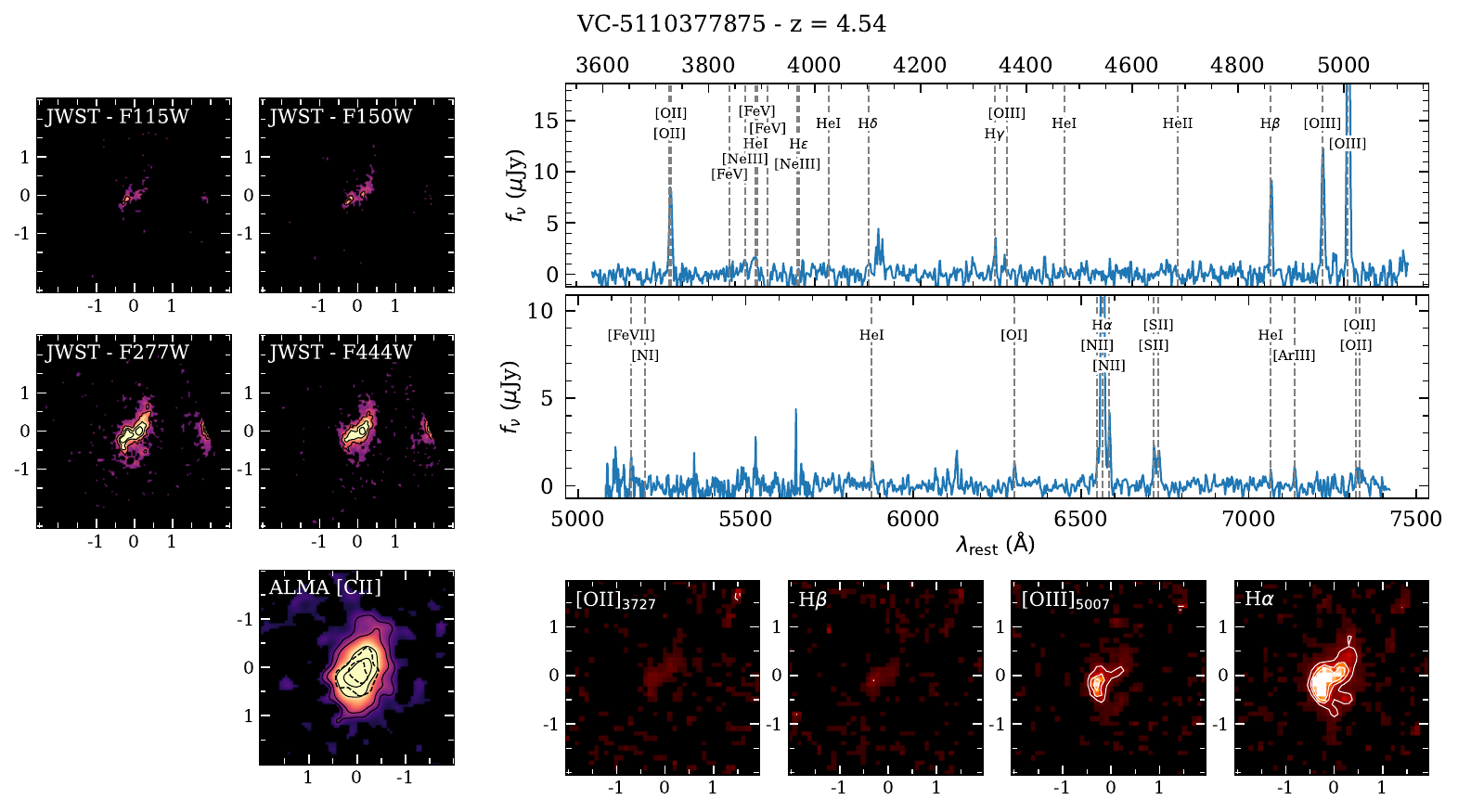}
\caption{Same as Figure~\ref{fig:eachgalaxy} but for targets {\it VC-5101244930} and {\it VC-5110377875}.
\label{fig:eachgalaxy9}}
\end{figure*}

Figures~\ref{fig:eachgalaxy} to~\ref{fig:eachgalaxy9} display various data taken by JWST, HST, and ALMA for each of the 18 ALPINE-CRISTAL JWST survey targets.

\newpage
\clearpage


\bibliography{bibli}{}
\bibliographystyle{aasjournalv7}



\allauthors
\end{document}